\documentclass[twocolumn,showpacs,showkeys,preprintnumbers,superscriptaddress,amsmath,floatfix,amssymb,secnumarabic,prb]{revtex4}

\usepackage[colorlinks=true]{hyperref}
\usepackage{graphicx}

\newcommand{\comment}[1]{}

\newcommand{\nm}{\, {\rm nm}}
\newcommand{\ev}{\, {\rm eV}}
\newcommand{\K}{\, {\rm K}}
\newcommand{\mev}{\, {\rm meV}}

\newcommand{\lr}[1]{ \left( #1 \right) }
\newcommand{\lrs}[1]{ \left[ #1 \right] }

\newcommand{\vev}[1]{ \langle \, #1 \, \rangle }
\newcommand{\cev}[1]{ \langle\langle \, #1 \, \rangle\rangle }

\newcommand{\tr}{ {\rm Tr} \, }
\newcommand{\re}{ {\rm Re} \, }

\newcommand{\ket}[1]{ \, | #1 \rangle }
\newcommand{\bra}[1]{ \langle #1 | \, }

\newcommand{\rot}{{\rm rot} \,}

\renewcommand{\det}[1]{ {\rm det} \left( #1 \right) }

\newcommand{\expa}[1]{ \exp{\left( #1 \right)} }

\newcommand{\logo}{\\ \vskip -18mm \leftline{\includegraphics[scale=0.3,clip=false]{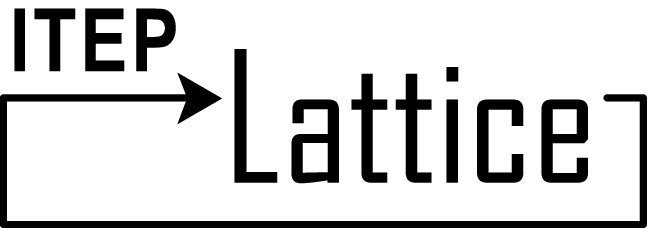}} \vskip 10mm}

\graphicspath{{./figures/paper/}}

\begin{document}
\sloppy
\preprint{ITEP-LAT/2012-06}

\title{Monte-Carlo study of the electron transport properties of monolayer graphene within the tight-binding model\logo}

\author{P. V. Buividovich}
\email{pavel.buividovich@physik.uni-regensburg.de}
\affiliation{Institute of Theoretical Physics, University of Regensburg, D-93053 Germany, Regensburg, Universit\"{a}tsstra{\ss}e 31}

\author{M. I. Polikarpov}
\email{polykarp@itep.ru}
\affiliation{ITEP, 117218 Russia, Moscow, B. Cheremushkinskaya str. 25}

\date{November 30, 2012}
\begin{abstract}
 We study the effect of Coulomb interaction between charge carriers on the properties of graphene monolayer, assuming that the strength of the interaction is controlled by the dielectric permittivity of the substrate on which the graphene layer is placed. To this end we consider the tight-binding model on the hexagonal lattice coupled to the non-compact gauge field. The action of the latter is also discretized on the hexagonal lattice. Equilibrium ensembles of gauge field configurations are obtained using the Hybrid Monte-Carlo algorithm. Our numerical results indicate that at sufficiently strong coupling, that is, at sufficiently small substrate dielectric permittivities $\epsilon \lesssim 4$, and at sufficiently small temperatures $T \lesssim 1 \cdot 10^4 \, \K$ the symmetry between simple sublattices of hexagonal lattice breaks down spontaneously and the low-frequency conductivity gradually decreases down to $20 - 30 \%$ of its weak-coupling value. On the other hand, in the weak-coupling regime (with $\epsilon \gtrsim 4$) the conductivity practically does not depend on $\epsilon$ and is close to the universal value $\sigma_0 = 1/4$.
\end{abstract}


\maketitle

\section*{Introduction}
\label{sec:introduction}

 Graphene, a two-dimensional crystal with hexagonal lattice formed by carbon atoms, has attracted a lot of attention in recent years both as a novel material with many unusual properties \cite{Geim:07:1, Novoselov:04:1} and as a unique laboratory which allows to study numerous quantum-field-theoretical phenomena in desktop-scale experiments (see \cite{Novoselov:05:1, Kim:05:1, Novoselov:09:1, Semenoff:11:1} for a review and further references).

 Electronic transport properties of graphene are of particular interest for industrial applications. Theoretical considerations within the tight-binding model of crystal lattice show that in the absence of interactions and at low energies charge carriers in graphene behave as massless Dirac fermions with an effective ``speed of light'' being equal to the Fermi velocity $v_F \approx c/300$ \cite{Wallace:47:1, Semenoff:84:1, Novoselov:09:1}. It turns out that the conductivity of these massless Dirac fermions takes a universal value $\sigma_0 = 1/4 \, e^2/\hbar$ in the limit of zero temperature and in the absence of interactions \cite{Miransky:02:1, Katsnelson:06:1, Drut:10:1, Drut:10:2}. This value, however, should strongly depend on the measurement procedure and on the geometry of a sample \cite{Katsnelson:06:1}.

 Due to the smallness of $v_F$, electromagnetic interactions between these Dirac fermions are well described by the instantaneous Coulomb potential. However, for the same reason the resulting effective field theory turns out to be strongly coupled with a coupling constant $\alpha = e^2/v_F \approx 300/137 \approx 2$. Thus one can expect that Coulomb interactions between electrons can significantly modify the properties of graphene such as the quasiparticle spectrum and the conductivity. For graphene on a substrate with dielectric permittivity $\epsilon$ Coulomb interaction is screened so that the coupling decreases by a factor $\frac{2}{\epsilon + 1}$. This provides a practical way to control the interaction strength.

 Experimental studies of the conductivity of suspended graphene \cite{Andrei:08:1, Bolotin:08:1}, for which the coupling constant is maximal, suggest the existence of a gap in the quasiparticle spectrum with the width of order of $10 \mev$ \cite{Andrei:08:1, Bolotin:08:1, Drut:10:1, Drut:10:2}. The opening of a gap due to strong Coulomb interactions for $\alpha \gtrsim 1$ is also supported by analytical calculations based on the solution of the gap equation \cite{Miransky:02:1, Leal:04:1, Gamayun:10:1} and on the strong-coupling expansion in lattice gauge theory \cite{Araki:10:1, Araki:10:2, Araki:12:1}. The transition to the gapped phase is likely to be of the second order \cite{Gamayun:10:1}. Within the effective theory of Dirac quasiparticles the opening of a gap in the spectrum is accompanied by a formation of the fermion chiral condensate $\vev{\bar{\psi} \psi}$ \cite{Semenoff:11:1, Son:07:1, Gamayun:10:1}. In terms of the original tight-binding lattice model, such condensate corresponds to the difference of the charge carrier densities on the two simple sublattices of the hexagonal lattice \cite{Semenoff:11:1, Araki:10:2, Araki:12:1}.

 On the other hand, more recent measurements \cite{Elias:11:1, Elias:12:1} indicate the absence of a gap in the quasiparticle spectrum of suspended graphene as well as logarithmic divergence of the Fermi velocity near the Fermi points, in agreement with analytical calculations based on renormalization-group techniques \cite{Shankar:94:1}, dynamical mean-field theory \cite{Jafari:09:1} and expansion in the large number of fermion flavors \cite{Son:07:1, Drut:08:1}.

 In view of such uncertainties both in the experimental measurements and analytical calculations, it seems natural to turn to the first-principle numerical methods, such as Monte-Carlo simulations on the lattice. This line of research has been actively pursued recently. In \cite{Lahde:09:1, Lahde:09:2, Lahde:09:3, Lahde:10:1, Lahde:11:1} the low-energy effective field theory of Dirac quasiparticles in graphene was studied numerically by using $2+1$-dimensional staggered lattice fermions coupled to $3+1$-dimensional non-compact Abelian lattice gauge field. A hint on the second-order semimetal-insulator phase transition at $\alpha_c = 1.11 \pm 0.06$ associated with the opening of a gap in the energy spectrum and spontaneous breaking of the chiral symmetry of Dirac fermions was found. In \cite{Hands:08:1, Hands:10:1} a similar model with staggered fermions and a contact interaction term instead of the Coulomb potential was studied, and a phase transition with respect to the coupling constant separating the gapless conducting weak-coupling phase and the gapped insulating strong-coupling phase was also observed. A study of the finite-temperature phase transition in this model was reported in \cite{Hands:11:1}. A phase transition of the Berezinskii-Kosterlitz-Thouless type was found at $T = 0.055\lr{2} \, \Delta_0$, where $\Delta_0$ is the width of the gap in the spectrum.

 A common strategy of the works \cite{Lahde:09:1, Lahde:09:2, Lahde:09:3, Lahde:10:1, Lahde:11:1, Hands:08:1, Hands:10:1, Hands:11:1} was to use the value of the fermion condensate as an order parameter. Recently, the results of the direct numerical measurements of the conductivity of graphene lattice effective field theory with staggered fermions were reported in \cite{Ulybyshev:12:1}. It was found that the DC conductivity obtained from the Green-Kubo relations indeed rapidly decreases when the fermion condensate is formed, in agreement with theoretical expectations.

 The effect of lattice artifacts of staggered fermions on the flavor symmetry breaking in graphene effective field theory was discussed in \cite{Giedt:11:1}, with the conclusion that lattice simulations with staggered fermions might yield somewhat lower value of the critical coupling constant than in the continuum graphene effective field theory. Thus the influence of such lattice artifacts could probably shift the critical coupling constant of the semimetal-insulator phase transition above the value of the coupling constant in suspended graphene. Such a shift would then explain the fact that the insulating state of the suspended monolayer graphene was not observed experimentally \cite{Elias:11:1, Elias:12:1}.

 Lattice regularization of the effective field theory of graphene used in \cite{Lahde:09:1, Lahde:09:2, Lahde:09:3, Lahde:10:1, Lahde:11:1, Hands:08:1, Hands:10:1, Hands:11:1, Giedt:11:1, Ulybyshev:12:1} involves two approximations: one first starts from the tight-binding lattice model of the electron transport in graphene \cite{Novoselov:09:1, Wallace:47:1, Metalidis:07:1} and derives the low-energy effective theory of Dirac fermions. This theory is then again discretized on the lattice by using suitable lattice fermions which reproduce the Dirac spectrum at low energies. However, since the original model is already formulated on the hexagonal lattice, it is tempting to circumvent these two approximations and perform direct simulations of the tight-binding model with Coulomb interactions included. This possibility has been recently discussed in \cite{Rebbi:12:1, Rebbi:11:1}. Such simulations, while technically being even simpler than simulations with staggered fermions, have several crucial advantages. First, they allow to study the patterns of spontaneous symmetry breaking which are specific for the hexagonal lattice, such as the Kekule distortion. Despite the fact that sublattice symmetry is not broken in this case, a gap in the spectrum might develop \cite{Nomura:09:1, Araki:12:1}. Second, since the lattice spacing is fixed in the tight-binding model, simulation results can be unambiguously compared with experimental data. Finally, all the symmetries of the tight-binding model are explicitly preserved. As discussed in Section \ref{sec:sublat_symm}, the latter has a $U\lr{1} \otimes U\lr{1}$ symmetry associated with the conservation of the total numbers of charge carriers with different spins as well as the discrete sublattice symmetry. Sublattice symmetry can be explicitly broken by the staggered potential $m$, which plays the role of the Dirac mass at low energies. In contrast, massive staggered fermions have only single global $U\lr{1}$ symmetry associated with total charge conservation \cite{Burden:87:1, Creutz:07:1, Giedt:11:1}. Thus simulations of the tight-binding model are free from lattice artifacts and can serve as a completely independent cross-check of simulations with staggered fermions. In particular, one can estimate the influence of lattice artifacts of staggered fermions on the simulation results.

 In this paper we report on the results of such direct lattice Monte-Carlo simulations of the tight-binding model of graphene on hexagonal lattice with Coulomb interactions. To account for the latter, we couple the tight-binding model to the $3+1$-dimensional non-compact Abelian lattice gauge field, as in \cite{Lahde:09:1, Lahde:09:2, Lahde:09:3, Lahde:10:1, Lahde:11:1, Hands:08:1, Hands:10:1, Hands:11:1, Giedt:11:1, Ulybyshev:12:1}. In contrast to the approach of \cite{Rebbi:12:1, Rebbi:11:1}, where the interactions are treated by applying the Hubbard-Stratonovich transformation, the resulting action is local and gauge fields outside of graphene plane can be efficiently sampled by a heat-bath algorithm \cite{MontvayMuenster}. As we demonstrate in Subsection \ref{subsec:em_field_action}, our discretization of electric field on the hexagonal lattice reproduces the continuum Coulomb potential with a very good precision. An additional advantage of the use of $3+1$-dimensional gauge fields is the possibility to study electromagnetic interactions of multilayered graphene, for example, Casimir forces.

 We focus on the study of spontaneous breaking of sublattice symmetry as well as on the direct measurements of electric conductivity of graphene. In agreement with the results of \cite{Lahde:09:1, Lahde:09:2, Lahde:09:3, Lahde:10:1, Lahde:11:1, Hands:08:1, Hands:10:1, Hands:11:1, Giedt:11:1, Ulybyshev:12:1}, we find that sublattice symmetry is spontaneously broken for coupling constants $\alpha \gtrsim 1$, which corresponds to the substrate dielectric permittivities $\epsilon \lesssim 4$, and for sufficiently small temperatures $T \lesssim 1.3 \cdot 10^4 \, \K$. At higher temperatures sublattice symmetry is not broken for the physical values of $\epsilon$, $\epsilon \ge 1$.

 It should be stressed, however, that in this paper we only consider the tight-binding model at finite temperature with the purpose of finding the range of lattice parameters which describe the low-temperature phase of this model. For this reason we also do not take into account the thermal fluctuations of the hexagonal lattice itself, which should become irrelevant for sufficiently low temperatures. As we will demonstrate below (see Subsection \ref{subsec:latt_params}), realistic lattice parameters correspond to quite high temperatures of the electron gas of order of $1 \ev \sim 10^4 \K$. Simulations at significantly lower temperatures are computationally very expensive, and it is important to give an upper bound for the temperatures which still describe the low-temperature phase in order to make an optimal choice of lattice parameters for practical simulations. As we will show in Subsection \ref{subsec:sublat_symm_num_res}, the critical temperature which separates the low- and the high-temperature phases of the tight-binding model can be estimated as $T_c \approx 1.3 \cdot 10^4 \K$. Thus one can hope that for the smallest temperature which we use in our simulations, $T = 8.8 \cdot 10^3 \K$, the low-temperature properties of the tight-binding model are already reproduced with a sufficiently good precision. Since at temperatures of order of $10^4 \K$ a real graphene monolayer would be destroyed due to thermal fluctuations of the lattice \cite{Katsnelson:11:2}, a detailed study of the finite-temperature phase transition within the tight-binding model (without taking into account the phononic degrees of freedom) seems to be of purely academic interest only.

 We extract the conductivity from the correlators of electric current densities with the help of the Green-Kubo relations. In order to estimate the conductivity in a model-independent way, we consider the low-frequency conductivity $\bar{\sigma}$ smeared over frequencies $w \lesssim T$. We observe that in the strong-coupling regime the low-frequency conductivity quickly decreases with $\epsilon$ down to $20-30 \%$ of its weak-coupling value at all temperatures which we have considered. On the other hand, in the weak-coupling regime ($\epsilon \gtrsim 4$) the conductivity practically does not depend on $\epsilon$.

 The outline of the paper is the following. In Section \ref{sec:lattice_action} we give a detailed description the geometry of hexagonal lattice and its extension to the $\lr{3 + 1}$-dimensional space, and describe the lattice actions for the gauge field and for fermions which we use in our simulations. The details of our simulation algorithm and the choice of lattice parameters are also discussed in this Section. In Section \ref{sec:sublat_symm} we discuss spontaneous breaking of sublattice symmetry. Our numerical measurements of the conductivity are summarized in Section \ref{sec:conductivity}. In Section \ref{sec:conclusions} we give some concluding remarks on our results. Some technical details and supplementary material (such as the calculation of the current-current correlator for the non-interacting tight-binding model) are relegated to the Appendices.

\section{Lattice action and simulation method}
\label{sec:lattice_action}

\subsection{Tight-binding model of graphene with electromagnetic interactions}
\label{subsec:action_defs}

 With a good precision the electronic properties of graphene can be described by the tight-binding Hamiltonian \cite{Wallace:47:1, Novoselov:09:1}
\begin{eqnarray}
\label{tb_hamiltonian}
\hat{H}_{tb} = -\kappa \, \sum\limits_{\sigma = \uparrow, \downarrow} \sum\limits_{<XY>}
\lr{\hat{a}^{\dag}_{\sigma, X} \, \hat{a}_{\sigma, Y}
+
\hat{a}^{\dag}_{\sigma, Y} \, \hat{a}_{\sigma, X}}  ,
\end{eqnarray}
where summation goes over all neighboring sites $X$, $Y$ on the hexagonal lattice with hexagon side $a = 0.142 \, \nm$ and $\kappa \approx 2.7 \ev$ is the hopping energy for carbon $\pi$ orbitals. $\hat{a}^{\dag}_{\sigma, X}$ and $\hat{a}_{\sigma, X}$ are the creation and annihilation operators for non-relativistic electrons with spin $\sigma = \uparrow, \downarrow$. This Hamiltonian describes electron hopping between nearest neighbor atoms only. The hopping energy for hopping between next to nearest neighbor atoms is much smaller than $\kappa$ \cite{Novoselov:09:1} and we neglect it here.

 Since in the ground state graphene is electrically neutral, there should be on average one electron per lattice site. We assume that the ground state of the free tight-binding Hamiltonian (\ref{tb_hamiltonian}) is fixed by the conditions \cite{Rebbi:12:1, Rebbi:11:1, Semenoff:11:1}
\begin{eqnarray}
\label{free_ground_state}
\hat{a}_{\uparrow, X} \, \ket{0} = 0, \quad
\hat{a}^{\dag}_{\downarrow, X} \, \ket{0} = 0 .
\end{eqnarray}
Thus there is one electron with spin $\sigma = \downarrow$ at each lattice site. For the ground state fixed by (\ref{free_ground_state}), it is convenient to define the creation and annihilation operators for ``particles'' and ``holes'' as \cite{Rebbi:12:1, Rebbi:11:1}
\begin{eqnarray}
\label{operator_redef}
 \hat{\psi}_{\uparrow, X} = \hat{a}_{\uparrow, X}  , \quad
 \hat{\psi}_{\downarrow, X} = \pm \hat{a}^{\dag}_{\downarrow, X}  .
\end{eqnarray}
In the definition of $\hat{\psi}_{\downarrow, X}$, we take the plus sign for lattice sites which belong to one simple rhombic sublattice of the graphene hexagonal lattice and the minus sign for lattice sites on another simple sublattice \cite{Rebbi:11:1, Rebbi:12:1}. Now the ground state satisfies the standard condition $\hat{\psi}_{\uparrow, X} \ket{0} = 0$, $\hat{\psi}_{\downarrow, X} \ket{0} = 0$ and the charge operator reads
\begin{eqnarray}
\label{charge_redef}
 \hat{q}_X =
 \hat{\psi}^{\dag}_{\uparrow, X} \, \hat{\psi}_{\uparrow, X}
 -
 \hat{\psi}^{\dag}_{\downarrow, X} \, \hat{\psi}_{\downarrow, X} .
\end{eqnarray}
In other words, we interpret the absence of electron at some lattice site as the positively charged hole, and valence electrons in graphene play the role of the Dirac sea. Obviously, the tight-binding Hamiltonian in terms of the new operators has the same form as (\ref{tb_hamiltonian}) with an additional shift of energy which does not affect physical results. It should be stressed that since in the partition function (and hence also in lattice Monte-Carlo simulations) we anyway sum over all possible states of the theory, the particular choice of the ``perturbative'' vacuum state (\ref{free_ground_state}) is only a matter of convenience.

 Coulomb interaction is described by the interaction Hamiltonian
\begin{eqnarray}
\label{interaction_hamiltonian}
 \hat{H}_I = \frac{1}{2} \, \sum\limits_{X,Y} e^2/r\lr{X, Y} \, \hat{q}_X \, \hat{q}_Y ,
\end{eqnarray}
where $r\lr{X, Y}$ is the distance between lattice sites $X$ and $Y$, $e^2 \approx 1/137$ is the electron charge. Throughout the paper we use the natural system of units with $c = \hbar = k_B = 1$. A common way to simulate theories with four-fermion interaction of the form (\ref{interaction_hamiltonian}) is to apply the Hubbard-Stratonovich transformation and to sample the fictitious Hubbard-Stratonovich field. For the long-ranged three-dimensional Coulomb potential (\ref{interaction_hamiltonian}) the resulting action is nonlocal \cite{Rebbi:12:1, Rebbi:11:1}.

 Here we adopt a different strategy and consider the tight-binding model (\ref{tb_hamiltonian}) coupled to the real electromagnetic field. This coupling is introduced by using the standard Peierls substitution within the tight-binding Hamiltonian (\ref{tb_hamiltonian}) \cite{Peierls:33:1, Hofstadter:76:1, Metalidis:07:1}:
\begin{eqnarray}
\label{peierls_substitution}
 \hat{a}^{\dag}_{\sigma, X} \, \hat{a}_{\sigma, Y}
 \rightarrow
 \hat{a}^{\dag}_{\sigma, X} \,
 \expa{i \hat{\theta}_{XY}}
 \, \hat{a}_{\sigma, Y}  ,
\end{eqnarray}
where $\hat{\theta}_{XY} \equiv -\hat{\theta}_{YX} = e \int\limits_{X}^{Y} dx^i \hat{A}_i$ is the operator of the integral of the electromagnetic vector potential $A_i$ along the lattice bond which joins the sites $X$ and $Y$.

 Furthermore, we use the operators $\hat{\theta}_{XY}$ and the momenta canonically conjugate to them to approximate the Hamiltonian of the electromagnetic field in continuous space
\begin{eqnarray}
\label{em_field_hamiltonian_continuum}
 \hat{H}_{em} = \frac{1}{8 \pi} \, \int d^3 \vec{r}
 \lr{
    \hat{E}^2\lr{\vec{r}}
 +  \lr{\rot \hat{A}\lr{\vec{r}} }^2
 }  ,
\end{eqnarray}
where $\hat{E}_i$, $i = 1,2,3$ is the operator of the electric field strength, and to construct the corresponding lattice action. It turns out that such a discretization of the electromagnetic field reproduces the continuum Coulomb potential with a very good precision (see Fig. \ref{fig:coulomb_test} in Subsection \ref{subsec:em_field_action}). The Hamiltonian (\ref{em_field_hamiltonian_continuum}) should be also supplemented with the Gauss law constraint
\begin{eqnarray}
\label{gauss_law_continuum}
\nabla \hat{E}\lr{\vec{r}} =
4 \pi \, e \, \sum\limits_X \, \hat{q}_X \, \delta\lr{\vec{r}, \vec{r}_X}  .
\end{eqnarray}

\begin{figure}[h!tb]
  \includegraphics[width=7cm]{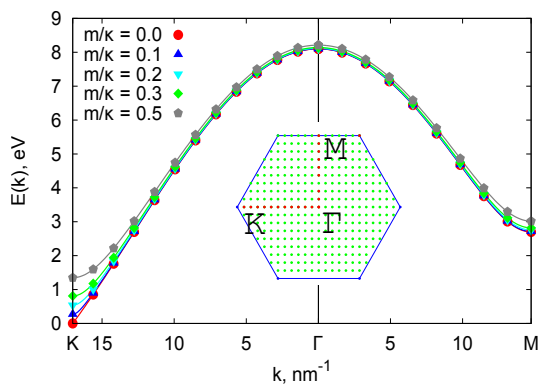}\\
  \caption{Dispersion relation $E\lr{\vec{k}}$ for the tight-binding Hamiltonian (\ref{tb_hamiltonian_peierls}) with the staggered potential of strength $m$ at different ratios $m/\kappa$. The points on the plot and the inset illustrate the filling of the graphene Brillouin zone with discrete lattice momenta for $18 \times 18$ lattice with periodic boundary conditions.}
  \label{fig:spectrum}
\end{figure}

 Let us also note that the spectrum of the tight-binding Hamiltonian (\ref{tb_hamiltonian}) has two zero-energy states which correspond to the Fermi points (see Appendix \ref{appsec:spectrum} for a more detailed discussion). These zero modes might result in certain singularities of the effective action of electromagnetic field, namely, in the appearance of zero modes of the fermionic hopping operator. These zero modes make standard numerical simulation methods such as Hybrid Monte-Carlo inapplicable \cite{MontvayMuenster, Lahde:09:1, Lahde:09:2, Hands:08:1}. In order to prohibit the existence of zero modes, it is convenient to introduce the staggered potential which is equal to $+ \, m_{\sigma} \, \hat{a}^{\dag}_{\sigma, X} \, \hat{a}_{\sigma, X}$ for lattice sites which belong to one simple sublattice of the hexagonal lattice and $- \, m_{\sigma} \, \hat{a}^{\dag}_{\sigma, X} \, \hat{a}_{\sigma, X}$ for sites of the other sublattice. As we will see from what follows, for the purpose of numerical simulations it is convenient to take $m_{\uparrow} = -m_{\downarrow} = m$. The addition of such potential to the tight-binding Hamiltonian (\ref{tb_hamiltonian}) opens a gap of width $2 \, m$ in its spectrum and prohibits the existence of zero modes (see Appendix \ref{appsec:spectrum}), however, at the cost of explicit breaking of sublattice symmetry. As we show in Appendix \ref{appsec:hopping}, this gap cannot be closed due to interactions, thus at nonzero $m$ Monte-Carlo simulations are possible for any value of the coupling constant $e^2$ in (\ref{interaction_hamiltonian}). At low energies the coefficient $m$ plays the role of the mass of Dirac quasiparticles in graphene \cite{Semenoff:84:1}. In order to describe the physics of massless fermions, we should extrapolate simulation results to $m = 0$. This situation is very similar to lattice QCD simulations, which are only possible at nonzero quark masses and the chiral limit is reached only by extrapolation. Dispersion relation $E\lr{\vec{k}}$ for the tight-binding Hamiltonian (\ref{tb_hamiltonian}) with the staggered potential of strength $m$ is shown on Fig.~\ref{fig:spectrum} at different ratios $m/\kappa$. The points on the plot and the inset illustrate the filling of the graphene Brillouin zone with discrete lattice momenta on the two-dimensional toric lattice made of $18 \times 18$ hexagons (see Subsection \ref{subsec:lattice_geometry} and Appendix \ref{appsec:spectrum} for more details).

 Finally, taking into account all the refinements of the tight-binding model discussed above and using the fermionic operators introduced in (\ref{operator_redef}), we arrive at the following Hamiltonian:
\begin{eqnarray}
\label{tb_hamiltonian_peierls}
\hat{H}_{tb} = -\kappa \, \sum\limits_{\sigma = \uparrow, \downarrow} \sum\limits_{<XY>}
\left(
\hat{\psi}^{\dag}_{\sigma, X} \,
\expa{\pm i \hat{\theta}_{XY}} \,
\hat{\psi}_{\sigma, Y}
+ \right. \nonumber \\ \left. +
\hat{\psi}^{\dag}_{\sigma, Y} \,
\expa{\pm i \hat{\theta}_{YX}} \,
\hat{\psi}_{\sigma, X} \right)
+ \nonumber \\ +
\sum\limits_{\sigma = \uparrow, \downarrow}  \sum\limits_{X_1} m \, \hat{\psi}^{\dag}_{\sigma, X_1} \hat{\psi}_{\sigma, X_1}
-
\sum\limits_{\sigma = \uparrow, \downarrow}  \sum\limits_{X_2} m \, \hat{\psi}^{\dag}_{\sigma, X_2} \hat{\psi}_{\sigma, X_2}
\end{eqnarray}
Since particles and holes have opposite charges, in the first term in (\ref{tb_hamiltonian_peierls}) one should take the plus sign before $\hat{\theta}_{XY}$ and $\hat{\theta}_{YX}$ for $\sigma = \uparrow$ and the minus sign for $\sigma = \downarrow$. In the second term, summation over $X_1$ and $X_2$ denotes summation over the sites of two simple sublattices of the hexagonal lattice.

 A starting point for lattice Monte-Carlo simulations is the path integral representation of the partition function and operator expectation values for the tight-binding model interacting with electromagnetic field:
\begin{eqnarray}
\label{partition_function1}
 \mathcal{Z} = \tr \expa{ - \hat{H}/T }  ,
\\
\label{vevs1}
 \vev{O_1\lr{\tau_1} \, \ldots \, O_n\lr{\tau_n} }
 = \nonumber \\ =
 \mathcal{Z}^{-1} \, \tr\lr{
 \hat{O}_1\lr{\tau_1} \ldots \hat{O}_n\lr{\tau_n}
 \expa{ - \hat{H}/T }}  ,
\end{eqnarray}
where $\hat{H} = \hat{H}_{tb} + \hat{H}_{em}$, $T$ is the temperature, the trace is taken over the joint Hilbert space of the fermions and the electromagnetic field, $\hat{O}_1, \ldots, \hat{O}_n$ are some operators and $\hat{O}\lr{\tau} = \expa{ - \tau \, \hat{H} } \, \hat{O} \, \expa{ \tau \, \hat{H}}$. As usual, exactly zero temperature cannot be reached in lattice Monte-Carlo simulations, but a reasonably small value can be achieved by using lattices with sufficiently large size in Euclidean time direction.

 We construct the lattice approximation to the path integral representation of the partition function (\ref{partition_function1}) in Subsections \ref{subsec:em_field_action} and \ref{subsec:fermionic_action} below. In Subsection \ref{subsec:em_field_action} we approximate the trace over the states of the electromagnetic field and in Subsection \ref{subsec:fermionic_action} - over the fermionic degrees of freedom. Before that, in Subsection \ref{subsec:lattice_geometry} we describe in details the geometry of the hexagonal lattice and its extension to the three-dimensional space.

\subsection{Geometry of hexagonal lattice and its extension to the three-dimensional space}
\label{subsec:lattice_geometry}

\begin{figure}[h!tb]
  \includegraphics[width=8.5cm]{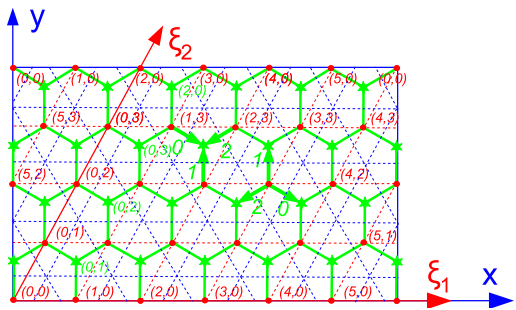}\\
  \caption{Cartesian and rhombic coordinate axes and coordinate grids for hexagonal lattice covering the torus of size $L_x \times L_y = 6 \times 4$ in the two-dimensional Euclidean space.}
  \label{fig:graphene_coordinates}
\end{figure}

 In order to perform lattice Monte-Carlo simulations, we should somehow compactify the hexagonal lattice on which the tight-binding model is defined. Here we consider lattices which have the topology of the torus. An example of such hexagonal lattice which consists of $L_x \times L_y = 6 \times 4$ hexagons is shown on Fig.~\ref{fig:graphene_coordinates}. Since hexagonal lattice can be thought of as a composition of two rhombic sublattices, it is convenient to classify the lattice sites which belong to these sublattices as either ``even'' or ``odd'' sites. All nearest neighbors of an even site are odd sites, and vice versa. On Fig.~\ref{fig:graphene_coordinates} even sites are marked with red circles, and odd sites - with green crosses.

 Lattice sites can be enumerated using the integer-valued coordinates $\xi_1 = 0 \ldots L_x - 1$, $\xi_2 = 0 \ldots L_y - 1$ which label the sites of one of sublattices, say, even sites. The corresponding coordinate axes and coordinate grid are shown on Fig.~\ref{fig:graphene_coordinates} with solid and dashed red lines. Numbers in parentheses near even lattice sites are their coordinates $\lr{\xi_1, \xi_2}$. Coordinate system for odd sites is the same as for the even sites, but its origin is shifted along the hexagon edge which is perpendicular to the $\xi_2$ axis. Altogether, we characterize each lattice site of the hexagonal lattice by two integer-valued coordinates $\xi = \lr{\xi_1, \, \xi_2}$ and a label $s = \alpha, \, \beta$, where $s = \alpha$ stands for even lattice sites and $s = \beta$ - for odd lattice sites. We also introduce the Cartesian coordinate system with coordinates $x$ and $y$ in the graphene plane, so that the $X$ coincides with $\xi_1$ axis of the rhombic coordinates. The axes of this coordinate system are shown on Fig.~\ref{fig:graphene_coordinates} with blue solid lines. Cartesian coordinates of the lattice site with rhombic coordinates $\lr{s, \xi}$ are
\begin{eqnarray}
\label{rhombic_to_cartesian}
 x = \sqrt{3} \, a \, \xi_1 +  \sqrt{3}/2 \, a \, \xi_2
 +  \sqrt{3}/2 \, a \, \delta_{s, \beta},
\nonumber \\
 y =                                   3/2 \, a \, \xi_2
 -         1/2 \, a \, \delta_{s, \beta}  ,
\end{eqnarray}
where $a$ is the lattice spacing, that is, the length of hexagon edge. Correspondingly, the area of the unit cell of the rhombic lattice in cartesian coordinates is equal to the hexagon area $3 \sqrt{3} a^2/2$.

  In order to embed our lattice into Euclidean space with torus topology, we identify the opposite sides of rectangle of size $\sqrt{3} \, L_x \, \times 3/2 \, L_y$ in Cartesian coordinates, as shown on Fig.~\ref{fig:graphene_coordinates}. Such identification implies the following identification of rhombic coordinates \cite{Hands:09:1}:
\begin{eqnarray}
\label{rhombic_coords_bc}
\lr{\xi_1 + L_x, \xi_2} \rightarrow \lr{\xi_1, \xi_2}  ,
\nonumber \\
\lr{\xi_1, \xi_2 + L_y} \rightarrow \lr{\xi_1 + L_y/2, \xi_2}  .
\end{eqnarray}

 We assume that the links of our hexagonal lattice are always directed from even sites to odd sites, as illustrated on Fig.~\ref{fig:graphene_coordinates}. Correspondingly, we label them by the coordinates $\xi = \lr{\xi_1, \xi_2}$ of the even site from which they originate and the direction number $b = 0, 1, 2$, such that the link with coordinates $\xi$ in direction $b$ goes from the site with coordinates $\lr{\alpha, \xi}$ to $\lr{\beta, \xi + \rho_b}$ (modulo the identification (\ref{rhombic_coords_bc})). Here we have introduced the following set of three vectors in rhombic coordinates:
\begin{eqnarray}
\label{hex_basis_vectors_rhombic}
 \rho_0 = \lr{0, 0}, \quad \rho_1 = \lr{-1, 1}, \quad \rho_2 = \lr{-1, 0}  .
\end{eqnarray}

\begin{figure}[h!tb]
  \includegraphics[width=6cm]{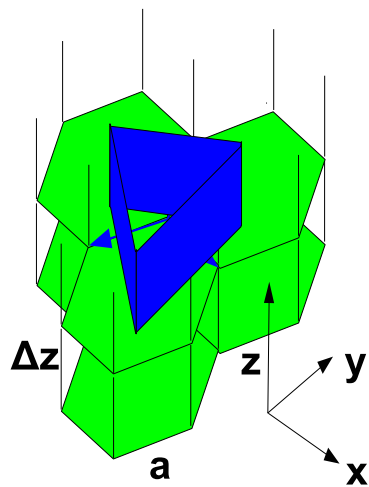}\\
  \caption{Direct product of two-dimensional hexagonal lattice with lattice spacing $a$ and rectangular lattice with lattice spacing $\Delta z$ which fill the three-dimensional space. Dark blue faces which form a triangular prism are the plaquettes of the dual lattice.}
  \label{fig:3d_graphene_lattice}
\end{figure}

 In order to describe electromagnetic fields, which propagate in the three-dimensional space, we also introduce the coordinate $z$ for the direction perpendicular to the graphene layer. In the following we assume that the latter is situated at $z = 0$. We denote the vectors in the three-dimensional space by arrows: $\vec{r} = \lr{x, y, z}$. We discretize the $z$ coordinate into the intervals of size $\Delta z$, so that the three-dimensional space is covered by a direct product of the hexagonal lattice in the $x, y$ plane and the regular rectangular lattice in the $z$ direction, as illustrated on Fig.~\ref{fig:3d_graphene_lattice}. Shifts along the links of this lattice are described by the vectors
\begin{eqnarray}
\label{hex_basis_vectors_cartesian}
 \vec{e}_0 = \lr{\sqrt{3}/2 \, a , -1/2 \, a, 0}  ,
 \quad
 \vec{e}_1 = \lr{0, a, 0}  ,
 \nonumber \\
 \vec{e}_2 = \lr{-\sqrt{3}/2 \, a, -1/2 \, a, 0}
 \quad
 \vec{e}_z = \lr{0, 0, {\Delta z}}  .
\end{eqnarray}

 It is also convenient to introduce the dual lattice with lattice sites which are situated above the centers of hexagons of the original lattice and which are shifted by $\Delta z/2$ along the $Z$ axis. The projection of the links of this dual lattice is shown on Fig.~\ref{fig:graphene_coordinates} with dashed lines. Now with each lattice link parallel to the graphene plane we can associate the rectangular plaquette of the dual lattice (with size $\sqrt{3} \, a  \times \Delta z$) which is orthogonal to it. Correspondingly, each link which goes in the $z$ direction is associated with some plaquette of the dual lattice which is parallel to the graphene plane and which has a form of equilateral triangle with side $\sqrt{3} \, a$. Plaquettes of such dual lattice are also shown on Fig.~\ref{fig:3d_graphene_lattice}.

\subsection{Lattice action for the electromagnetic field}
\label{subsec:em_field_action}

 We discretize the Hamiltonian (\ref{em_field_hamiltonian_continuum}) on the three-dimensional lattice described above in Subsection \ref{subsec:lattice_geometry}. As discussed in Subsection \ref{subsec:action_defs}, upon discretization the vector potential is replaced by its integrals along the lattice links:
\begin{eqnarray}
\label{vector_potential_discretization}
\theta_b\lr{\xi, z} = e \, \int\limits_{0}^{1} du \, e_b^i \cdot A_i\lr{\vec{x}\lr{\alpha, \xi, z} + u \, \vec{e}_b }
\nonumber \\
\theta_z\lr{s, \xi, z} = e \, \int\limits_{0}^{1} du \, e_z^i \cdot A_i\lr{\vec{x}\lr{s, \xi, z} + u \, \vec{e}_z}  .
\end{eqnarray}
From now on, we replace the abstract labels $X$, $Y$ of lattice sites used in Subsection \ref{subsec:action_defs} by the coordinates $s, \xi, z$ introduced in Subsection \ref{subsec:lattice_geometry}. Since for different $s, \, \xi, \, z$ we take vector potential in different points, all variables (\ref{vector_potential_discretization}) should be considered as independent. Note also that while the operators $\hat{\theta}_b\lr{\xi, z}$ are associated only with even lattice sites, the operators $\hat{\theta}_z\lr{s, \xi, z}$ are associated with both even and odd sites.

 The momentum operators canonically conjugate to $\hat{\theta}_b\lr{\xi, z}$ and $\hat{\theta}_z\lr{\xi, z}$ can be constructed as operators of electric field flux through the plaquettes $p^{*}$ of the dual lattice which are dual to the corresponding links:
\begin{eqnarray}
\label{vector_momenta_discretization0}
\hat{\pi}_b\lr{\xi, z} = \frac{1}{4 \pi  e} \, \int\limits_{p^{*} \perp \vec{e}_b} d\sigma_i \, \hat{E}^i
\quad, b = 0, 1, 2 , \nonumber \\
\hat{\pi}_z\lr{\xi, z} = \frac{1}{4 \pi  e} \, \int\limits_{p^{*} \perp \vec{e}_z} d\sigma_i \, \hat{E}^i  ,
\end{eqnarray}
where $d \vec{\sigma}$ is the element of area on the dual plaquettes. Since by construction the operators $\hat{\pi}_b$, $\hat{\pi}_z$ satisfy the canonical commutation relations with the operators $\hat{\theta}_b$, $\hat{\theta}_z$, they can be represented as differential operators $\hat{\pi}_b\lr{\xi, z} = -i \, \frac{\partial}{\partial \, \theta_b\lr{\xi, z}}$, $\hat{\pi}_z\lr{s, \xi, z} = -i \, \frac{\partial}{\partial \, \theta_z\lr{s, \xi, z}}$ on the Hilbert space of functions of the variables $\theta_b\lr{\xi, z}$, $\theta_z\lr{s, \xi, z}$. For notational convenience, let us also introduce the redundant set of field and momenta operators $\hat{\theta}_b\lr{s, \xi, z}$ and $\hat{\pi}_b\lr{s, \xi, z}$ associated with each lattice site, either odd or even:
\begin{eqnarray}
\label{redundant_variable_set}
\hat{\theta}_b\lr{\alpha, \xi, z} \equiv  \hat{\theta}_b\lr{\xi, z}
\nonumber \\
\hat{\theta}_b\lr{\beta, \xi, z}  \equiv - \hat{\theta}_b\lr{\xi - \rho_b, z}
\nonumber \\
\hat{\pi}_b\lr{\alpha, \xi, z} \equiv  \hat{\pi}_b\lr{\xi, z}
\nonumber \\
\hat{\pi}_b\lr{\beta, \xi, z} \equiv - \hat{\pi}_b\lr{\xi - \rho_b, z}
\end{eqnarray}

 Let us first consider the discretization of the electric part of the Hamiltonian (\ref{em_field_hamiltonian_continuum}). Below we will see that since the charge carriers in graphene move with velocities which are much smaller than the speed of light, the magnetic term $\lr{\rot \hat{A}}^2$ in the Hamiltonian can be neglected. To the leading order in $a$ and $\Delta z$ we can write
\begin{eqnarray}
\label{vector_momenta_discretization1}
\pi_b\lr{\xi, z} \approx  \frac{\sqrt{3} \, \Delta z}{4 \pi e} \, \vec{e}_b \cdot \vec{E}\lr{\vec{r}\lr{\alpha, \xi, z}}
\nonumber \\
\pi_z\lr{s, \xi, z} \approx \frac{3 \sqrt{3} a^2 }{16 \pi \, e \, \Delta z} \, \vec{e}_z \cdot \vec{E}\lr{\vec{r}\lr{s, \xi, z}} ,
\end{eqnarray}
where we have taken into account that the areas of the plaquettes $p^{*}$ dual to lattice links in the graphene plane and perpendicular to it are equal to $\sqrt{3} a \, \Delta z$ and $\frac{3 \sqrt{3} a^2}{4}$, respectively. Using the identity
\begin{eqnarray}
\label{hex_basis_vectors_identity}
\sum\limits_{b=0}^{2} \vec{e}_b \otimes \vec{e}_b
 =
\frac{3 \, a^2}{2} \, \lr{I - \frac{\vec{e}_z \otimes \vec{e}_z}{\lr{\Delta z}^2}}
\end{eqnarray}
we can now express the square of the electric field in graphene plane as
\begin{eqnarray}
\label{vector_momenta_discretization2}
 \hat{E}_x^2\lr{\vec{r}\lr{\alpha, \xi, z}} + \hat{E}_y^2\lr{\vec{r}\lr{\alpha, \xi, z}}
 = \nonumber \\ =
 \frac{32 \pi^2 \, e^2}{9 \, a^2 \, {\Delta z}^2} \sum\limits_{b=0}^{2} \hat{\pi}_b^2\lr{\xi, z} .
\end{eqnarray}
The integral over the three-dimensional space in (\ref{em_field_hamiltonian_continuum}) can be also approximated by the sums over the vertices of the hexagonal lattice:
\begin{eqnarray}
\label{integral_approximation}
 \int d^3\vec{r} \, f\lr{\vec{r}}
 \approx 
 \sum\limits_{z}
 \sum\limits_{\xi}
 \frac{3 \sqrt{3} \, a^2 \, {\Delta z}}{2}
 f\lr{\vec{r}\lr{\alpha, \xi, z}}
 \approx \nonumber \\ \approx
 \sum\limits_{s = \alpha, \beta}
 \sum\limits_{z}
 \sum\limits_{\xi}
 \frac{3 \sqrt{3} \, a^2 \, {\Delta z}}{4}
 f\lr{\vec{r}\lr{s, \xi, z}} .
\end{eqnarray}
Finally, we arrive at the following discretization of the electric part of the Hamiltonian (\ref{em_field_hamiltonian_continuum}):
\begin{eqnarray}
\label{em_field_hamiltonian_discrete}
 \frac{1}{8 \, \pi} \,
 \int d^3\vec{r} \, \hat{E}^2\lr{\vec{r}}
 \approx
 \sum\limits_{z, \xi}
 \left(
 \frac{2 \pi \, e^2}{\sqrt{3} {\Delta z}}
 \sum\limits_{b=0}^{2} \hat{\pi}_b^2\lr{\xi, z}
 \right. + \nonumber \\ + \left.
 \frac{8 \pi e^2 \, {\Delta z}}{3 \sqrt{3} \, a^2}
 \sum\limits_{s=\alpha, \beta} \hat{\pi}_z^2\lr{s, \xi, z}
 \right) .
\end{eqnarray}
By similar reasoning one can show that the discretization of the magnetic part of the Hamiltonian (\ref{em_field_hamiltonian_continuum}) should contain the square of some combination of the operators $\hat{\theta}_b\lr{\xi, z}$ and $\hat{\theta}_z\lr{s, \xi, z}$ multiplied by factors of order $\frac{1}{8 \pi \, e^2 \, \Delta z}$ and $\frac{1}{8 \pi \, e^2 \, a}$.

 Let us now take the trace over the states of electromagnetic field in the partition function (\ref{partition_function1}). To this end we use the standard Feynman-Kac transformation and rewrite
\begin{eqnarray}
\label{feynman_kac_em}
 \expa{ - \hat{H}/T}
 \approx
 \prod\limits_{\tau/{\Delta \tau}=0}^{L_{\tau}-1} \expa{ - \hat{H} \, {\Delta \tau}}  ,
\end{eqnarray}
where ${\Delta \tau} = \lr{T L_{\tau}}^{-1}$, and insert the identity operators $\hat{I} = \int \mathcal{D}\vec{A}\lr{\vec{r}} \ket{\vec{A}\lr{\vec{r}}} \, \bra{\vec{A}\lr{\vec{r}}} $ decomposed into the complete set of eigenvectors $\ket{\vec{A}\lr{\vec{r}}}$ of the operators $\hat{A}_{i}$ between these factors. Upon discretization (\ref{em_field_hamiltonian_discrete}), the Hilbert space of the discretized theory is equivalent to the space of all functions of link variables $\theta_b\lr{\xi, z}$ and $\theta_z\lr{s, \xi, z}$, and the decomposition of identity reads:
\begin{eqnarray}
\label{identity_decomposition_discrete}
 \hat{I} =
 \prod\limits_{z, \xi} \,
 \lr{\prod\limits_{b=0}^{2} \int d\theta_b\lr{\xi, z} } \,
 \lr{\prod\limits_{s=\alpha, \beta} \int d\theta_z\lr{s, \xi, z} } \,
 \nonumber \\
 \ket{ \theta_b\lr{\xi, z}, \theta_z\lr{s, \xi, z} } \,
 \bra{ \theta_b\lr{\xi, z}, \theta_z\lr{s, \xi, z} }  ,
\end{eqnarray}
where $\ket{ \theta_b\lr{\xi, z}, \theta_z\lr{s, \xi, z} }$ are eigenvectors of the operators $\hat{\theta}_b\lr{\xi, z}$ and $\hat{\theta}_z\lr{s, \xi, z}$. For the sake of brevity, we will denote the integrals over $\theta$ in (\ref{identity_decomposition_discrete}) as $\mathcal{D} \theta$, and the corresponding eigenvectors as $\ket{\theta}$.

 The decomposition of identity (\ref{identity_decomposition_discrete}), however, contains non-physical states which violate the constraint (\ref{gauss_law_continuum}). In order to get rid of such states, one should also insert the projectors $\mathcal{P}$ on the physical Hilbert space between the exponents in (\ref{feynman_kac_em}). Integrating the constraint (\ref{gauss_law_continuum}) over the unit cell of the dual lattice which encloses the lattice site with coordinates $\lr{s, \xi, z}$ and taking into account the definitions (\ref{vector_momenta_discretization0}), we see that the discretized version of the constraint (\ref{gauss_law_continuum}) becomes
\begin{eqnarray}
\label{gauss_law_discrete}
 \sum\limits_{b=0}^{2}\hat{\pi}_b\lr{s, \xi, z} + \hat{\pi}_z\lr{s, \xi, z}
 - \nonumber \\ -
 \hat{\pi}_z\lr{s, \xi, z - \Delta z} = \hat{q}\lr{s, \xi} \, \delta\lr{z, 0}  ,
\end{eqnarray}
where $\hat{q}\lr{s, \xi}$ is the charge operator (\ref{charge_redef}) at lattice site with coordinates $\lr{s, \xi}$ and we have taken into account that graphene layer is placed at $z = 0$.

 It is convenient to rewrite the projection operator as an integral over the Lagrange multiplier field $\phi\lr{s, \xi, \tau, z}$, which becomes the electric potential field in the path integral formalism. The matrix element of the $\tau$-th factor in (\ref{feynman_kac_em}) between the states $\bra{\theta\lr{\tau}}$ and $\ket{\theta\lr{\tau + {\Delta \tau}}}$ can be now written as
\begin{widetext}
\begin{eqnarray}
\label{em_matrix_element1}
 \bra{\theta\lr{\tau}}
 \hat{\mathcal{P}} \,
 \expa{ - \hat{H}_{em} \, {\Delta \tau}} \,
 \ket{\theta\lr{\tau + \Delta \tau}}
 =
 \prod\limits_{s, \xi, z} \, \int d\phi\lr{s, \xi, \tau, z}
 \nonumber \\
 \bra{\theta\lr{\tau}}
 \expa{
  \sum\limits_{s, \xi, z}
  i \phi\lr{s, \xi, \tau, z}
  \lr{ \sum\limits_{b=0}^{2}\hat{\pi}_b\lr{s, \xi, z} + \hat{\pi}_z\lr{s, \xi, z}
  - \hat{\pi}_z\lr{s, \xi, z - \Delta z} - \hat{q}\lr{s, \xi} \, \delta\lr{z, 0} }
 }
 \times \nonumber\\ \times
 \expa{
  - \frac{2 \pi \, e^2 \, {\Delta \tau}}{\sqrt{3} \, {\Delta z}}
  \sum\limits_{\xi, z, b} \hat{\pi}_b^2\lr{\xi, z}
  -
  \frac{8 \pi \, e^2 \, {\Delta z} \, {\Delta \tau}}{3 \sqrt{3} \, a^2}
  \sum\limits_{s, \xi, z} \hat{\pi}_z^2\lr{s, \xi, z}
 } \,
 \ket{\theta\lr{\tau + \Delta \tau}} \,
 \times \nonumber \\ \times
 \expa{
  - \sum\limits_{\xi, \tau, z} \frac{\Delta \tau}{8 \pi} \lr{\rot A\lrs{\theta\lr{\xi, \tau + {\Delta \tau}, z}}}^2 \,
 }  ,
\end{eqnarray}
\end{widetext}
where $\rot A\lrs{\theta\lr{\tau + \Delta \tau}}$ is the lattice discretization of the rotor of the vector potential and we have temporarily omitted the fermionic part of the Hamiltonian $\hat{H}$. Following the standard procedure, we have also approximated the exponent of the Hamiltonian by a product of exponents of the kinetic and potential terms. Evaluating the remaining matrix element which contains the exponentials of momentum operators, we arrive at the following expression:
\begin{widetext}
\begin{eqnarray}
\label{em_matrix_element2}
 \bra{\theta\lr{\tau}}
 \hat{\mathcal{P}} \,
 \expa{ - \hat{H}_{em} \, {\Delta \tau}} \,
 \ket{\theta\lr{\tau + \Delta \tau}}
 =
 \prod\limits_{s, \xi, z} \int d\phi\lr{s, \xi, \tau, z} \,
 \nonumber \\
 \exp\left(
  - \frac{\sqrt{3} \, \Delta z}{8 \pi \, e^2 \, \Delta \tau}
  \sum\limits_{b, \xi, z}
  \lr{\phi\lr{\alpha, \xi, \tau, z}-\phi\lr{\beta, \xi+\rho_b, \tau, z} + \theta_b\lr{\xi, \tau, z} - \theta_b\lr{\xi, \tau + {\Delta \tau}, z}}^2
  - \right. \nonumber \\ - \left.
  \frac{3 \sqrt{3} a^2}{32 \pi \, e^2 \, {\Delta z} \, {\Delta \tau}}
  \sum\limits_{s, \xi, z}
  \lr{\phi\lr{s, \xi, \tau, z}-\phi\lr{s, \xi, \tau, z+\Delta z} +
      \theta_z\lr{s, \xi, \tau, z} - \theta_z\lr{s, \xi, \tau + {\Delta \tau}, z}}^2
  \right)
 \times \nonumber \\ \times
 \expa{
  -  i \sum\limits_{s, \xi} \phi\lr{s, \xi, \tau, z=0} \, \hat{q}\lr{s, \xi}
   - \sum\limits_{\xi, \tau, z} {\Delta \tau} \, \frac{\lr{\rot A\lrs{\theta\lr{\xi, \tau + {\Delta \tau}, z}}}^2}{8 \pi} \,
 } ,
\end{eqnarray}
\end{widetext}

 Now we can collect all such factors into the path integral over the lattice fields $\theta_b\lr{\xi, \tau, z}$, $\theta_z\lr{s, \xi, \tau, z}$ and $\phi\lr{s, \xi, \tau, z}$. Before writing down the final result for the lattice action, let us consider the proportion between the electric (first exponential) and the magnetic (last term in the second exponential) terms in the path integral weight (\ref{em_matrix_element2}). The coefficients before finite differences of lattice fields in the electric and the magnetic terms in (\ref{em_matrix_element2}) are of order $\frac{\kappa \, a}{4 \pi \, e^2 \, \kappa \, {\Delta \tau}}$ and $\frac{\kappa \, {\Delta \tau}}{4 \pi \, e^2 \, \kappa \, a}$, respectively. In lattice simulations, we should choose $\Delta \tau$ such that $\kappa {\Delta \tau} \ll 1$, but $\kappa {\Delta \tau} L_{\tau} \gg 1$. For realistic simulations, $L_{\tau} \sim 10^1$, thus $\kappa \, {\Delta \tau} \sim 10^{-1}$. Taking into account that $\kappa \, a = 1.946 \cdot 10^{-3}$ and $e^2 \approx 1/137$, we conclude that the coefficients before the electric and the magnetic terms in the lattice action are of order of $10^{-1}$ and $10^3$, respectively. Therefore, in Monte-Carlo simulations the fluctuations of the spatial component of the gauge field $\theta_b\lr{\xi, \tau, z}$ with nontrivial field strength are suppressed by two orders of magnitude in comparison with fluctuations of the electric potential $\phi\lr{s, \xi, z, \tau}$, and one can disregard them, assuming that $\rot A\lrs{\theta\lr{\xi, \tau + {\Delta \tau}, z}}$ is effectively equal to zero. By a gauge transformation one can then set all the spatial link variables $\theta_b\lr{\xi, \tau, z}$ to zero.

 Physically the dominance of the electric part of the action means that we adjust $\Delta \tau$ so that the ratio $a / {\Delta \tau}$ is comparable with characteristic velocity of charge carriers in graphene $v_F = 3/2 \kappa a \approx 1/300$. Since this velocity is much less than the speed of light, with a good approximation we can describe electromagnetic interactions between charge carriers by instantaneous Coulomb potential, and neglect the magnetic fields created by them.

 Finally, we should take into account that graphene is placed on a substrate with dielectric permittivity $\epsilon$. A physical way to account for the substrate would be to modify the lattice action only for plaquettes which are inside the medium. Such modification might be advantageous for studying multi-layered graphene or Casimir interactions with graphene sheets. However, since in this paper we are interested only in the simplest geometry in which the substrate fills half of the three-dimensional space, we simply replace $e^2$ by $\frac{2 e^2}{\epsilon + 1}$ in all expressions above.

 We thus arrive at the following discretized path integral representation for the trace $\tr_{em}$ over the states of the electromagnetic field in the partition function (\ref{partition_function1}):
\begin{widetext}
\begin{eqnarray}
\label{em_trace1}
 \tr_{em} \expa{- \hat{H}/T }
 =
 \prod\limits_{s, \xi, \tau, z} \int d\phi\lr{s, \xi, \tau, z} \,
  \expa{ -S_{em}\lrs{\phi\lr{s, \xi, \tau, z}}}
 \times \nonumber \\ \times
 \prod\limits_{\tau/{\Delta \tau}=0}^{L_{\tau}-1} \expa{
   -  \hat{H}_{tb} \, {\Delta \tau}
 + i \sum\limits_{s, \xi} \hat{q}\lr{s, \xi} \, \phi\lr{s, \xi, \tau, z=0}
} ,
\end{eqnarray}
where the lattice action for the electrostatic potential $\phi\lr{s, \xi, \tau, z}$ is
\begin{eqnarray}
\label{em_action_discrete}
 S_{em}\lrs{\phi\lr{s, \xi, \tau, z}} =
 \frac{\beta_{hex}}{2} \, \sum\limits_{b, \xi, \tau, z}
  \lr{\phi\lr{\alpha, \xi, \tau, z} - \phi\lr{\beta, \xi+\rho_b, \tau, z}}^2
 + \nonumber \\ +
  \frac{\beta_z}{2} \, \sum\limits_{s, \xi, \tau, z}
  \lr{\phi\lr{s, \xi, \tau, z} - \phi\lr{s, \xi, \tau, z + {\Delta z}}}^2
\end{eqnarray}
and
\begin{eqnarray}
\label{em_coupling_constants}
 \beta_{hex}
  =
 \frac{\sqrt{3} \, \Delta z}{4 \pi e^2 {\Delta \tau}} \, \frac{\epsilon + 1}{2}
  =
 \frac{\sqrt{3}}{4 \pi e^2} \, \lr{\frac{\Delta z}{a}} \,
 \frac{\lr{\kappa a}}{\lr{\kappa {\Delta \tau}}} \, \frac{\epsilon + 1}{2}
 \nonumber \\
 \beta_z
  =
 \frac{3 \sqrt{3} a^2}{16 \pi \, e^2 \, {\Delta z} \, {\Delta \tau}} \,
 \frac{\epsilon + 1}{2}
  =
 \frac{3 \sqrt{3}}{16 \pi \, e^2} \, \lr{\frac{\Delta z}{a}}^{-1} \,
 \frac{\lr{\kappa a}}{\lr{\kappa {\Delta \tau}}} \, \frac{\epsilon + 1}{2}  .
\end{eqnarray}
\end{widetext}
For further convenience, we have represented the inverse lattice coupling constants $\beta_{hex}$ and $\beta_{z}$ in terms of dimensionless combinations of lattice parameters ${\Delta z}/a$, $\kappa \, {\Delta \tau}$ and $\kappa \, a = 1.946 \cdot 10^{-3}$.

 Since the fluctuations of the spatial components $\theta_b\lr{\xi, z}$ and $\theta_z\lr{s, \xi, z}$ of lattice gauge field can be neglected, one can also remove the operators $\expa{\pm i \hat{\theta}_{XY}}$ from the tight-binding Hamiltonian (\ref{tb_hamiltonian_peierls}) in (\ref{em_trace1}).

\begin{figure}[h!tb]
  \includegraphics[width=8cm]{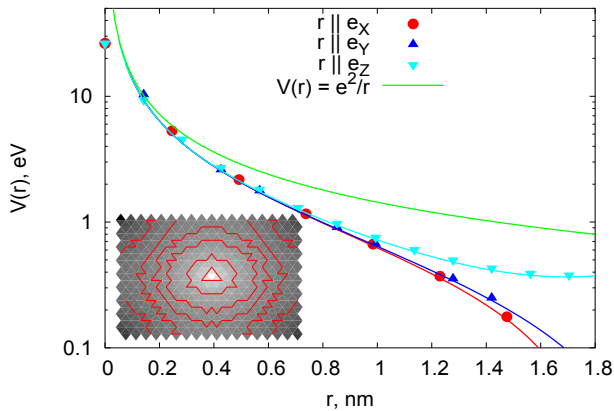}\\
  \caption{A comparison of the interaction potential of two charges on the $24 \times 24 \times 24$ lattice obtained from the discretized action (\ref{em_action_discrete}) (points) with the potential obtained from the solution of the continuum Laplace equation with two point sources with opposite charges on the torus of appropriate size and with the infinite-space Coulomb potential $V\lr{r} = e^2/r$ (solid lines). The inset shows the level surfaces of electrostatic potential on the hexagonal lattice.}
  \label{fig:coulomb_test}
\end{figure}

 From (\ref{em_action_discrete}) one can see that the field $\phi\lr{s, \xi, z, \tau}$ is non-propagating, thus it does not describe any dynamics of the electromagnetic field, but only the electrostatic interaction. The interaction potential of two charges at distance $r$ is now different from the Coulomb potential $e^2/r$ due to discretization errors of order $O\lr{a^2/r^2}$ and $O\lr{a \, {\Delta z}/r^2}$. These errors can be systematically reduced by constructing improved actions with finite differences which involve not only nearest neighbors, but also lattice sites separated by two and more lattice spacings.

 On Fig.~\ref{fig:coulomb_test} we compare the interaction potential of two static charges obtained from the discretized action (\ref{em_action_discrete}) on the $24 \times 24 \times 24$ lattice with ${\Delta z} = a$ with the potential obtained from the solution of the Laplace equation with two point sources with opposite charges on the torus of appropriate size $\lr{\sqrt{3} \, 24 \, a} \times \lr{3/2 \, 24 \, a} \times \lr{24 \, a}$, as well as with the Coulomb potential in the infinite space $V\lr{r} = e^2/r$. The inset illustrates the level surfaces of electrostatic potential on the hexagonal lattice. Fig.~\ref{fig:coulomb_test} shows that our lattice discretization indeed reproduces the continuum electrostatic potential with a good precision. For all lattices which we have used for simulations discretization errors do not exceed few percents. Our lattice regularization of the electrostatic interaction also unambiguously fixes the value of the on-site interaction potential to $u_0 = 26.1 \ev$.

 According to \cite{Wehling:11:1}, the bare Coulomb interaction with $V\lr{r} = e^2/r$ between electrons on $\pi$ orbitals is in fact additionally screened by a factor $\sim 2$ even in suspended graphene due to the influence of electrons on other orbitals. Such screening can be roughly accounted for by multiplying the coupling constants (\ref{em_coupling_constants}) by this factor. Note that in this case our value of the screened on-site interaction potential $u_0^{\prime} \sim u_0/2$ is quite close to the value $u_0 \approx 10 \ev$ obtained in \cite{Wehling:11:1}. However, a consistent treatment of this screening of Coulomb potential requires many technical complications, which are certainly beyond the approximations used in this paper. For this reason, here we do not take it into account. We further discuss possible effect of such screening in the concluding Section \ref{sec:conclusions}.

\subsection{Lattice action for the fermion fields}
\label{subsec:fermionic_action}

 We now take the remaining trace over the states of the fermionic field in the partition function (\ref{partition_function1}). To this end we insert the decomposition of the identity operator in the fermionic Hilbert space between the operator exponentials in (\ref{em_trace1}):
\begin{eqnarray}
\label{fermionic_identity_decomposition}
 \hat{I} = \prod\limits_{\sigma, s, \xi}\int d\bar{\eta}_{\sigma}\lr{s, \xi, \tau} \, d\eta_{\sigma}\lr{s, \xi, \tau} \,
 \nonumber \\
 e^{-\bar{\eta}_{\sigma}\lr{s, \xi, \tau} \, \eta_{\sigma}\lr{s, \xi, \tau} } \,
 \ket{\eta_{\sigma}\lr{s, \xi, \tau}} \bra{\eta_{\sigma}\lr{s, \xi, \tau}}  .
\end{eqnarray}
Here $\ket{ \eta_{\sigma}\lr{s, \xi, \tau} }$ are the fermionic coherent states \cite{MontvayMuenster} defined in terms of the Grassman-valued field $\eta_{\sigma}\lr{s, \xi, \tau}$:
\begin{eqnarray}
\label{fermionic_coherent_states}
 \ket{\eta_{\sigma}\lr{s, \xi, \tau}} = \expa{ \sum\limits_{s, \xi} \eta_{\sigma}\lr{s, \xi, \tau} \psi^{\dag}_{\sigma}\lr{s, \xi} } \, \ket{0}  .
\end{eqnarray}

 In the coordinates $\lr{s, \xi}$ introduced above, the tight-binding Hamiltonian (\ref{tb_hamiltonian_peierls}) can be written as
\begin{eqnarray}
\label{tb_hamiltonian_abstract}
\hat{H}_{tb} =
\sum\limits_{\sigma=\uparrow, \downarrow} \,
\sum\limits_{s, \xi, s', \xi'} h_{s, \xi; s' \xi'} \,
\hat{\psi}_{\sigma}^{\dag}\lr{s, \xi} \, \hat{\psi}_{\sigma}\lr{s', \xi'} ,
\end{eqnarray}
where we have again replaced the abstract site indices $X$, $Y$ with the coordinates $\lr{s, \xi}$ introduced in Subsection \ref{subsec:lattice_geometry}. $h_{s, \xi; s' \xi'}$ are the matrix elements of the single-particle Hamiltonian, which acts on single-particle wave functions as follows:
\begin{eqnarray}
\label{one_part_ham_action}
\lrs{h \psi}\lr{\alpha, \xi} = -\kappa \sum\limits_{b=0}^{2} \psi\lr{\beta, \xi + \rho_b} + m \, \psi\lr{\alpha, \xi}
\nonumber \\
\lrs{h \psi}\lr{\beta, \xi}  = -\kappa \sum\limits_{b=0}^{2} \psi\lr{\alpha, \xi - \rho_b} - m \, \psi\lr{\beta,  \xi}  .
\end{eqnarray}

 We consider now the matrix element of one of the operator exponentials in (\ref{em_trace1}) between the states $\ket{\eta_{\sigma}\lr{s, \xi, \tau}}$ and $\ket{\eta_{\sigma}\lr{s, \xi, \tau + \Delta \tau}}$ and apply the identity \cite{MontvayMuenster}
\begin{widetext}
\begin{eqnarray}
\label{coherent_state_identity}
 \bra{\eta} \expa{ \sum\limits_{i,j} A_{ij} \hat{\psi}^{\dag}_i \hat{\psi}_j } \ket{\eta'}
 = 
\expa{ \sum\limits_{i,j} \lr{e^A}_{ij} \bar{\eta}_i \, \eta_j' } .
\end{eqnarray}
Including also the electrostatic potential $\phi\lr{s, \xi, \tau}$ introduced above, we obtain
\begin{eqnarray}
\label{ferm_matrix_el1}
\bra{\eta_{\sigma}\lr{s, \xi, \tau}}
\exp\left(
 -{\Delta \tau} \, \sum\limits_{\sigma, s, \xi, s', \xi'} h_{s, \xi; s' \xi'} \, \hat{\psi}_{\sigma}^{\dag}\lr{s, \xi} \, \hat{\psi}_{\sigma}\lr{s', \xi'}
 + \right. \nonumber \\ \left. +
 i \sum\limits_{\sigma, s, \xi} \pm \phi\lr{s, \xi, \tau, z=0} \, \hat{\psi}_{\sigma}^{\dag}\lr{s, \xi} \, \hat{\psi}_{\sigma}\lr{s, \xi}
 \right)
\ket{\eta_{\sigma}\lr{s, \xi, \tau + \Delta \tau}}
= \nonumber \\ =
\expa{
 \sum\limits_{\sigma, s, \xi, s', \xi'}
 \bar{\eta}_{\sigma}\lr{s, \xi, \tau}
 \lrs{\expa{-h {\Delta \tau} \pm i \phi\lr{\tau}}}\lr{s, \xi; s', \xi'}
 \eta_{\sigma}\lr{s', \xi', \tau + {\Delta \tau}}
} ,
\end{eqnarray}
where $\expa{-h {\Delta \tau} \pm i \phi\lr{\tau}}$ is the matrix exponent of the one-particle operator $h\lr{s, \xi; s', \xi'} \, {\Delta \tau} \pm i \phi\lr{s, \xi, \tau, z=0} \delta\lr{s, \xi; s', \xi'}$. Since the fields $\hat{\psi}_{\sigma}\lr{s, \xi}$ have opposite charges for $\sigma = \uparrow$ and $\sigma = \downarrow$, we take the plus sign before the electrostatic potential $\phi\lr{s, \xi, \tau, z=0}$ in (\ref{ferm_matrix_el1}) for $\sigma = \uparrow$ and the minus sign - for $\sigma = \downarrow$.

 Evaluating this exponential to the first order in ${\Delta \tau}$, we obtain:
\begin{eqnarray}
\label{ferm_matrix_exp1}
 \lrs{\expa{-h {\Delta \tau} \pm i \phi\lr{\tau}}}\lr{s, \xi; s', \xi'} =
 \delta_{s s'} \, \delta\lr{\xi, \xi'} \, e^{\pm i \phi\lr{s, \xi, \tau, z=0}}
 - \nonumber \\ -
 {\Delta \tau} \int\limits_{0}^{1} du \,
 e^{\pm i \lr{1-u} \, \phi\lr{s, \xi, \tau, z=0}}
 h\lr{s, \xi; s', \xi'}
 e^{\pm i u \, \phi\lr{s', \xi', \tau, z=0}}  .
\end{eqnarray}
\end{widetext}
Here we have used the matrix identity
\begin{eqnarray}
\label{matrix_exp_identity1}
 e^{A + B} = e^A \, \mathcal{P}\expa{ \int\limits_{0}^{1} du \, e^{-u A} \, B \, e^{u A}  } ,
\end{eqnarray}
where $\mathcal{P}$ denotes the path-ordering of the second exponential with respect to the $u$ integration variable. As discussed in Appendix \ref{appsec:hopping}, within the approximations that we make in this work one can replace integral over $u$ in (\ref{ferm_matrix_exp1}) by a value of the integrand at any $u \in \lrs{0, 1}$. For definiteness, we approximate the matrix exponential in (\ref{ferm_matrix_el1}) as
\begin{eqnarray}
\label{ferm_matrix_exp2}
 \lrs{\expa{-h {\Delta \tau} \pm i \phi\lr{\tau}}}\lr{s, \xi; s', \xi'}
  = \nonumber \\ =
 e^{\pm i \phi\lr{s, \xi, \tau, z=0}}
 \lr{\delta_{s s'} \, \delta\lr{\xi, \xi'} - {\Delta \tau} \, h\lr{s, \xi; s', \xi'}}
\end{eqnarray}
Note that such a choice is different from the approximation discussed in \cite{Rebbi:12:1, Rebbi:11:1}.

 Now we insert the approximation (\ref{ferm_matrix_exp2}) into the matrix element (\ref{ferm_matrix_el1}). Finally, we should take the product of such matrix elements for all $\tau$ to obtain the weight of the integral over $\eta\lr{s, \xi, \tau}$. Taking into account the form of the Hamiltonian $h\lr{s, \xi; s', \xi'}$, we obtain the following lattice action for the fermion fields
\begin{widetext}
\begin{eqnarray}
\label{fermion_lattice_action}
 S_{tb}\lrs{ \eta_{\sigma}\lr{s, \xi, \tau} } =
 \sum\limits_{\sigma, s, \xi, \tau, s', \xi', \tau'}
 \bar{\eta}_{\sigma}\lr{s, \xi, \tau}
 M_{\sigma}\lrs{s, \xi, \tau; s', \xi', \tau'}
 \eta_{\sigma}\lr{s', \xi', \tau'}
 = \nonumber \\ =
 \sum\limits_{\sigma, s, \xi, \tau}
 \bar{\eta}_{\sigma}\lr{s, \xi, \tau}
 \lr{ \eta_{\sigma}\lr{s, \xi, \tau} -
 e^{\pm i \phi\lr{s, \xi, \tau, z=0}} \, \eta_{\sigma}\lr{s, \xi, \tau + {\Delta \tau}, z=0}}
+ \nonumber \\ +
 \kappa {\Delta \tau} \, \sum\limits_{\sigma, \xi, \tau, b}
 \bar{\eta}_{\sigma}\lr{\alpha, \xi, \tau}
 e^{\pm i \phi\lr{\alpha, \xi, \tau, z=0}}
 \eta_{\sigma}\lr{\beta, \xi + \rho_b, \tau + {\Delta \tau}}
+ \nonumber \\ +
 \kappa {\Delta \tau} \, \sum\limits_{\sigma, \xi, \tau, b}
 \bar{\eta}_{\sigma}\lr{\beta, \xi, \tau}
 e^{\pm i \phi\lr{\beta, \xi, \tau, z=0}}
 \eta_{\sigma}\lr{\alpha, \xi - \rho_b, \tau + {\Delta \tau}}
+ \nonumber \\ +
 m \, {\Delta \tau} \sum\limits_{\sigma, \xi, \tau}
 \bar{\eta}_{\sigma}\lr{\alpha, \xi, \tau}
 e^{\pm i \phi\lr{\alpha, \xi, \tau, z=0}}
 \eta_{\sigma}\lr{\alpha, \xi, \tau + {\Delta \tau}}
- \nonumber \\ -
 m \, {\Delta \tau} \sum\limits_{\sigma, \xi, \tau}
 \bar{\eta}_{\sigma}\lr{\beta, \xi, \tau}
 e^{\pm i \phi\lr{\beta, \xi, \tau, z=0}}
 \eta_{\sigma}\lr{\beta, \xi, \tau + {\Delta \tau}}  .
\end{eqnarray}
\end{widetext}
Here we have introduced the fermion hopping matrices $M_{\sigma}$ with matrix elements $M_{\sigma}\lrs{s, \xi, \tau; s', \xi', \tau'}$, which are the functions of the electrostatic potential field $\phi\lr{s, \xi, \tau, z=0}$ in the graphene plane. A crucial observation is that due to the symmetry between particles and holes (which correspond to two different components of spin $\sigma$ with our choice of the ground state (\ref{free_ground_state})) the matrices $M_{\uparrow}$ and $M_{\downarrow}$ are complex conjugate:
\begin{eqnarray}
\label{hopping_matrix_conjugacy}
 M_{\downarrow}\lr{s, \xi, \tau; s', \xi', \tau'}
 =
 \overline{M}_{\uparrow}\lr{s, \xi, \tau; s', \xi', \tau'}
\end{eqnarray}

 We note here that in contrast to fermionic actions commonly used in the context of lattice gauge theories, such as staggered fermions \cite{MontvayMuenster, DeGrandDeTarLQCD}, our fermionic action (\ref{fermion_lattice_action}) does not suffer from the doubling of fermion flavors (see Appendix \ref{appsec:spectrum} for the proof). The reason is that we use the non-symmetric discretization of the time derivative $\partial_{\tau} \eta_{\sigma}\lr{s, \xi, \tau} \approx \lr{\eta_{\sigma}\lr{s, \xi, \tau + {\Delta \tau}} - \eta_{\sigma}\lr{s, \xi, \tau} }/{\Delta \tau}$. For lattice discretizations of relativistic field theories, such lattice derivative would violate cubic symmetry group of the lattice, which is the remainder of the Lorentz invariance. However, in our case there is no Lorentz invariance, and Euclidean time and spatial coordinates enter the action in essentially different ways. Thus we do not break any symmetry by using the non-symmetric finite difference for the lattice derivative. Interestingly, a similar path integral representation of the partition function of the tight-binding model (\ref{tb_hamiltonian_peierls}) with the symmetric lattice derivative in the time direction has been considered recently in \cite{Araki:12:1}, and the two fermionic doublers which appear due to such discretization were interpreted as the two non-relativistic spin components with $\sigma = \uparrow, \downarrow$.

 Finally, integrating over the fields $\eta_{\sigma}\lr{s, \xi, \tau}$ and taking into account the relation between fermion hopping matrices (\ref{hopping_matrix_conjugacy}), we arrive at the following representation of the partition function (\ref{partition_function1}) in terms of the lattice path integral over the electrostatic potential field $\phi\lrs{s, \xi, \tau, z}$:
\begin{widetext}
\begin{eqnarray}
\label{path_int_final}
 \mathcal{Z} =
 \int \mathcal{D}\bar{\eta}_{\sigma}\lr{s, \xi, \tau} \mathcal{D}\eta_{\sigma}\lr{s, \xi, \tau} \mathcal{D}\phi\lr{s, \xi, \tau, z} \,
 \expa{ - \sum\limits_{\sigma} \bar{\eta}_{\sigma} M_{\sigma}\lrs{\phi\lr{s, \xi, \tau, z=0}} \eta_{\sigma} - S_{em}\lrs{\phi\lr{s, \xi, \tau, z}}  }
= \nonumber \\ =
\int \mathcal{D}\phi\lr{s, \xi, \tau, z} \, |\det{M_{\uparrow}\lrs{\phi\lr{s, \xi, z=0, \tau} }} |^2
 \expa{ -S_{em}\lrs{\phi\lr{s, \xi, \tau, z}} }
\end{eqnarray}
\end{widetext}
As usual, taking the trace over fermionic states involves one additional permutation of Grassman-valued fields $\eta_{\sigma}\lr{s, \xi, \tau}$, thus anti-periodic boundary conditions in Euclidean time $\tau$ with period $\lr{k T}^{-1}$ should be imposed on them. In practice, this amounts to the replacement $\phi\lr{s, \xi, \tau, z=0} \rightarrow \phi\lr{s, \xi, \tau, z=0} + \pi$ within the fermionic part of the action in (\ref{path_int_final}) on a single time slice at $\tau/{\Delta \tau} = \lr{L_{\tau} - 1}$ and $z = 0$.

\subsection{Lattice Monte-Carlo simulations}
\label{ref:simulation_methods}

 Path integral weight in the partition function (\ref{path_int_final}) is manifestly positive, thus functional integration can be performed numerically by a Monte-Carlo method. Configurations of the electrostatic potential field $\phi\lr{s, \xi, \tau, z}$ should be therefore sampled with the weight (\ref{path_int_final}). Since this weight includes nonlocal determinant of the fermion hopping matrix $M_{\sigma}\lrs{\phi\lr{s, \xi, \tau, z=0}}$, the most suitable simulation method is the Hybrid Monte-Carlo algorithm \cite{MontvayMuenster, DeGrandDeTarLQCD}.

 We use the so-called $\Phi$-algorithm \cite{Gottlieb:87:1, DeGrandDeTarLQCD}, in which the squared modulus of the determinant of $M_{\sigma}\lrs{\phi\lr{s, \xi, \tau, z=0}}$ in (\ref{path_int_final}) is represented in terms of the complex-valued pseudo-fermion field $\chi\lr{s, \xi, \tau}$:
\begin{eqnarray}
\label{fermion_determinant_pseudofermions}
 | \det{M}_{\uparrow} |^2 = \int \mathcal{D}\bar{\chi} \, \mathcal{D}\chi \,
 \expa{ - \bar{\chi} \, \frac{1}{M_{\uparrow} M_{\uparrow}^{\dag}} \, \chi }  .
\end{eqnarray}
At the beginning of each Molecular Dynamics trajectory, we generate the random pseudo-fermion field $\chi$ according to the weight $P\lrs{\chi} \sim \expa{ - \bar{\chi} \, \lr{M_{\uparrow} M_{\uparrow}^{\dag}}^{-1} \, \chi }$ and then perform the Molecular Dynamics evolution of the electrostatic potential field $\phi\lr{s, \xi, \tau, z}$ with the force
\begin{eqnarray}
\label{md_force}
F\lrs{\phi}\lr{s, \xi, \tau, z} = - \frac{\partial}{\partial \phi\lr{s, \xi, \tau, z}} \, S_{em}\lrs{ \phi }
 - \nonumber \\ -
\, \delta_{z, 0} \,  \bar{\chi} \,
\frac{\partial}{\partial \phi\lr{s, \xi, \tau, z=0}} \,  \lr{M_{\uparrow}\lrs{\phi} M_{\uparrow}^{\dag}\lrs{\phi}}^{-1} \, \chi .
\end{eqnarray}
The corresponding equations of motion are solved by using the Sexton-Weingarten integrator \cite{Sexton:92:1, DeGrandDeTarLQCD}. In order to improve the ergodicity of the algorithm, the number of integrator steps is drawn from the Poisson distribution (see, e.g. \cite{DelDebbio:96:1}). The mean value of this distribution is automatically tuned during the thermalization process so that the acceptance rate of the algorithm lies in the range $0.6 \ldots 0.9$. The integrator step size is then changed in such a way that the total trajectory length is equal to one. We use the standard Conjugate Gradient algorithm to invert the operator $M_{\uparrow}\lrs{\phi} M_{\uparrow}^{\dag}\lrs{\phi}$ in (\ref{md_force}). After the Molecular Dynamics evolution we perform the usual accept-reject step. All random numbers are generated by using the \texttt{ranlux} random number generator \cite{Luscher:93:1} with double precision.

 We also additionally speed up our algorithm by applying local heatbath updates \cite{MontvayMuenster, DeGrandDeTarLQCD} to the variables $\phi\lr{s, \xi, \tau, z}$ with $z \neq 0$ between Hybrid Monte-Carlo updates. For both updates, the path integral weight (\ref{path_int_final}) is the stationary probability distribution. In addition, Hybrid Monte-Carlo updates satisfy the detailed balance condition, and heatbath updates satisfy the local detailed balance \cite{MontvayMuenster}. By combining the corresponding transition probabilities it is easy to see that the path integral weight (\ref{path_int_final}) is still the stationary probability distribution for the successive application of both updates, despite the fact that the detailed balance condition is no longer satisfied. We perform $20$ global heatbath updates between successive Hybrid Monte-Carlo updates. Within each global heatbath update, we select at random $L_x \times L_y \times  L_{\tau} \times \lr{L_z - 1}$ lattice sites outside of the graphene plane and apply local heatbath updates to them. This procedure, while consuming less than $10\%$ of the total CPU time, significantly decreases the autocorrelation time of the algorithm. We have estimated the latter for the physical observables such as the mean plaquette, the Polyakov loop and the chiral condensate as well as for purely algorithmic parameters such as the number of iterations of the CG algorithm and the energy difference for the Molecular Dynamics trajectories. We have found that for all observables and for all lattice parameters which we have used the autocorrelation time does not exceed $5$ full Monte-Carlo updates (which comprise both Hybrid Monte Carlo and heatbath updates).

\subsection{The choice of lattice parameters}
\label{subsec:latt_params}

 In practice, the simulations can only be performed for the finite lattice sizes $L_x$, $L_y$, $L_z$, $L_{\tau}$ and at finite nonzero values of ${\Delta \tau}$ and $m$. The results should be then extrapolated to the limits ${\Delta \tau} \rightarrow 0$ with fixed temperature in physical units $T = \lr{L_{\tau} \, {\Delta \tau}}^{-1}$ and $m \rightarrow 0$, ${\Delta z} \rightarrow 0$, $L_x \rightarrow \infty$, $L_y \rightarrow \infty$, $L_z \rightarrow \infty$. The latter limit should be taken in such a way that $\lr{m \, L_x}^{-1} \rightarrow 0$, $\lr{m \, L_y}^{-1} \rightarrow 0$, $\lr{{\Delta z} \, L_z}^{-1} \rightarrow 0$. These limits are completely analogous to thermodynamic and chiral limits in lattice QCD simulations.

\begin{table}
  \centering
  \begin{tabular}{|c|c|c|c|c|}
    \hline
    $\kappa \, {\Delta \tau}$ & $L_x \times L_y \times L_{\tau} \times L_z$ & $T$\\
    \hline
    $0.2000$ & $18 \times 18 \times 18 \times 18$ & $0.28 \, \kappa = 0.76 \ev = 8.8 \cdot 10^3 \K$ \\
    $0.1500$ & $24 \times 24 \times 24 \times 24$ & $0.28 \, \kappa = 0.76 \ev = 8.8 \cdot 10^3 \K$ \\
    $0.1333$ & $18 \times 18 \times 18 \times 18$ & $0.42 \, \kappa = 1.13 \ev = 1.3 \cdot 10^4 \K$ \\
    $0.1000$ & $24 \times 24 \times 24 \times 24$ & $0.42 \, \kappa = 1.13 \ev = 1.3 \cdot 10^4 \K$ \\
    $0.1000$ & $18 \times 18 \times 18 \times 18$ & $0.56 \, \kappa = 1.51 \ev = 1.8 \cdot 10^4 \K$ \\
    $0.0750$ & $24 \times 24 \times 24 \times 24$ & $0.56 \, \kappa = 1.51 \ev = 1.8 \cdot 10^4 \K$ \\
    \hline
  \end{tabular}
  \caption{The parameters of lattices which we have used in our simulations.}
  \label{tab:latt_params}
\end{table}

 The parameters of lattices which we have used in our simulations are summarized in Table \ref{tab:latt_params}. In order to locate the low-temperature phase of the tight-binding model (\ref{tb_hamiltonian_peierls}), we have considered three different temperatures ($T/\kappa = 0.28$, $T/\kappa = 0.42$ and $T/\kappa = 0.56$). As discussed in the Introduction, as long as we consider only phenomena which involve electronic degrees of freedom and which are characterized by typical energies of order of $\kappa$, the low-temperature limit can be studied with a good precision even by considering the temperatures which are much higher than the room temperature.

 Finite-volume effects are controlled by performing simulations at fixed temperature in physical units but at different spatial volumes ($18^3$ and $24^3$). We use lattices with sizes which are multiples of three, because the Dirac points are only covered by discrete lattice momenta on such lattices (see Appendix \ref{appsec:spectrum}). For all lattices we also assume that ${\Delta z} = a$. Indeed, since the Coulomb interaction potential is anyway approximated with an error of order $a^2$, it does not make sense to take ${\Delta z} \ll a$.

 For each set of lattice parameters described in Table \ref{tab:latt_params}, we consider four different values of the staggered potential $m$ (which also plays the role of the Dirac mass at low energies): $m/\kappa = 0.1$, $m/\kappa = 0.2$, $m/\kappa = 0.3$ and $m/\kappa = 0.5$. With such values of $m$, the induced gap in the spectrum of the free tight-binding model (\ref{tb_hamiltonian_peierls}) is still much smaller than the energy scale $E \sim \kappa$ at which deviations from the low-energy linear dispersion relation $E\lr{k} = v_F \, k$ become important (see Fig. \ref{fig:spectrum}). On the other hand, with such choice of parameters the gap width is comparable to the temperature, thus one can expect that finite-temperature effects might be quite significant. For fixed values of $\kappa \, {\Delta \tau}$ and $m/\kappa$, we change the strength of Coulomb interaction by adjusting the coupling constants $\beta_z$ and $\beta_{hex}$ in (\ref{em_action_discrete}) according to (\ref{em_coupling_constants}). We consider the values of substrate dielectric permittivity uniformly covering the range $\epsilon = 1.0 \ldots 10.0$ with step ${\Delta \epsilon} = 0.5$. For each data point at fixed values of $\epsilon$, $\kappa \, {\Delta \tau}$, $L_x$, $L_y$, $L_z$, $L_{\tau}$ and $m$ we have generated $100$ statistically independent configurations of the electrostatic potential field $\phi\lr{s, \xi, \tau, z}$.

\begin{figure}[h!tb]
  \includegraphics[width=6cm, angle=-90]{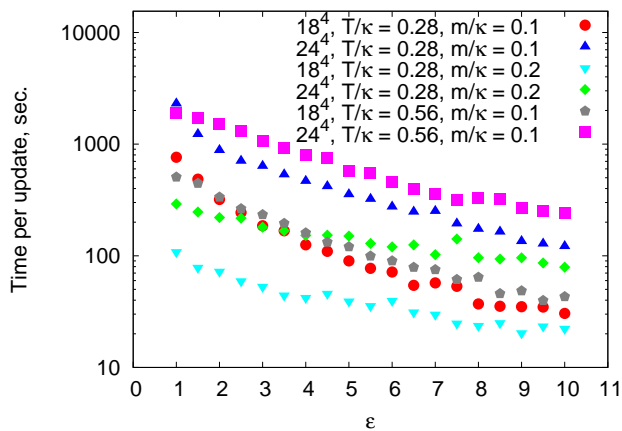}\\
  \caption{Total CPU time (for a 2.4 GHz Intel Xeon CPU) required for one Hybrid Monte-Carlo update as a function of the substrate dielectric permittivity $\epsilon$ for different lattice parameters.}
  \label{fig:hmc_performance}
\end{figure}

 To illustrate the performance of our algorithm, on Fig.~\ref{fig:hmc_performance} we plot the total CPU time required for one Hybrid Monte-Carlo update (plus $20$ heatbath updates) as a function of substrate dielectric permittivity $\epsilon$ in (\ref{em_coupling_constants}) for different lattices and for different values of the Dirac mass $m$. One can see that the algorithm significantly slows down as we move to smaller $\epsilon$ and $m$, that is, deeper in the non-perturbative regime with large coupling constant and small energy gap. The situation is similar to that in lattice QCD, where simulations at small quark masses also suffer from significant slow-down sometimes called the ``Berlin Wall'' \cite{Bernard:02:1}.

\section{Spontaneous breaking of sublattice symmetry}
\label{sec:sublat_symm}

\subsection{Basic definitions and lattice observables}
\label{subsec:sublat_symm_defs}

 In this Section we study the spontaneous breaking of sublattice symmetry within the tight-binding model of graphene (\ref{tb_hamiltonian_peierls}) with electromagnetic interactions. Before discussing the relevant order parameters, let us consider more closely the symmetries of this model.

 In the absence of interactions, the Hamiltonian (\ref{tb_hamiltonian_peierls}) has a global $U\lr{2}$ flavor symmetry. For graphene at half-filling, it is explicitly broken down to $U\lr{1} \otimes U\lr{1}$ by the Coulomb interaction term. This $U\lr{1} \otimes U\lr{1}$ symmetry ensures the conservation of the total numbers of charge carriers with different spins and cannot be broken neither by the staggered potential nor by the Coulomb interactions. In contrast, sublattice symmetry is a discrete symmetry, which can be realized as reflections with respect to the planes which are perpendicular to one of the basis vectors $\vec{e}_a$ (see (\ref{hex_basis_vectors_cartesian})) of the hexagonal lattice and which intersect lattice links in the direction $a$ in the middle. It can be broken either explicitly, by introducing the staggered potential, or spontaneously, due to a strong enough Coulomb interaction. Close to the Dirac points this discrete symmetry is enhanced to a continuous chiral symmetry of Dirac fermions, so that the overall global symmetry group is enhanced to $U\lr{4}$. Thus, strictly speaking, Goldstone's theorem is not applicable to spontaneous breaking of sublattice symmetry, but Goldstone bosons might still appear as effective degrees of freedom at low energies \cite{Araki:10:1}. Finally, we note that the symmetries of the tight-binding model are quite different from those of the staggered fermionic action, which was used for numerical simulations of the effective field theory of graphene in \cite{Lahde:09:1, Lahde:09:2, Lahde:09:3, Lahde:10:1, Lahde:11:1, Hands:08:1, Hands:10:1, Hands:11:1, Giedt:11:1, Ulybyshev:12:1}. Staggered fermions have a $U\lr{1}$ symmetry associated with charge conservation as well as a $U\lr{1}$ chiral symmetry which is explicitly broken by the mass term, thus for staggered fermions discrete sublattice symmetry is replaced by a continuous symmetry at all energy scales. On the other hand, only the total charge of both flavors is conserved, but not the charges of each flavor \cite{Burden:87:1, Creutz:07:1, Lahde:10:1, Giedt:11:1}.

 An obvious order parameter for the spontaneous breaking of the discrete sublattice symmetry is the difference of the particle number densities on the two sublattices of the hexagonal lattice $\Delta_N$:
\begin{eqnarray}
\label{npart_diff_def}
 \vev{\Delta_N}
 = - \frac{T}{L_x \, L_y} \frac{\partial \log \mathcal{Z}}{\partial m}
 = \nonumber \\ =
 \frac{1}{\mathcal{Z} L_x L_y} \tr\lr{ \hat{\Delta}_N e^{-\beta \hat{H}} } ,
 \nonumber \\
 \hat{\Delta}_N = \sum\limits_{\xi, \sigma} \,
 \lr{\hat{\psi}^{\dag}_{\sigma}\lr{\alpha, \xi} \hat{\psi}_{\sigma}\lr{\alpha, \xi}
 -
 \hat{\psi}^{\dag}_{\sigma}\lr{\beta, \xi} \hat{\psi}_{\sigma}\lr{\beta, \xi} }
\end{eqnarray}
A simple calculation within the Dirac approximation shows that in the absence of interactions and at sufficiently small values of the staggered potential $m$ $\vev{\Delta_N}$ is a linear function of $m$. It is also convenient to introduce the susceptibility of $\Delta_N$ as
\begin{eqnarray}
\label{npart_susc_def}
 \chi_N = \kappa \, \frac{\partial \, \vev{\Delta_N}}{\partial m}|_{m \rightarrow 0}
\end{eqnarray}
At sufficiently small temperatures, when only the linear part of the spectrum contributes to the expectation values, $\vev{\Delta}_N$ and $\chi_N$ can be expressed in terms of the chiral condensates and chiral susceptibilities of the two flavors of Dirac quasiparticles. By analogy with chiral symmetry breaking in gauge theories, one can expect that at a second-order phase transition the susceptibility (\ref{npart_susc_def}) should diverge.

 To obtain the expression for the expectation value (\ref{npart_diff_def}) on the lattice, we insert the operator $\hat{\Delta}_N = \sum\limits_{\xi, \sigma} \, \lr{\hat{\psi}^{\dag}_{\sigma}\lr{\alpha, \xi} \hat{\psi}_{\sigma}\lr{\alpha, \xi}
 -  \hat{\psi}^{\dag}_{\sigma}\lr{\beta, \xi} \hat{\psi}_{\sigma}\lr{\beta, \xi} }$ between $0$'th and $L_{\tau}$'th factors in the Feynman-Kac representation (\ref{em_trace1}) of the partition function (\ref{partition_function1}). After bringing the operators $\hat{\psi}_{\sigma}\lr{s, \xi}$ and $\hat{\psi}^{\dag}_{\sigma}\lr{s, \xi}$ to the normal order, the integral over the fermionic coherent states can be easily taken. We then obtain the following expression for $\vev{\Delta_N}$ in terms of the fermionic hopping matrix $M_{\uparrow}$:
\begin{eqnarray}
\label{npart_diff_latt}
 \vev{\Delta_N} = \frac{2}{L_x \, L_y \, L_{\tau} }
 \sum\limits_{\xi, \tau}
 \re \vev{M_{\uparrow}^{-1}\lr{\alpha, \xi, \tau; \alpha, \xi, \tau}}
 - \nonumber \\ -
 \frac{2}{L_x \, L_y \, L_{\tau} }
 \sum\limits_{\xi, \tau}
 \re \vev{M_{\uparrow}^{-1}\lr{\beta, \xi, \tau; \beta, \xi, \tau}}  ,
\end{eqnarray}
where the brackets $\vev{ \ldots }$ on the r.h.s. denote averaging over the electrostatic potential field $\phi\lr{s, \xi, \tau, z}$ with the weight (\ref{path_int_final}) and $M_{\uparrow}^{-1}$ is understood as the matrix inversion of the fermion hopping matrix $M_{\uparrow}$ in (\ref{fermion_lattice_action}).

 By analogy with chiral condensate measurements in lattice QCD, one can expect that the expectation value $\vev{ \Delta_N }$ in the limit $m \rightarrow 0$ should be distorted by large finite-volume corrections as well as by systematic errors of extrapolation to $m = 0$. The reason is that the ``chiral'' limit $m \rightarrow 0$ and the thermodynamic limit $L_x \rightarrow \infty$, $L_y \rightarrow \infty$ do not commute (see e.g. \cite{DeGrandDeTarLQCD}, Sec. 15.2.1), and strictly speaking $\vev{ \Delta_N }$ at $m = 0$ should be zero in any finite volume. The numerical value of the condensate therefore strongly depends on the way of extrapolating it to the limit $m \rightarrow 0$. Therefore we also consider the dispersion of $\Delta_N$, which is finite in finite volume and is thus much less affected by these numerical uncertainties:
\begin{eqnarray}
\label{npart_diff_square}
\cev{\Delta_N^2} = \frac{1}{\mathcal{Z} \, L_x \, L_y}
 \tr\lr{ \hat{\Delta}_N^2 e^{-\beta \hat{H}} } - L_x L_y \vev{\Delta_N}^2 .
\end{eqnarray}
Similarly to the susceptibility $\chi_N$, $\cev{\Delta_N^2}$ should significantly increase or diverge at the phase transition. The expression (\ref{npart_diff_square}) contains contributions both from connected and from disconnected fermionic diagrams:
\begin{widetext}
\begin{eqnarray}
\label{npart_diff_square_latt}
 \cev{\Delta_N^2} = \cev{\Delta_N^2}_{conn.} + \cev{\Delta_N^2}_{disc.}
 \nonumber \\
 \cev{\Delta_N^2}_{conn.} =
 \frac{2}{L_x \, L_y \, L_{\tau}} \, \sum\limits_{\xi, \tau} \, \re \left( \right.
 \sum\limits_{s} \,
 \vev{M_{\uparrow}^{-1}\lr{s, \xi, \tau; s, \xi, \tau}}
 - \nonumber \\ -
 \sum\limits_{s, \xi'} \,
 \vev{
  M_{\uparrow}^{-1}\lr{s, \xi, \tau; s, \xi', \tau} \,
  M_{\uparrow}^{-1}\lr{s, \xi', \tau; s, \xi, \tau}
 }
 + 
 \sum\limits_{\xi'} \,
 \vev{
  M_{\uparrow}^{-1}\lr{\alpha, \xi, \tau; \beta, \xi', \tau} \,
  M_{\uparrow}^{-1}\lr{\beta, \xi', \tau; \alpha, \xi, \tau}
 }
 + \nonumber \\ +
 \sum\limits_{\xi'} \,
 \vev{
  M_{\uparrow}^{-1}\lr{\beta, \xi, \tau; \alpha, \xi', \tau} \,
  M_{\uparrow}^{-1}\lr{\alpha, \xi', \tau; \beta, \xi, \tau}
 }
 \left. \right) ,
 \nonumber \\
 \cev{\Delta_N^2}_{disc.} = \frac{4}{L_x L_y L_{\tau}} \, \sum\limits_{\tau} \,
 \vev{\lr{\sum\limits_{\xi} \lr{M_{\uparrow}^{-1}\lr{\alpha, \xi, \tau; \alpha, \xi, \tau} - M_{\uparrow}^{-1}\lr{\beta, \xi, \tau; \beta, \xi, \tau} } }^2}
 - \nonumber \\ -
 \frac{4}{L_x L_y} \,
 \vev{\frac{1}{L_{\tau}} \, \sum\limits_{\tau, \xi} \lr{M_{\uparrow}^{-1}\lr{\alpha, \xi, \tau; \alpha, \xi, \tau} - M_{\uparrow}^{-1}\lr{\beta, \xi, \tau; \beta, \xi, \tau} } }^2  ,
\end{eqnarray}
\end{widetext}
where again the brackets $\vev{ \ldots }$ on the r.h.s. denote averaging over the electrostatic potential field with the weight (\ref{path_int_final}) and the first summand in $\cev{\Delta_N^2}_{conn.}$ arises due to an additional exchange of fermion operators $\hat{\psi}_{\sigma}\lr{s, \xi}$ and $\hat{\psi}_{\sigma}^{\dag}\lr{s, \xi}$ which is necessary in order to bring them to the normal order within the expectation value of the four-fermion operator. We have found that the disconnected contribution is much noisier than the connected one (see Fig. \ref{fig:condensate_dispersion_conndisc_k0.2000_s18_t18} below), so that the computer time required to estimate it with sufficient precision becomes prohibitively large. For this reason, we have considered only the connected part $\cev{\Delta_N^2}_{conn.}$.  It should be also noticed that the connected contribution $\cev{ \Delta_N^2 }_{conn.}$ does not in general correspond to the expectation value of any operator in canonical formalism, that is, it cannot be represented in the form $\mathcal{Z}^{-1} \, \tr{\lr{ \hat{O} e^{-\beta \hat{H}} }}$ with some operator $\hat{O}$. This is in contrast with lattice QCD, where the connected contributions to the correlators of two-fermion operators can be interpreted in terms of physical meson states consisting of quarks with different flavors. The reason is that in our case the two fermion flavours have opposite charges and thus couple differently to the electrostatic potential field $\phi$.


\subsection{Simulation results}
\label{subsec:sublat_symm_num_res}

\begin{figure*}[h!tb]
  \includegraphics[width=6cm,angle=-90]{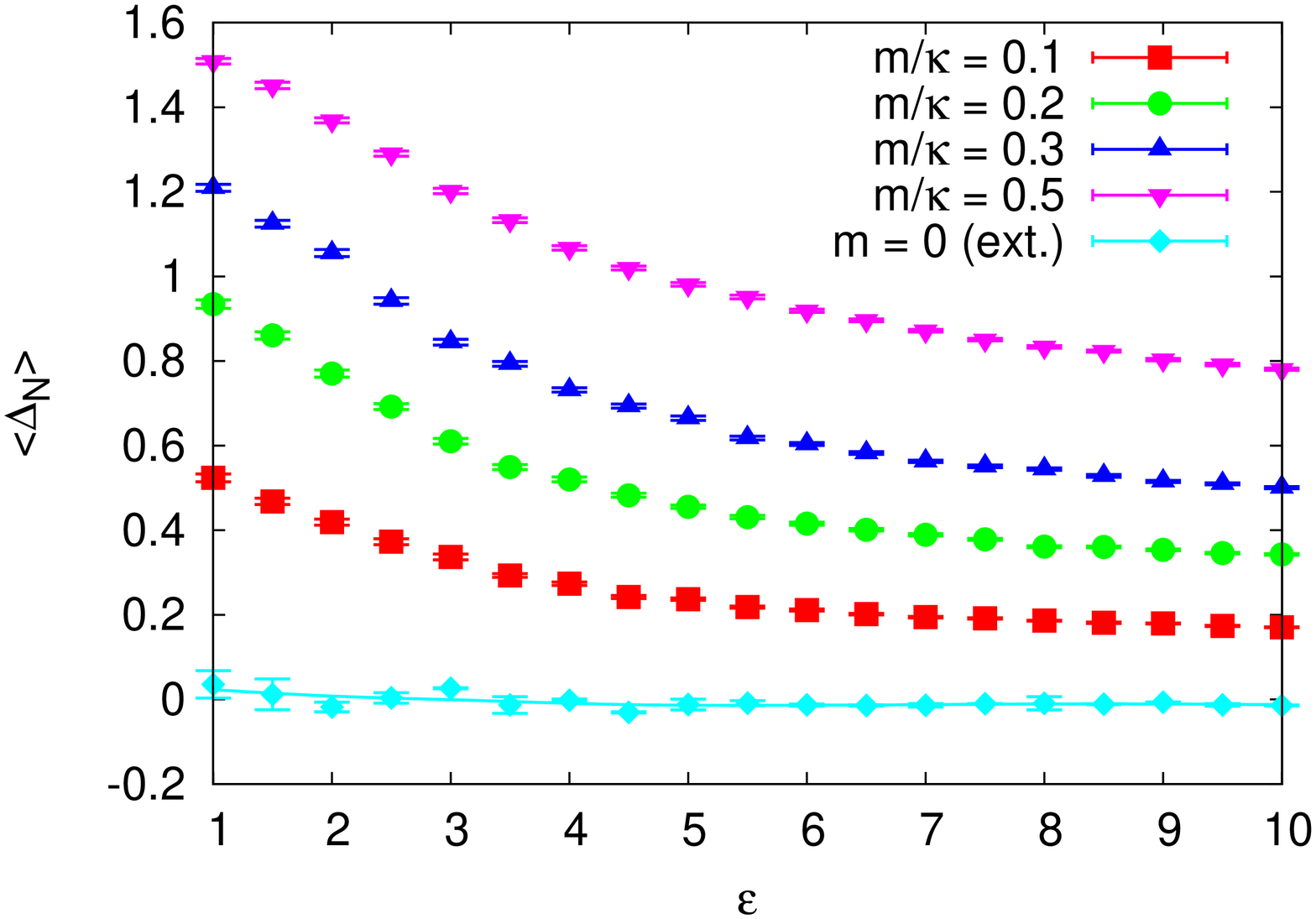}
  \includegraphics[width=6cm,angle=-90]{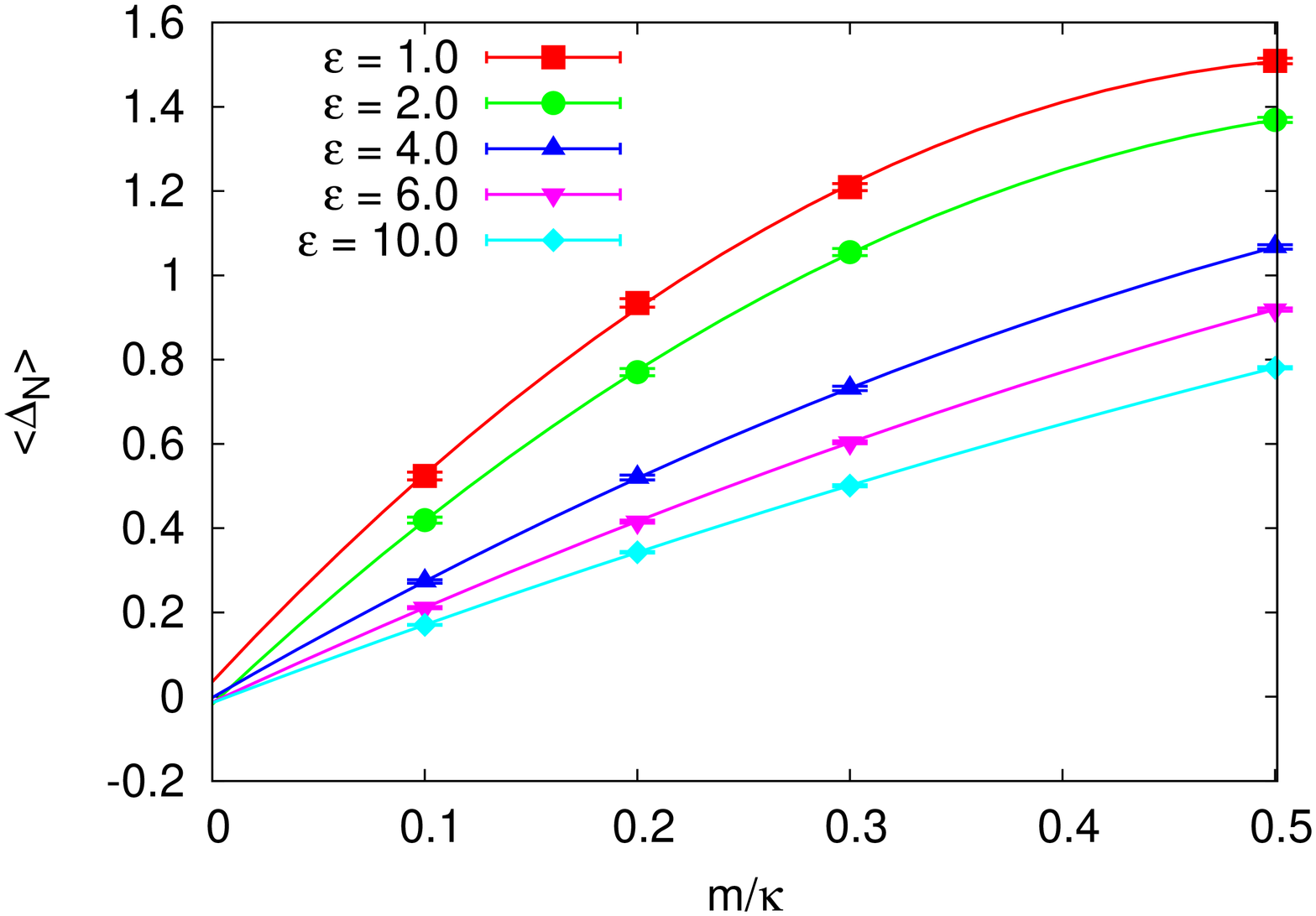}\\
  \includegraphics[width=6cm,angle=-90]{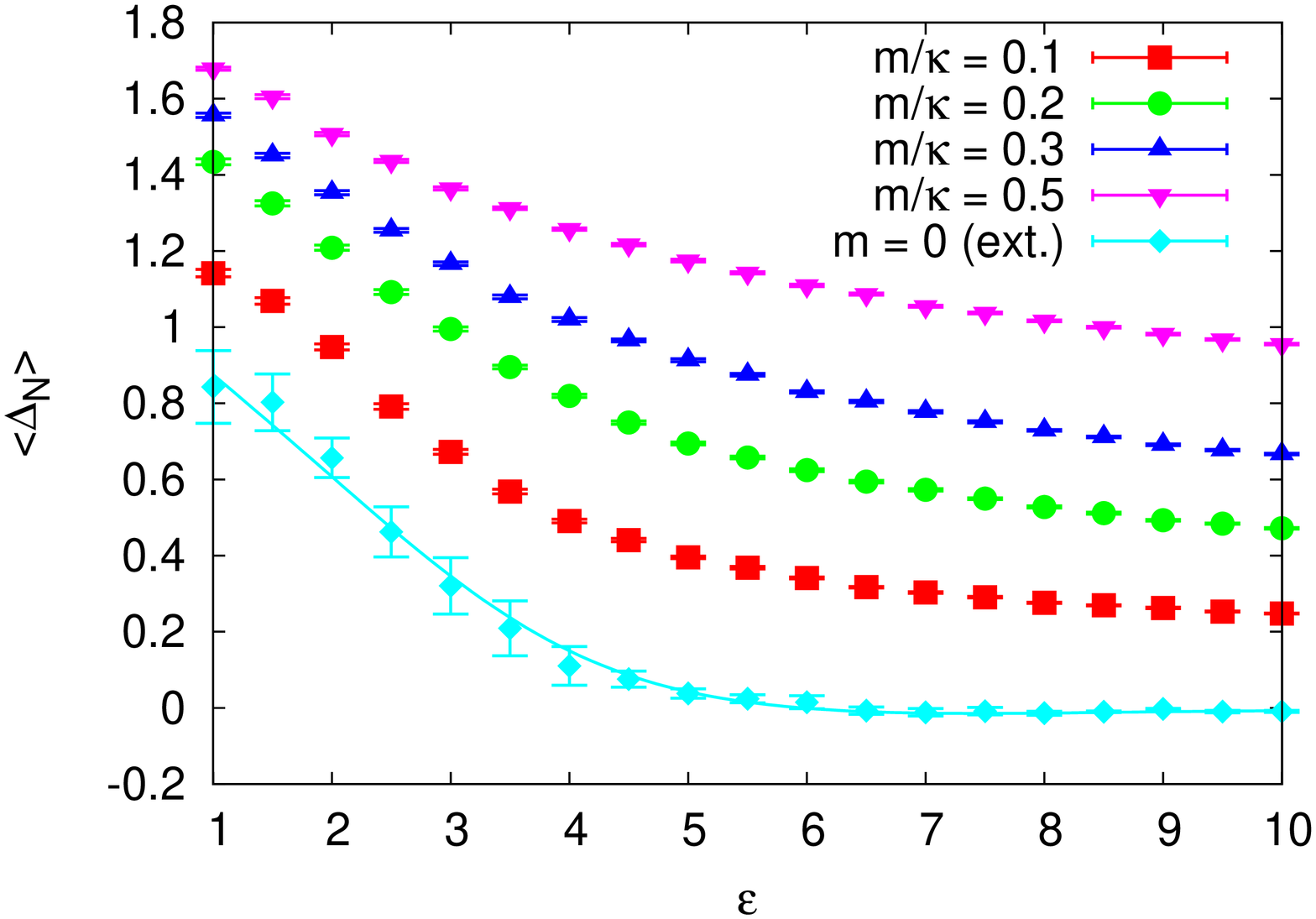}
  \includegraphics[width=6cm,angle=-90]{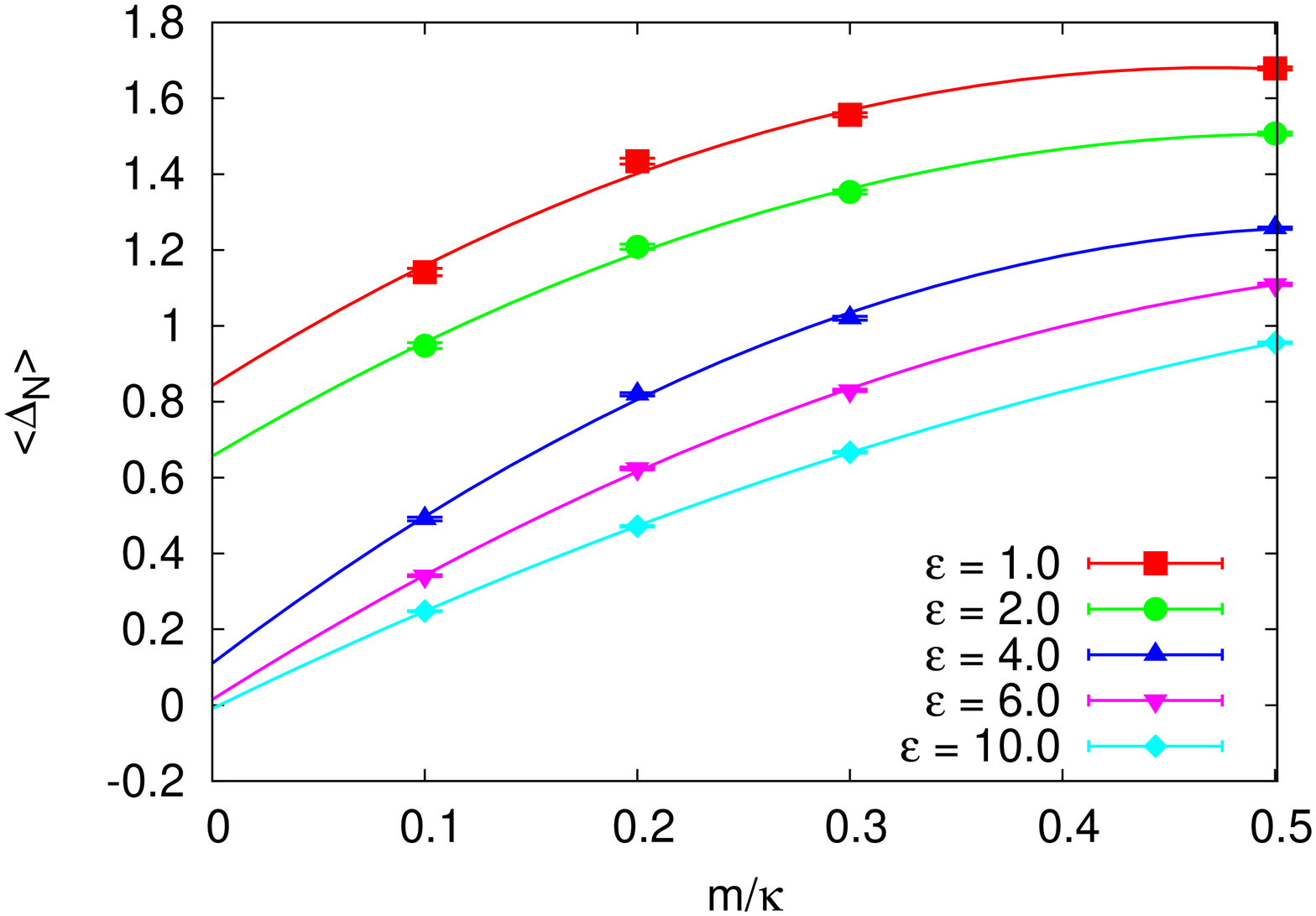}\\
  \caption{Differences of particle number densities on simple sublattices in graphene as a function of substrate dielectric permittivity $\epsilon$ (on the left) and staggered potential $m$ (on the right) on the $24^4$ lattice. Above: at $T/\kappa = 0.56$ ($\kappa \, {\Delta \tau} = 0.075$). Below: at $T/\kappa = 0.28$ ($\kappa \, {\Delta \tau} = 0.15$). Points with solid lines through them on the plots on the left are the results of extrapolation to the limit $m \rightarrow 0$. These solid lines are weighted splines and are shown to guide the eye. Solid lines on the plots on the right are the quadratic fits which were used for extrapolation. }
  \label{fig:condensate}
\end{figure*}

 On Fig. \ref{fig:condensate} we plot the expectation value $\Delta_N$ of the difference of particle number densities on even and odd lattice sites as a function of the substrate dielectric permittivity $\epsilon$ at fixed values of $m/\kappa$ (plots on the left) and as a function of $m/\kappa$ at fixed values of $\epsilon$. We present the results for the $24^4$ lattice at the temperature $T/\kappa = 0.56$ ($\kappa \, {\Delta \tau} = 0.075$, plots at the top) and at $T/\kappa = 0.28$ ($\kappa \, {\Delta \tau} = 0.15$, plots at the bottom). $\vev{\Delta_N}$ gradually increases as we move into the strong-coupling region (small $\epsilon$) or to larger values of $m/\kappa$. In the weak-coupling region (large $\epsilon$) $\vev{\Delta_N}$ is almost a linear function of $m$, while in the strong-coupling limit this dependence becomes essentially nonlinear.

\begin{figure}[h!tb]
  \includegraphics[width=6cm,angle=-90]{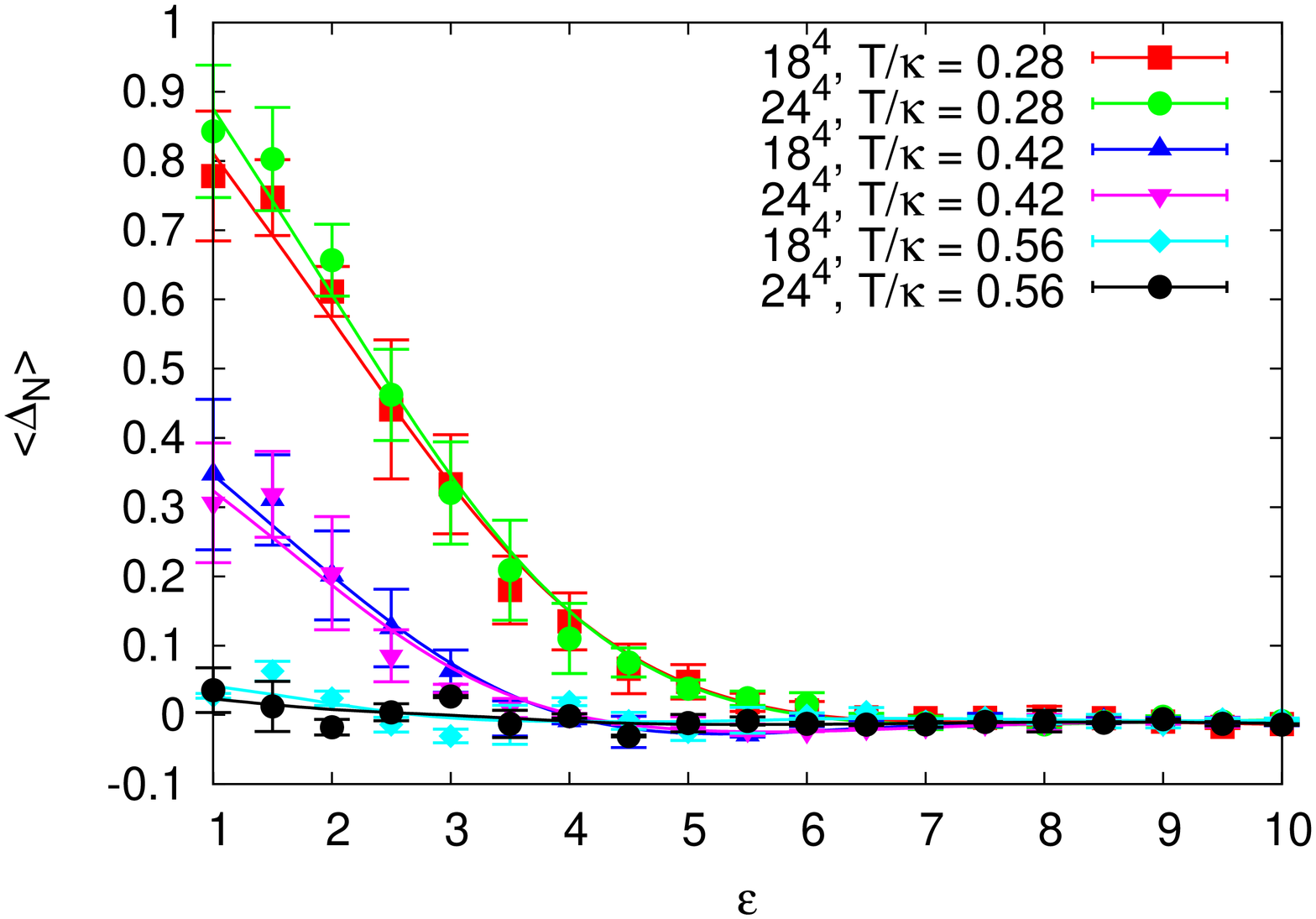}\\
  \caption{Extrapolation of the difference of particle number densities on simple sublattices $\Delta_N$ to the limit $m \rightarrow 0$ as a function of substrate dielectric permittivity $\epsilon$ at different lattice parameters. Solid lines are the weighted splines which are plotted to guide the eye.}
  \label{fig:condensate_m0}
\end{figure}

 In order to extrapolate $\vev{\Delta_N}$ to the limit $m \rightarrow 0$, we fit the dependence of $\vev{\Delta_N}$ on $m/\kappa$ at fixed $\epsilon$ with a quadratic polynomial and use the value of this polynomial at $m = 0$ as an estimate of $\vev{\Delta_N}|_{m \rightarrow 0}$. The result of such extrapolation and the corresponding fits are shown on Fig. \ref{fig:condensate} with solid lines. All fits have $\chi^2/d.o.f.$ of order unity. On Fig. \ref{fig:condensate_m0} we also compare the dependence of the extrapolated values of $\vev{\Delta_N}$ on $\epsilon$ for different lattice parameters. While at higher temperature ($T/\kappa = 0.56$) the result of extrapolation is equal to zero within error range, at $T/\kappa = 0.42$ and $T/\kappa = 0.28$ one can clearly see that $\vev{\Delta_N}$ remains finite in the limit $m \rightarrow 0$ for $\epsilon \lesssim 4$ and grows as the temperature decreases. This indicates that sublattice symmetry of the tight-binding model (\ref{tb_hamiltonian_peierls}) is spontaneously broken due to Coulomb interaction at $T \lesssim 0.42 \, \kappa$ and at $\epsilon \lesssim 4$. For the $24^4$ lattice at $T/\kappa = 0.28$ the extrapolated value $\Delta_N$ is somewhat larger than for the $18^4$ lattice at the same temperature, as it should be for spontaneous symmetry breaking.

\begin{figure}[h!tb]
  \includegraphics[width=6cm,angle=-90]{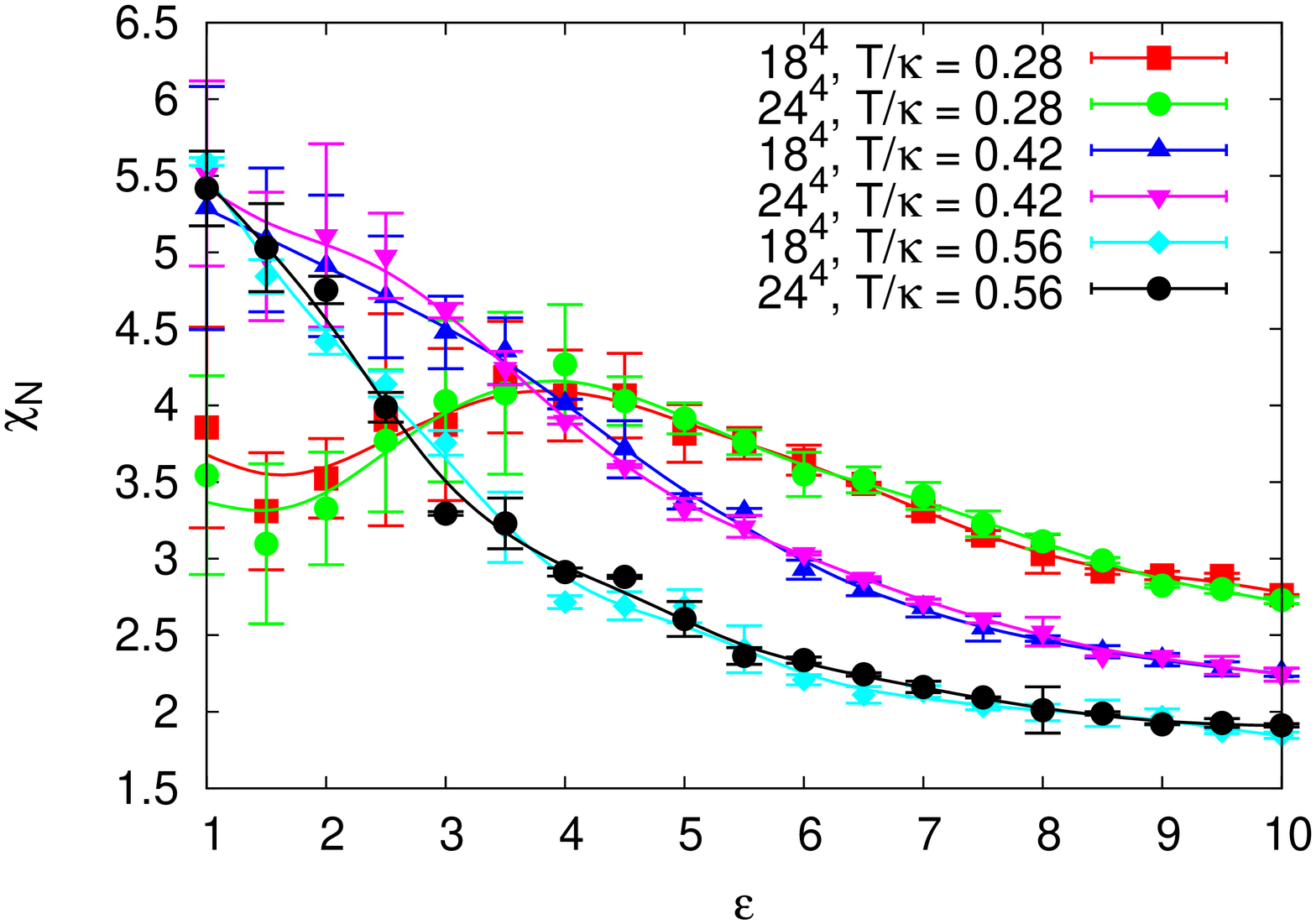}\\
  \caption{Susceptibility $\chi_N$ of the difference of particle number densities on simple sublattices $\Delta_N$ as a function of substrate dielectric permittivity $\epsilon$ at different lattice parameters. Solid lines are the weighted splines which are plotted to guide the eye.}
  \label{fig:susceptibility}
\end{figure}

 In order to get further insight into the nature of the transition to the spontaneously broken phase, on Fig. \ref{fig:susceptibility} we plot the susceptibility $\chi_N$ as a function of substrate dielectric permittivity $\epsilon$ for different lattice parameters. The susceptibility was also obtained from the quadratic fits of the dependence of $\vev{ \Delta_N }$ on $m/\kappa$ as the first derivative of the fitting polynomial at $m = 0$. At $T/\kappa = 0.56$ and $T/\kappa = 0.42$ $\chi_N$ monotonically grows as $\epsilon$ decreases, reaching its maximal value at $\epsilon = 1$, that is, for the strongest Coulomb interaction. In contrast, at $T/\kappa = 0.28$ $\chi_N$ becomes a non-monotonic function of $\epsilon$ with a characteristic peak at $\epsilon \approx 4$. For the $24^4$ lattice this peak is somewhat sharper than for the $18^4$ lattice.

\begin{figure}[h!tb]
  \includegraphics[width=6cm, angle=-90]{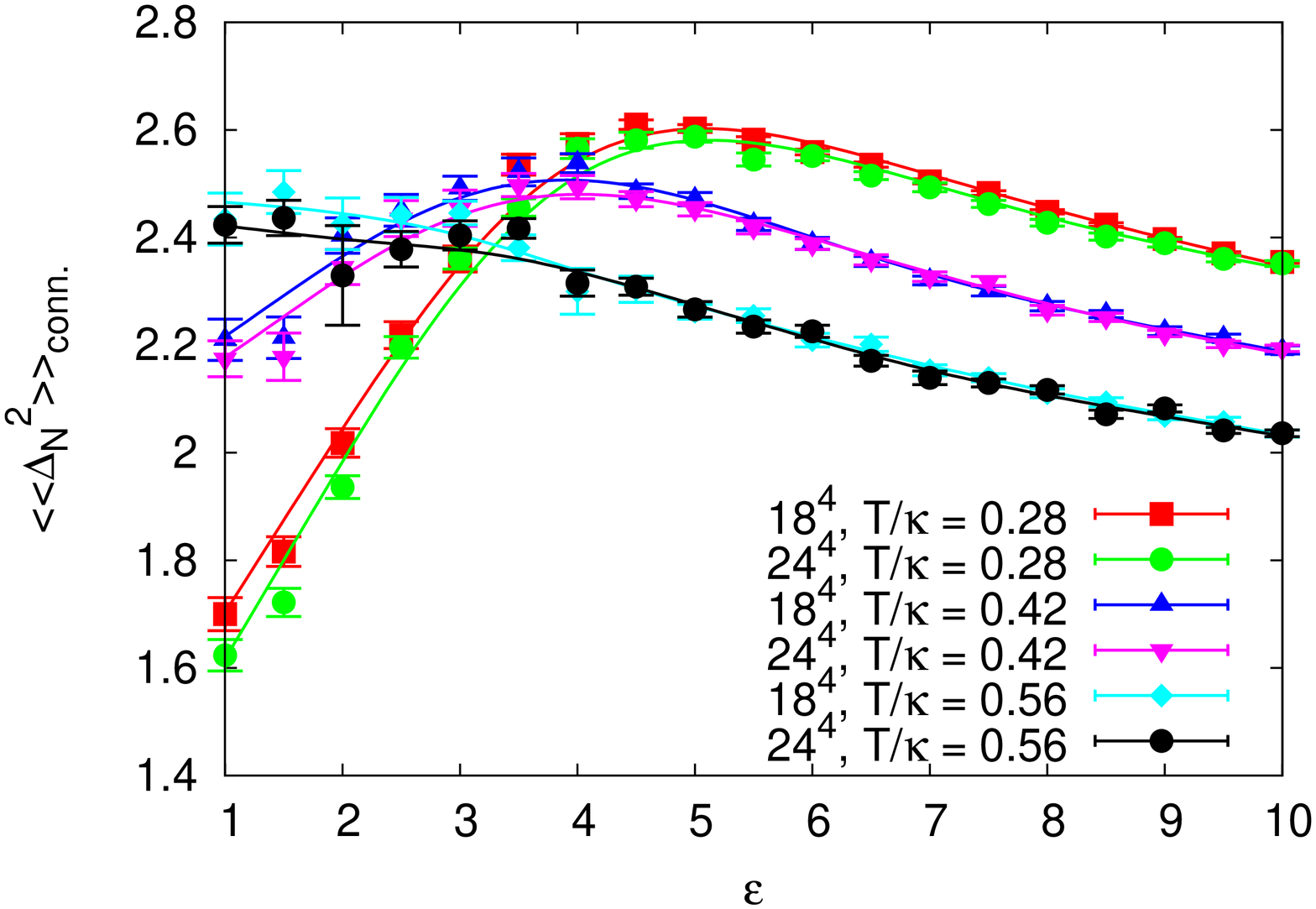}\\
  \caption{Connected part of the dispersion of the difference of particle number densities on two simple sublattices $\cev{ \Delta_N^2 }_{conn.}$ as a function of substrate dielectric permittivity $\epsilon$ at $m/\kappa = 0.1$ on different lattices. Solid lines are the weighted splines which are plotted to guide the eye.}
  \label{fig:condensate_dispersion}
\end{figure}

 To present an additional evidence of the existence of this peak which is independent of any fitting procedure, on Fig. \ref{fig:condensate_dispersion} we plot the connected part of the dispersion of the difference of particle number densities on two simple sublattices $\cev{ \Delta_N^2 }_{conn.}$, which was directly calculated on different lattices according to (\ref{npart_diff_square_latt}) at $m = 0.2 \, \kappa$. $\cev{ \Delta_N^2 }_{conn.}$ as a function of $\epsilon$ also has a distinct peak at $4 \lesssim \epsilon \lesssim 5$ for $T/\kappa = 0.28$, and a somewhat less pronounced peak at $3 \lesssim \epsilon \lesssim 4$ at $T/\kappa = 0.42$. Interestingly, for $\cev{ \Delta_N^2 }_{conn.}$ the height of the peaks practically does not depend on the lattice size.

\begin{figure}[h!tb]
  \includegraphics[width=6cm, angle=-90]{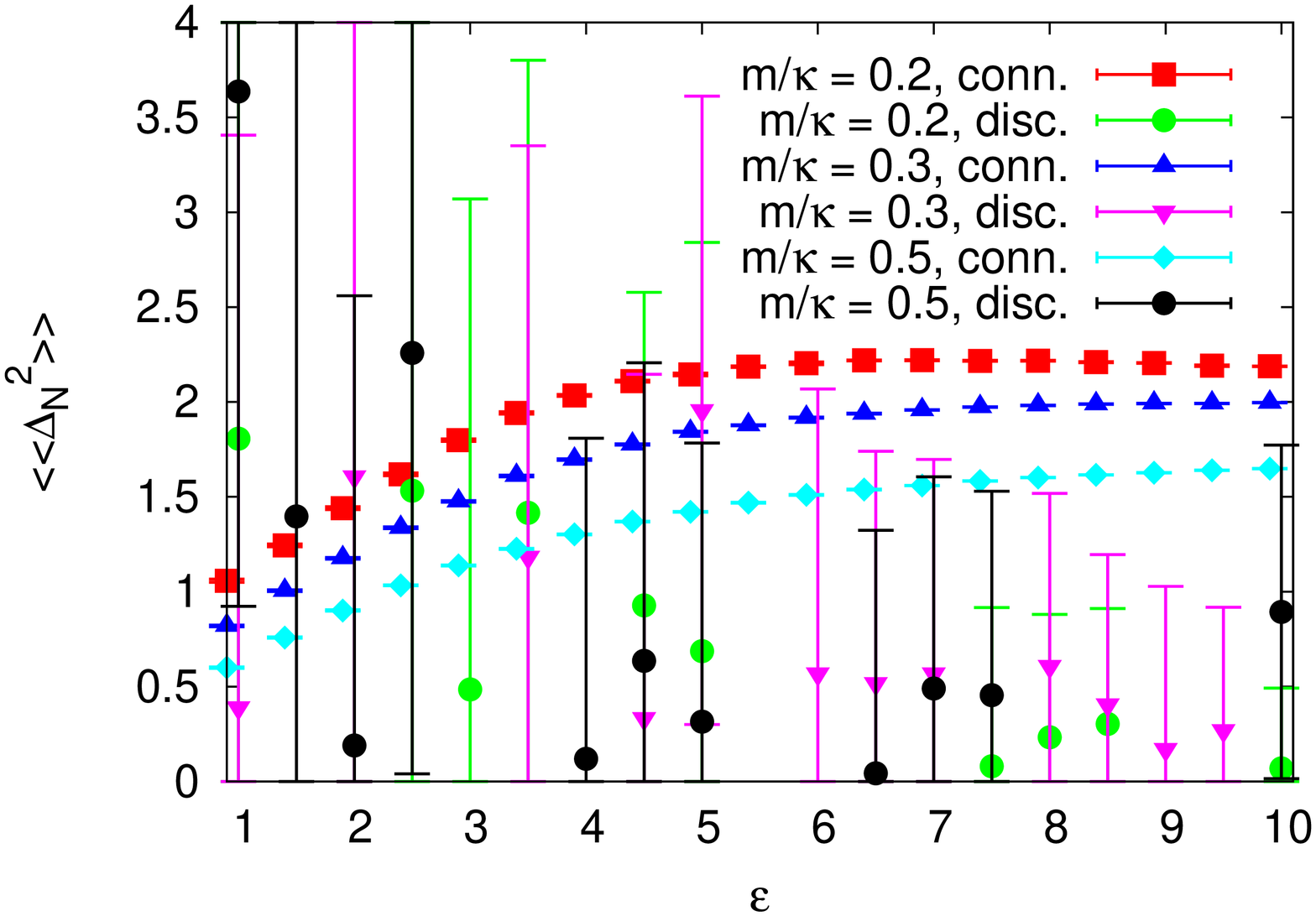}\\
  \caption{A comparison of connected and disconnected contributions to the dispersion of the difference of particle number densities (\ref{npart_diff_square_latt}) on simple sublattices $\cev{\Delta_N^2}$ for the $18^4$ lattice at $\kappa {\Delta \tau} = 0.2$.}
  \label{fig:condensate_dispersion_conndisc_k0.2000_s18_t18}
\end{figure}

 Our reason for considering only the connected part of $\cev{ \Delta_N^2 }$ is that the disconnected part turns out to be much noisier. In order to illustrate this observation, on Fig. \ref{fig:condensate_dispersion_conndisc_k0.2000_s18_t18} we plot both the connected and disconnected contributions for the $18^4$ lattice with $\kappa {\Delta \tau} = 0.2$ and $m/\kappa = 0.2, \, 0.3, \, 0.5$ as a function of $\epsilon$. The disconnected contribution was calculated using $500$ Gaussian stochastic estimators (see e.g. \cite{DeGrandDeTarLQCD}, Sec. 11.1), which required several hundreds core-hours per data point (for a 2.4 GHz Intel Xeon CPU) in the strong-coupling regime and at the smalles value of $m$ ($m/\kappa = 0.1$). While this amount of computer time is already close to the time required to generate the field configurations, the numerical errors of the disconnected part are still much larger than those of the connected part, especially in the strong-coupling regime. Thus the reliable estimate of the former would be prohibitively expensive, and we do not consider it here. Probably some progress in this direction could be made by using graphic cards (GPUs) \cite{Alexandrou:12:1} or by applying some more refined measurement procedure. We leave such developments as a direction for further work. Since we only use $\cev{ \Delta_N^2 }$ for a more precise location of the transition point, one can expect that the connected contribution alone can also be a good estimator. Indeed, if there is a second-order phase transition in the tight-binding model (\ref{tb_hamiltonian_peierls}), than all order parameters have singularities at a single value of $\epsilon$, and if there is a crossover, than the critical value is not well-defined anyway.

 We conclude that the behavior of $\Delta_N$ and $\chi_N$ at $T = 0.28 \, \kappa$ is suggestive of a second-order quantum phase transition with respect to substrate dielectric permittivity $\epsilon$ at the critical value $4 \lesssim \epsilon_c \lesssim 5$. At $T = 0.56 \, \kappa$ sublattice symmetry is always restored in the limit $m \rightarrow 0$ and this phase transition is obviously absent. The interpretation of the lattice data at $T = 0.42 \, \kappa$ is not quite straightforward: while the extrapolation of $\Delta_N$ to the limit $m \rightarrow 0$ yields nonzero result and the dispersion $\cev{ \Delta_N^2 }_{conn.}$ as a function of $\epsilon$ still has the characteristic peak, the susceptibility $\chi_N$ is a monotonic function of $\epsilon$. It is likely that at this temperature we are close to the endpoint of the second-order phase transition line in the parametric space $\lr{\epsilon, T}$, where the second-order phase transition disappears or turns into a crossover. The fact that at $T = 0.42 \, \kappa$ the peak of $\cev{ \Delta_N^2 }_{conn.}$ is situated at smaller $\epsilon$ than at $T = 0.28 \, \kappa$ suggests that for higher temperatures the critical value of $\epsilon$ becomes somewhat smaller. As discussed in the Introduction, in this paper we are interested in the low-temperature phase of the theory. Therefore we conclude that the expected pattern of spontaneous symmetry breaking in the low-temperature phase is observed for $T \lesssim 0.28 \, \kappa$, and do not study the presumable finite-temperature phase transition at $T = T_c \approx 0.42 \, \kappa$. Finally, we note that the volume-independence of $\cev{ \Delta_N^2 }_{conn.}$ at $m/\kappa = 0.1$ might indicate that for nonzero $m$ the second-order phase transition at $T \lesssim 0.28 \, \kappa$ and $\epsilon \approx 4$ also turns into a crossover.

 Combining all the data, we estimate the critical value of the substrate dielectric permittivity as $\epsilon_c = 4 \pm 1$. Expressing the effective QED coupling constant $\alpha$ in terms of $\epsilon$ as $\alpha = \frac{\alpha_0}{v_F} \, \frac{2}{\epsilon + 1}$, $\alpha_0 \approx 1/137$, we find that this value corresponds to $\alpha_c = 0.9 \pm 0.2$. This estimate is in agreement with the results of simulations of graphene effective field theory with staggered fermions \cite{Lahde:09:1, Lahde:09:2, Lahde:09:3, Lahde:11:1, Ulybyshev:12:1}, where the second-order phase transition to the phase with spontaneously broken chiral symmetry was observed at $\alpha_c = 1.11 \pm 0.06$. We further discuss the phase structure of the tight-binding model (\ref{tb_hamiltonian_peierls}) in the concluding Section \ref{sec:conclusions}.

\section{Graphene conductivity from the Green-Kubo relations}
\label{sec:conductivity}

\subsection{Basic definitions and lattice observables}
\label{subsec:conductivity_defs}

 In this Section we study numerically the conductivity of graphene monolayer, that is, the linear response of the electric current to the applied homogeneous electric field. In order to define the operator of electric current within the tight-binding model (\ref{tb_hamiltonian_peierls}), we consider the time evolution $\frac{d}{d t} \, \hat{q}\lr{s, \xi} = -i \lrs{\hat{q}\lr{s, \xi}, \, \hat{H} }$ of the charge operator (\ref{charge_redef}). This leads to the charge conservation equation of the form
\begin{eqnarray}
\label{charge_conservation}
 \frac{d}{d t} \, \hat{q}\lr{s, \xi} =
 \sum\limits_{b} \hat{J}_{b}\lr{s, \xi}  ,
\end{eqnarray}
where $t$ is the real (Minkowski) time and $\hat{J}_{a}\lr{s, \xi}$ is the operator of the electric current flowing through the lattice link which goes in direction $a$ and originates from lattice site with coordinates $\lr{s, \xi}$. It is equal to the difference of the currents $\hat{J}_{\sigma, a}\lr{s, \xi}$ of ``particles'' and ``holes'':
\begin{eqnarray}
\label{current_def}
 \hat{J}_b\lr{\xi} = \hat{J}_{\uparrow, b}\lr{\xi} - \hat{J}_{\downarrow, b}\lr{\xi}
 \nonumber \\
 \hat{J}_{\sigma, b}\lr{\xi} =
 i \kappa \,
 \hat{\psi}_{\sigma}^{\dag}\lr{\beta, \xi + \rho_b} \,
 e^{\mp i \hat{\theta}_b\lr{\xi}}
 \hat{\psi}_{\sigma}\lr{\alpha, \xi}
 - \nonumber \\ -
 i \kappa \,
 \hat{\psi}_{\sigma}^{\dag}\lr{\alpha, \xi}
 e^{\pm i \hat{\theta}_b\lr{\xi}}
 \hat{\psi}_{\sigma}\lr{\beta, \xi + \rho_b},
 \nonumber \\
 \hat{J}_{b}\lr{\alpha, \xi} \equiv \hat{J}_{b}\lr{\xi},
 \quad
 \hat{J}_{b}\lr{\beta, \xi} \equiv \hat{J}_{b}\lr{\xi - \rho_b}.
\end{eqnarray}

 To study the conductivity of graphene within the linear response theory, we have to introduce the classical time-dependent background electromagnetic field in the tight-binding Hamiltonian (\ref{tb_hamiltonian_peierls}). This amounts to replacing the operators $\hat{\theta}_{XY}$ in (\ref{tb_hamiltonian_abstract}) with the corresponding classical variables (\ref{vector_potential_discretization}). As it should be, the electric current operator $\hat{J}_b\lr{s, \xi}$ is then equal to the derivative of the Hamiltonian (\ref{tb_hamiltonian_peierls}) over the classical link variables $\theta_b\lr{s, \xi}$. Correspondingly, the linear response of the electric current to a small variation $\delta \theta_b\lr{\xi, t}$ of the classical link variables is given by
\begin{eqnarray}
\label{linear_response_def}
 \vev{\hat{J}_b\lr{\xi, t}}
 =
 \sum\limits_{c, \xi'} \int\limits_{-\infty}^{+\infty} dt' \,
 G_{R \, bc}\lr{\xi, t; \xi', t'} \, {\delta \theta_c\lr{\xi', t'}}  .
\end{eqnarray}
Here $\hat{J}_b\lr{\xi, t}$ is the current operator in the Heisenberg representation and $G_{R \, bc}\lr{\xi, t; \xi', t'}$ is the retarded current-current correlator
\begin{eqnarray}
\label{retarded_propagator_def}
 G_{R \, bc}\lr{\xi, t; \xi', t'} = i \, \theta\lr{t - t'}
 \times \nonumber \\ \times
 \tr\lr{\lrs{ \hat{J}_b\lr{\xi, t}, \hat{J}_c\lr{\xi', t'} } \, e^{- \hat{H}/T}}
\end{eqnarray}
with $\theta\lr{t - t'}$ denoting the Heaviside step function.

 Let us now consider the infinitesimal spatially homogeneous time-dependent electric field $\delta \vec{E}\lr{t} = \frac{\partial}{\partial t} \, \delta \vec{A}\lr{t}$. According to the definition (\ref{vector_potential_discretization}), the corresponding variation of the classical link variables is ${\delta \theta_b\lr{\xi, t}} = a \, \lr{\vec{e}_{b} \cdot {\delta \vec{A}\lr{t}}}$. Performing the Fourier transform ${\delta \vec{A}\lr{w}} = {\delta \vec{E}\lr{w}}/w = \int dt e^{-i w t} \, {\delta \vec{A}\lr{t}}$ and taking into account the spatial homogeneity, we can also write the relation (\ref{linear_response_def}) as
\begin{eqnarray}
\label{linear_response_fourier}
 \vev{\hat{J}_b\lr{w}} = \frac{a}{w} \, \sum\limits_{c} G_{R \, b c}\lr{w} \,
 \lr{\vec{e}_c \cdot {\delta \vec{E}\lr{w}}} ,
\end{eqnarray}
where
\begin{eqnarray}
\label{retarded_propagator_fourier}
 G_{R \, b c}\lr{w} = \sum\limits_{\xi} \, \int\limits_{-\infty}^{\infty} dt \,
 e^{-i w t} \,
 G_{R \, b c}\lr{0, 0; \xi, t}  .
\end{eqnarray}

 The conductivity of graphene is defined as the coefficient relating the total charge transported through unit length per unit time and the applied electric field \cite{Geim:07:1, Miransky:02:1, Katsnelson:06:1, Stauber:08:1}. Since the canonical dimensionality of the electric field strength is $L^{-2}$ (where $L$ is the unit length), $\sigma\lr{w}$ is a dimensionless quantity. For conversion to the SI system of units, it should be multiplied by $e^2/\hbar$.

 For simplicity we assume that the electric field ${\delta \vec{E}\lr{w}}$ is parallel to one of the lattice link vectors $\vec{e}_{b_0}$. Taking into account that the side of the plaquette of the dual lattice which is perpendicular to $\vec{e}_{b_0}$ is equal to $\sqrt{3} \, a$ and averaging over all equivalent directions $b_0$, we arrive at the following expression for the AC conductivity of graphene:
\begin{eqnarray}
\label{ac_conductivity}
 \sigma\lr{w} = \frac{G_{R \, bc}\lr{w} \, T_{bc}}{3 \, \sqrt{3} \, w}  ,
\end{eqnarray}
where
\begin{eqnarray}
\label{t_tensor_def}
 T_{bc} = \vec{e}_b \cdot \vec{e}_c = 3/2 \, \delta_{bc} - 1/2
\end{eqnarray}
and we assume summation over the repeated indices $b, c$.

 In practice, the AC conductivity $\sigma\lr{w}$ can be extracted from the Euclidean current-current correlator, which we define as
\begin{eqnarray}
\label{corr_euclide}
 G\lr{\tau} = \frac{1}{3 \sqrt{3} \, L_x L_y} \, \sum\limits_{\xi, \xi'} \, T_{bc}
\times \nonumber \\ \times \,
\tr\lr{e^{\tau \hat{H}} \, \hat{J}_b\lr{\xi} \, e^{-\tau \hat{H}} \, \hat{J}_c\lr{\xi'} \, e^{- \hat{H}/T} }
\end{eqnarray}
with the help of the Green-Kubo relations \cite{Kadanoff:63:1, Hosoya:84:1, Asakawa:01:1, Aarts:07:1}:
\begin{eqnarray}
\label{corr_eq}
 G\lr{\tau} = \int\limits^{\infty}_{0}
 \frac{dw}{2 \pi} \, K\lr{w, \tau} \, \sigma\lr{w} ,
\end{eqnarray}
where the thermal kernel $K\lr{w, \tau}$ is \cite{Aarts:07:1}
\begin{eqnarray}
\label{kernel}
 K\lr{w, \tau} = \frac{2 w \, \cosh\lr{w \lr{\tau - \frac{1}{2 T}} }}{\sinh\lr{\frac{w}{2 T}}}  .
\end{eqnarray}

 In Appendix \ref{appsec:free_spectral_func} we derive explicit expressions for the Euclidean current-current correlator $G^{\lr{0}}\lr{\tau}$ and the AC conductivity $\sigma^{\lr{0}}\lr{w}$ for the tight-binding model (\ref{tb_hamiltonian_peierls}) in the absence of Coulomb interaction. For $w \ll \kappa$, $\sigma^{\lr{0}}\lr{w}$ can be approximated as
\begin{eqnarray}
\label{free_conductivity_simple}
 \sigma^{\lr{0}}\lr{w}
 \approx
 \Xi \, \delta\lr{w}
 + \nonumber \\ +
 \frac{\theta\lr{w - 2 m}}{4} \, \lr{1 + \frac{4 m^2}{w^2}} \,
 \tanh\lr{\frac{w}{4 \, T}} ,
\end{eqnarray}
with $\Xi$ being some constant. The $\delta$-function singularity at $w = 0$ is a common feature of all ideal crystals which arises due to the absence of scattering of charge carriers. Thus, strictly speaking, $\sigma^{\lr{0}}\lr{w}$ has no well-defined zero-frequency limit. A commonly quoted universal value $\sigma_0 = 1/4$ (in units of $e^2/\hbar$) is obtained from (\ref{free_conductivity_simple}) at $w \gg T$, $w \gg m$ (but still $w \ll \kappa$) \cite{Katsnelson:06:1, Stauber:08:1}. Frequency dependence of $\sigma^{\lr{0}}\lr{w}$ is illustrated on Fig. \ref{fig:free_ac_conductivity} (see Appendix \ref{appsec:free_spectral_func}).

 In order to obtain the expression for the discretized correlator (\ref{corr_euclide}) on the lattice, we insert the current operators (\ref{current_def}) between $0$'th and $L_{\tau}$'th and between $\lr{\tau/{\Delta \tau}}$'th and $\lr{\tau/{\Delta \tau} + 1}$'th factors in the Feynman-Kac representation (\ref{em_trace1}) of the partition function (\ref{partition_function1}). For the time being we assume that $\tau \neq 0$. Repeating the derivation of the fermionic lattice action presented in Subsection \ref{subsec:fermionic_action}, we arrive at the following expression for the discretized correlator (\ref{corr_euclide}) in terms of the fermionic path integral:
\begin{eqnarray}
\label{corr_euclide_path_int}
 G\lr{\tau} = \mathcal{Z}^{-1}
 \int \mathcal{D}\bar{\eta}_{\sigma} \mathcal{D}\eta_{\sigma} \mathcal{D}\phi \,
 \frac{T_{bc}}{3 \sqrt{3} L_x L_y}
 \times \nonumber \\ \times
 \sum\limits_{\sigma, \sigma'} \,
 \lr{\bar{\eta}_{\sigma}\lr{0} j_{\sigma, b} \eta_{\sigma}\lr{0}} \,
 \lr{\bar{\eta}_{\sigma'}\lr{\tau} j_{\sigma', c} \eta_{\sigma'}\lr{\tau} } \,
 \times \nonumber \\ \times
 \expa{ - \sum\limits_{\sigma} \bar{\eta}_{\sigma} M_{\sigma}\lrs{\phi} \eta_{\sigma} - S_{em}\lrs{\phi}  }  ,
\end{eqnarray}
where we have omitted the arguments of the field variables for the sake of brevity and $j_{\sigma, b}$ is the one-particle operator of the total current of particles with spin $\sigma$ on the whole lattice, which is defined by the identity
\begin{eqnarray}
\label{one_part_current_def}
 \sum\limits_{\xi} \, \hat{J}_{\sigma, b}\lr{\xi}
 = \nonumber \\ =
 \sum\limits_{s, \xi, s', \xi'}
 j_{\sigma, b}\lr{s, \xi; s', \xi'} \hat{\psi}_{\sigma}^{\dag}\lr{s, \xi} \, \hat{\psi}_{\sigma}\lr{s', \xi'}  .
\end{eqnarray}
Correspondingly, the symbol $\eta_{\sigma}\lr{\tau}$ in (\ref{corr_euclide_path_int}) denotes the one-particle wave function $\eta_{\sigma}\lr{s, \xi, \tau}$.

 Integrating over the fermion fields and taking into account that $M_{\downarrow} = \bar{M}_{\uparrow}$, we obtain the following expression for the correlator $G\lr{\tau}$ in terms of the fermionic hopping matrix $M_{\uparrow}\lr{s, \xi, \tau; s', \xi', \tau'}$:
\begin{widetext}
\begin{eqnarray}
\label{corr_euclide_lattice}
 G\lr{\tau}
 =
 - \frac{2 \, T_{bc}}{3 \sqrt{3} L_x L_y} \,
\vev{ \re\tr\lr{j_{\uparrow, b} \, M_{\uparrow}^{-1}\lr{0, \tau} \, j_{\uparrow, c} \, M_{\uparrow}^{-1}\lr{\tau, 0}} }
 + \nonumber \\ +
 \vev{ \frac{4 \, T_{bc}}{3 \sqrt{3} L_x L_y} \, \re\tr\lr{j_{\uparrow, b} \, M_{\uparrow}^{-1}\lr{0, 0}} \,
 \re\tr\lr{j_{\uparrow, c} \, M_{\uparrow}^{-1}\lr{\tau, \tau}} } ,
\end{eqnarray}
\end{widetext}
where $\vev{\ldots}$ denotes averaging over the electrostatic potential field $\phi\lr{s, \xi, \tau, z}$ with the weight (\ref{path_int_final}) and $M_{\uparrow}^{-1}\lr{\tau, \tau'}$ is treated as a one-particle operator. Correspondingly, the trace in (\ref{corr_euclide_lattice}) is taken over one-particle states. The first and the second summand in (\ref{corr_euclide_lattice}) are the contributions of connected and disconnected fermion diagrams, respectively.

 When $\tau = 0$, an additional interchange of field operators $\hat{\psi}_{\sigma}\lr{s, \xi}$ and $\hat{\psi}^{\dag}_{\sigma'}\lr{s', \xi'}$ is required in order to bring them in the normal order and to apply the Feynman-Kac transformation. This leads to an additional contact term at $\tau = 0$, so that
\begin{eqnarray}
\label{corr_euclide_lattice_t0}
 G\lr{0}
 =
 \frac{2 \, T_{bc}}{3 \sqrt{3} L_x L_y} \,
\vev{ \re\tr\lr{j_{\uparrow, b} \, j_{\uparrow, c} \, M_{\uparrow}^{-1}\lr{0, 0}}}
 - \nonumber \\ -
 \frac{2 \, T_{bc}}{3 \sqrt{3} L_x L_y} \vev{ \re\tr\lr{
 j_{\uparrow, b} M_{\uparrow}^{-1}\lr{0, 0} j_{\uparrow, c} M_{\uparrow}^{-1}\lr{0, 0} } }
\end{eqnarray}

\subsection{Simulation results}
\label{subsec:conductivity_num_res}

\begin{figure}[h!tb]
  \includegraphics[width=8cm]{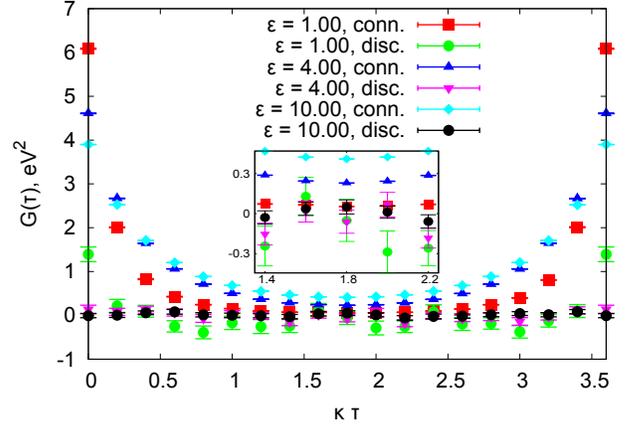}\\
  \caption{A comparison of connected and disconnected contributions to the Euclidean current-current correlator (\ref{corr_euclide}) (first and second summands in (\ref{corr_euclide_lattice})) for the $18^4$ lattice with $\kappa \, {\Delta \tau} = 0.2$ and $m/\kappa = 0.2$. The inset shows both contributions to the correlators $G\lr{\tau}$ for $\tau$ close to $\beta/2$ in a larger scale.}
  \label{fig:correlators_conndisc_mk0.2000_k0.2000_s18_t18}
\end{figure}

 First we estimate the connected and disconnected parts of the correlator (\ref{corr_euclide}), that is, the first and the second summands in (\ref{corr_euclide_lattice}). Both contributions are shown on Fig. \ref{fig:correlators_conndisc_mk0.2000_k0.2000_s18_t18} for the $18^4$ lattice with $\kappa \, {\Delta \tau} = 0.2$ and $m/\kappa = 0.2$. Disconnected contributions were estimated using $500$ Gaussian stochastic estimators \cite{DeGrandDeTarLQCD}. One can readily see that the disconnected contribution is much smaller than the connected one, and the relative statistical errors are much larger. From Fig. \ref{fig:correlators_conndisc_mk0.2000_k0.2000_s18_t18} one can also see that the relative importance of this disconnected contribution is somewhat higher for $\tau$ close to $\beta/2$ and/or for smaller values of $\epsilon$. As discussed in Subsection \ref{subsec:sublat_symm_num_res}, estimating the disconnected contributions with sufficient precision in the strong-coupling regime by using our current measurement methods would require prohibitively large computer time. For these reasons, we disregard them in what follows and leave their detailed study as a direction for future investigations.

\begin{figure*}[h!tb]
  \includegraphics[width=6cm, angle=-90]{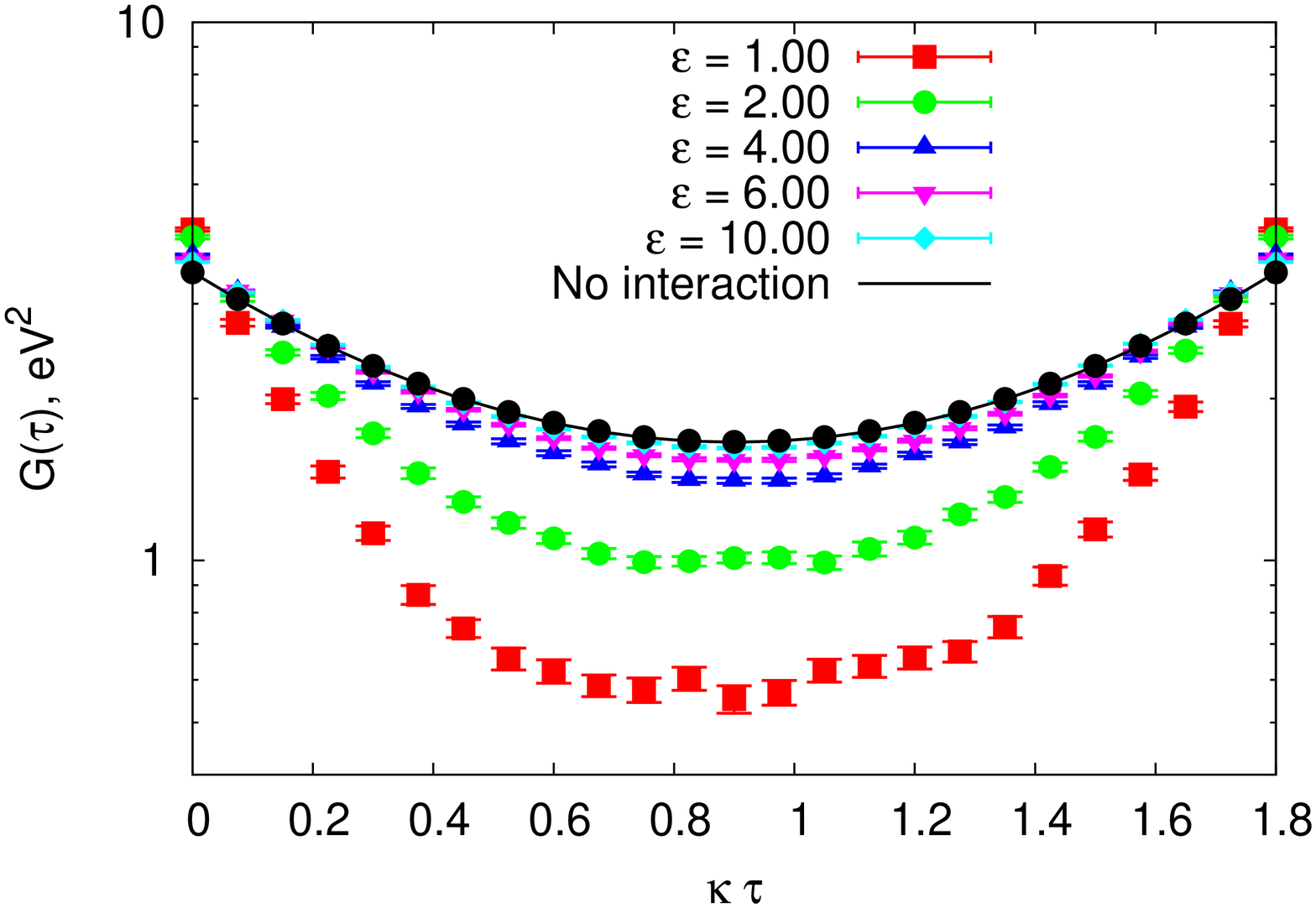}
  \includegraphics[width=6cm, angle=-90]{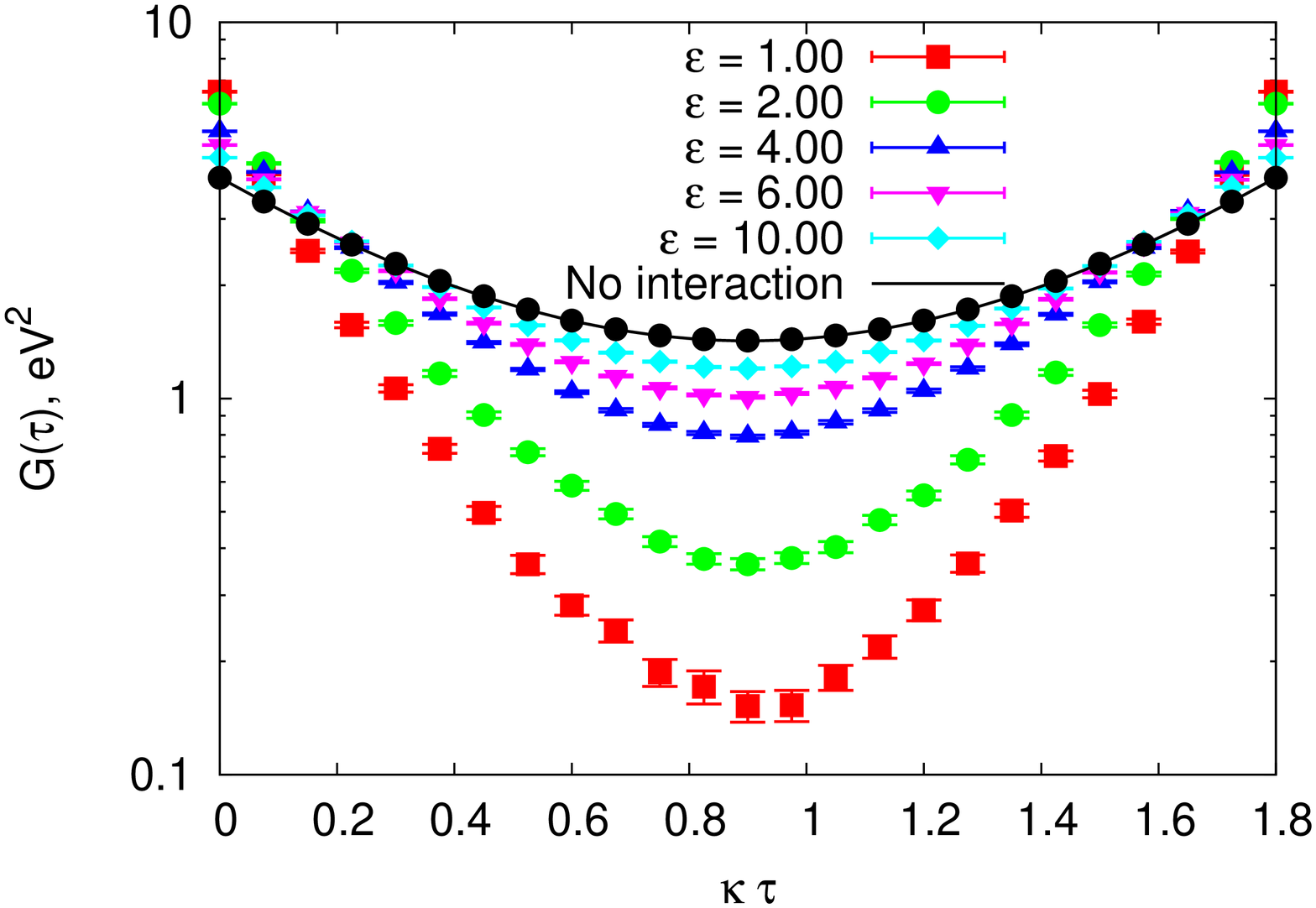}\\
  \includegraphics[width=6cm, angle=-90]{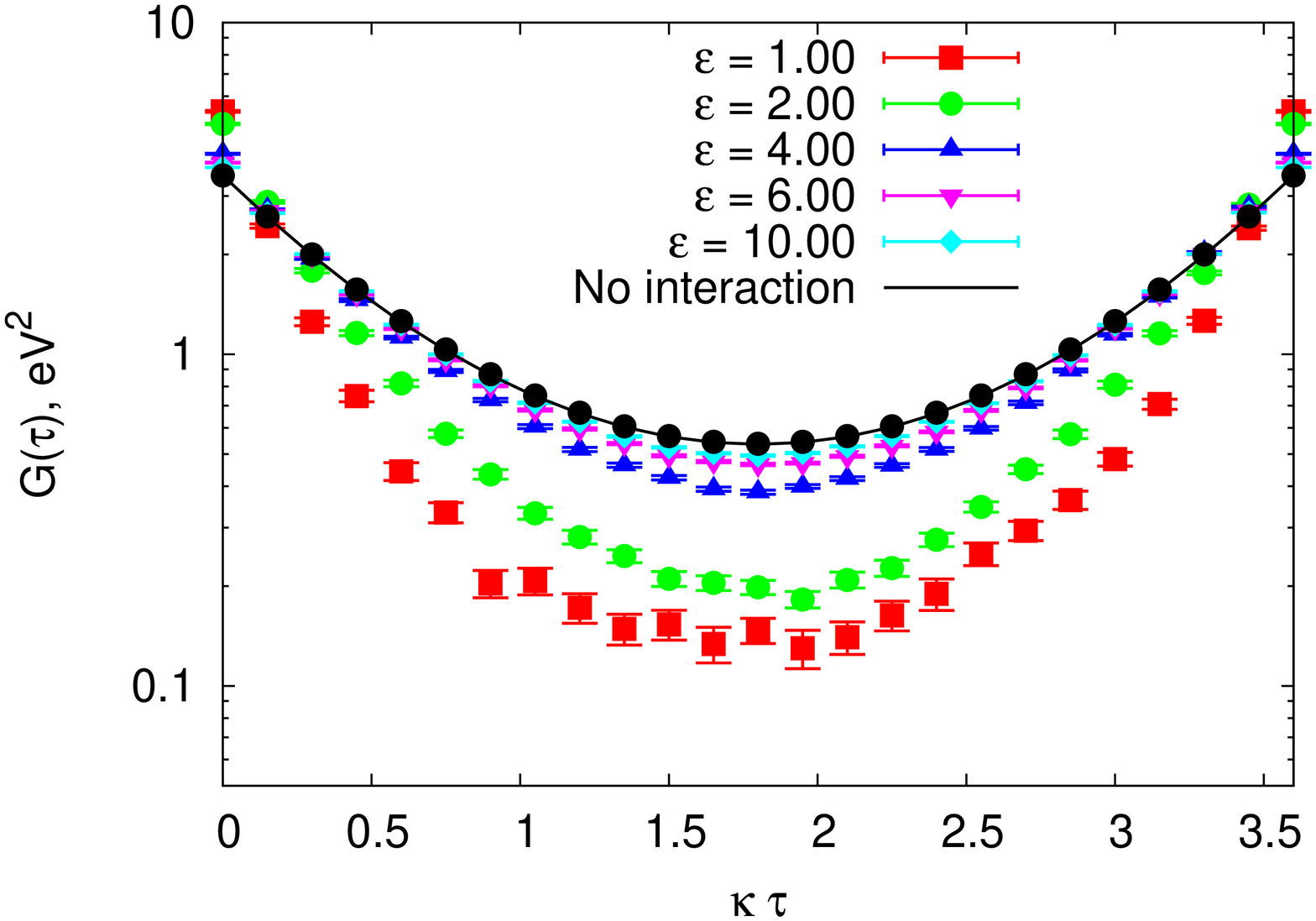}
  \includegraphics[width=6cm, angle=-90]{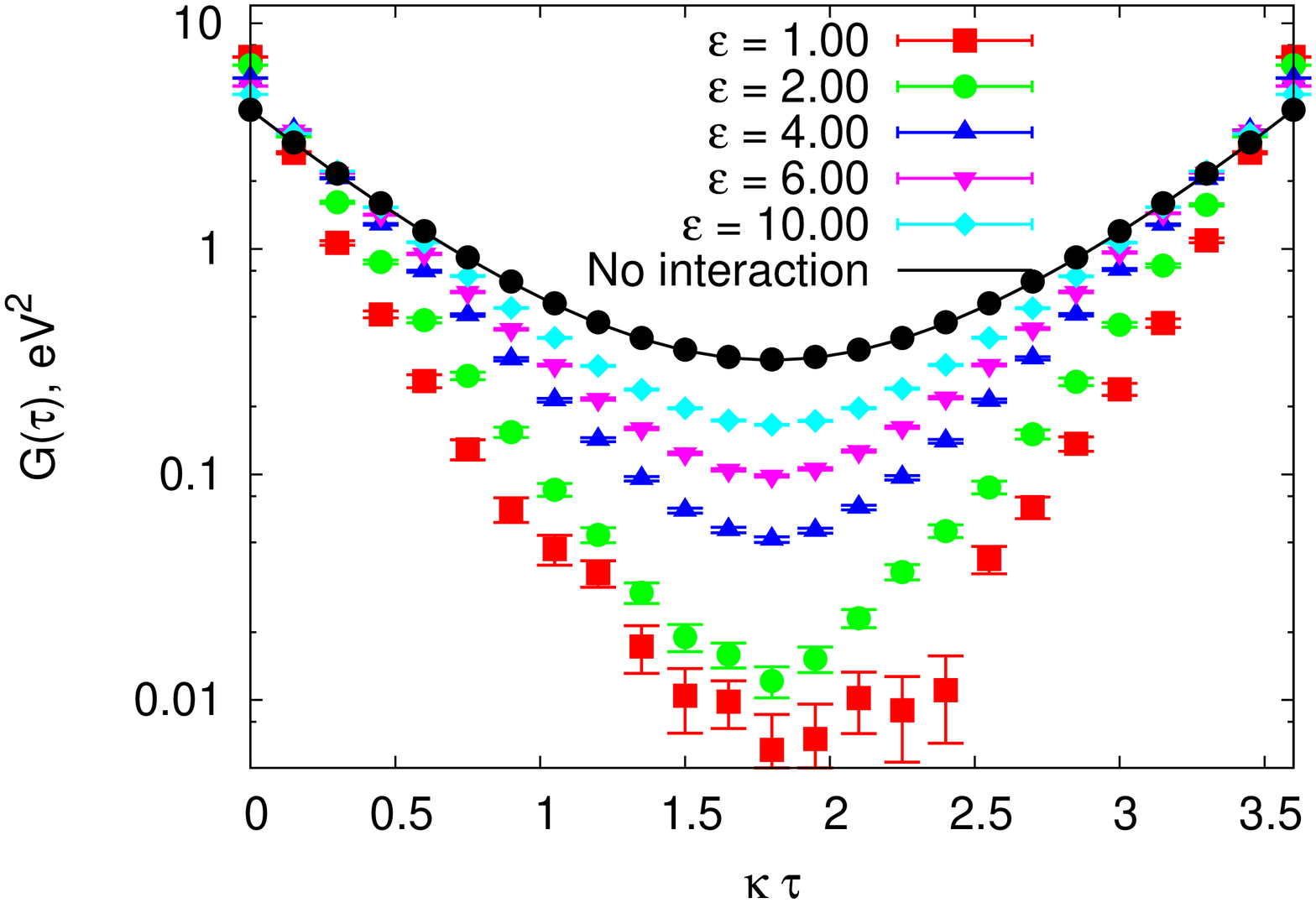}\\
  \caption{Euclidean current-current correlators (\ref{corr_euclide}) on the $24^4$ lattice at different values of substrate dielectric permittivity $\epsilon$. Above on the left: for $T/\kappa = 0.56$ ($\kappa \, {\Delta \tau} = 0.075$) and $m/\kappa = 0.1$. Above on the right: for $T/\kappa = 0.56$ ($\kappa \, {\Delta \tau} = 0.075$) and $m/\kappa = 0.5$. Below on the left: for $T/\kappa = 0.28$ ($\kappa \, {\Delta \tau} = 0.15$) and $m/\kappa = 0.1$. Below on the right: for $T/\kappa = 0.28$ ($\kappa \, {\Delta \tau} = 0.15$) and $m/\kappa = 0.5$.}
  \label{fig:correlators}
\end{figure*}

 Connected contributions to Euclidean current-current correlators (\ref{corr_euclide}) on the $24^4$ lattice at the temperature $T = 0.56 \, \kappa$ ($\kappa \, {\Delta \tau} = 0.075$), $m/\kappa = 0.1, \, 0.5$ and $T = 0.28 \, \kappa$ ($\kappa \, {\Delta \tau} = 0.15$), $m/\kappa = 0.1, \, 0.5$ are plotted on Fig. \ref{fig:correlators} for different values of the substrate dielectric permittivity $\epsilon$. One can see that as $\epsilon$ decreases and the Coulomb interaction becomes stronger, the correlators decay much faster, which, according to (\ref{corr_eq}), indicates that the AC conductivity $\sigma\lr{w}$ becomes smaller in the low-frequency region. This effect becomes more prominent at lower temperature or at larger values of the staggered potential $m$.

 For reference, on Fig. \ref{fig:correlators} we also plot the current-current correlators for the non-interacting tight-binding model (see Appendix \ref{appsec:free_spectral_func} for an explicit expression). The free correlator obtained from the expression (\ref{free_current_corr}) with continuous Euclidean time $\tau$ is plotted with black solid line. The corresponding numerical result which was calculated according to (\ref{corr_euclide_lattice}) and (\ref{corr_euclide_lattice_t0}) with $M_{\uparrow}$ given by (\ref{fermion_lattice_action}) and with $\phi\lr{s, \xi, \tau, z} = 0$ is shown with black circles. A comparison of the results of these two calculations suggests that the effect of discretization of Euclidean time $\tau$ on the current-current correlators should be rather small.

 A commonly used method to invert the integral equation (\ref{corr_eq}) and to estimate the AC conductivity $\sigma\lr{w}$ from the values of $G\lr{\tau}$ in a discrete set of lattice points is the Maximum Entropy Method (MEM) \cite{Asakawa:01:1, Aarts:07:1}. However, in practice we have found that for our data MEM does not produce stable results for $\sigma\lr{w}$ in the low-frequency limit ($w \lesssim T$). In particular, it does not reproduce the free AC conductivity $\sigma^{\lr{0}}\lr{w}$ when supplied with the free Euclidean correlator $G^{\lr{0}}\lr{\tau}$. This fact can be probably explained by the singularity and discontinuity of $\sigma^{\lr{0}}\lr{w}$ at small frequencies, which cannot be reproduced by the smooth basis functions used in MEM \cite{Asakawa:01:1, Aarts:07:1} (we have used the modified thermal kernel (\ref{kernel}) introduced in \cite{Aarts:07:1}). In fact, MEM tends to simply smear the function $\sigma\lr{w}$ at low frequencies, so that the numerically obtained function $\sigma\lr{w}$ is smooth and shows no signatures of the gap. In this situation the values of $\sigma\lr{w}$ at low frequencies are quite meaningless and cannot be compared to the universal limiting value $\sigma_0 = 1/4$ \cite{Katsnelson:06:1, Stauber:08:1}. Thus we conclude that MEM does not give a reliable estimate of the AC conductivity of graphene at small frequencies. The situation could be probably improved by more advanced modifications and tuning of the method, which are out of scope of the present paper.

\begin{figure*}[h!tb]
  \includegraphics[width=6cm, angle=-90]{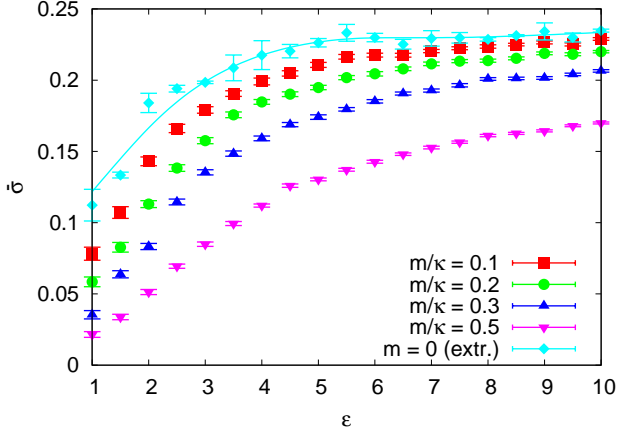}
  \includegraphics[width=6cm, angle=-90]{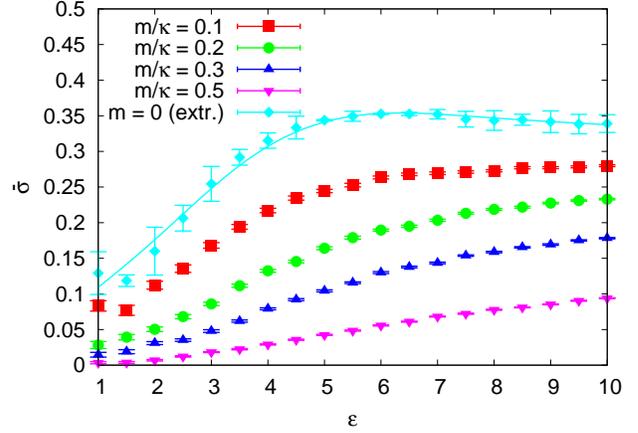}\\
  \caption{Smeared low-frequency conductivity (\ref{conductivity_midpoint}) in units of $e^2/\hbar$ as a function of substrate dielectric permittivity $\epsilon$ at different values of the ratio $m/\kappa$. On the left: on $24^4$ lattice at $T = 0.56 \, \kappa$ ($\kappa \, {\Delta \tau} = 0.075$), on the right: on $24^4$ lattice at $T = 0.28 \, \kappa$  ($\kappa \, {\Delta \tau} = 0.15$). Points with solid line through them is the extrapolation to the limit $m \rightarrow 0$. Solid lines are the weighted splines which are plotted to guide the eye.}
  \label{fig:conductivity}
\end{figure*}

 In order to obtain an estimate of the conductivity which is free of the ambiguities introduced by MEM, let us consider the Euclidean correlator (\ref{corr_euclide}) at $\tau = \beta/2$. According to (\ref{corr_euclide_path_int}) and (\ref{kernel}), its value can be represented as
\begin{eqnarray}
\label{corr_midpoint}
 G\lr{\beta/2} = \int\limits^{\infty}_{0}
 \frac{dw}{2 \pi} \, \frac{2 w}{\sinh\lr{\frac{w}{2 T}}} \, \sigma\lr{w}  .
\end{eqnarray}
The weight factor $\frac{2 w}{\sinh\lr{\frac{w}{2 T}}}$ is finite at $w \rightarrow 0$ and decays exponentially at $w \gtrsim T$. The integral in (\ref{corr_midpoint}) is thus saturated in the region with $w \lesssim T$. It is thus natural to introduce the conductivity $\bar{\sigma}$ smeared over small frequencies as
\begin{eqnarray}
\label{conductivity_midpoint}
 \bar{\sigma} = \mathcal{N}^{-1} \int\limits^{\infty}_{0}
 \frac{dw}{2 \pi} \, \frac{2 w}{\sinh\lr{\frac{w}{2 T}}} \, \sigma\lr{w}
 = \frac{1}{\pi \, T^2} \, G\lr{\beta/2} ,
\end{eqnarray}
where $\mathcal{N}$ is the normalization factor: $\mathcal{N}=\int\limits^{\infty}_{0}  \frac{dw}{2 \pi} \, \frac{2 w}{\sinh\lr{\frac{w}{2 T}}} = \pi \, T^2$. Analytical calculation of $\bar{\sigma}$ within the Dirac approximation to the non-interacting tight-binding model (see Appendix \ref{appsec:free_spectral_func}) shows that it is quite close to the limiting value $\sigma_0 = 1/4$ and does not depend on temperature in the limit $m \rightarrow 0$.

\begin{figure}[h!tb]
  \includegraphics[width=6cm, angle=-90]{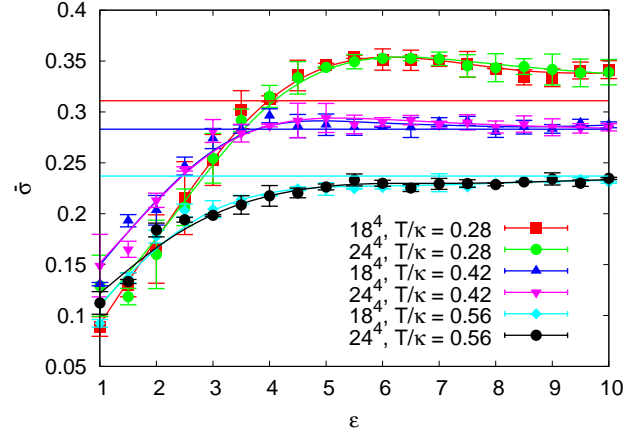}\\
  \caption{Smeared low-frequency conductivity $\bar{\sigma}$ in units of $e^2/\hbar$ after extrapolation to the limit $m \rightarrow 0$ as a function of substrate dielectric permittivity $\epsilon$ for different lattice parameters. Solid horizontal lines on the plot correspond to the results of analytic calculation of $\bar{\sigma}^{\lr{0}}$ in the non-interacting tight-binding model with $m = 0$. Solid lines which are plotted through data points are the weighted splines which are shown to guide the eye.}
  \label{fig:conductivity_m0}
\end{figure}

 Smeared low-frequency conductivity $\bar{\sigma}$ for the $24^4$ lattice at the temperature $T = 0.56 \, \kappa$ ($\kappa \, {\Delta \tau} = 0.075$) and at $T = 0.28 \, \kappa$ ($\kappa \, {\Delta \tau} = 0.15$) and with different values of $m/\kappa$ is plotted on Fig. \ref{fig:conductivity} as a function of substrate dielectric permittivity $\epsilon$. At nonzero $m$ $\bar{\sigma}$ gradually decreases with $\epsilon$. As $m$ becomes smaller, $\bar{\sigma}$ becomes almost a constant function for $\epsilon \gtrsim 4$, but changes faster at $\epsilon \lesssim 4$. In order to extrapolate the conductivity to the limit $m \rightarrow 0$, we fit its dependence on $m$ at fixed $\epsilon$ with quadratic polynomial and use the value of this polynomial at $m = 0$. All the fits yield $\chi^2/d.o.f.$ of order of unity. Results of such extrapolation for different lattices are summarized on Fig. \ref{fig:conductivity_m0}. Solid horizontal lines on the plot correspond to the results of analytic calculation of $\bar{\sigma}^{\lr{0}}$ in the non-interacting tight-binding model with $m = 0$ (see Appendix \ref{appsec:free_spectral_func}).

 The extrapolated conductivity $\bar{\sigma}$ is practically constant and close to its value in the non-interacting tight-binding model at $\epsilon \gtrsim 4$ for $T = 0.56 \, \kappa$ and for $T = 0.42 \, \kappa$, but quickly decreases with $\epsilon$ at $\epsilon \lesssim 4$. At $T = 0.28 \, \kappa$, the behavior of the conductivity is essentially the same, but the critical value of $\epsilon$ at which $\bar{\sigma}$ starts decreasing is somewhat higher, $\epsilon \approx 5$. One can also note a slight decrease in the conductivity at larger $\epsilon$, which is in agreement with perturbative calculations \cite{Rosenstein:12:1}. Remarkably, the critical values of $\epsilon$ which separate the regimes of constant and decreasing conductivity coincide with the critical values which were obtained in Section \ref{sec:sublat_symm} from the analysis of spontaneous breaking of sublattice symmetry at the corresponding temperatures. At $\epsilon = 1$, when the strength of Coulomb interaction is maximal, the smeared conductivity is still finite and comprises $20 - 30 \%$ of its weak-coupling value. Such behaviour of the conductivity should be contrasted with the results obtained from simulations of the graphene lattice effective field theory \cite{Ulybyshev:12:1}, for which the low-frequency conductivity in the strong-coupling phase decreased by at least one order of magnitude as compared to its value in the weak-coupling phase. It seems therefore that the semimetal-insulator phase transition associated with spontaneous symmetry breaking is somewhat softer for the tight-binding model than for the graphene effective field theory.

 The fact that the numerically obtained weak-coupling limit of $\bar{\sigma}\lr{m \rightarrow 0}$ for $\kappa \, {\Delta \tau} = 0.15$ and $\kappa \, {\Delta \tau} = 0.20$ deviates from the result of analytical calculation (solid horizontal line on Fig. \ref{fig:conductivity_m0}) suggests that our extrapolation procedure tends to overestimate the value of the conductivity, with extrapolation error being as large as $10 \%$ at $\kappa \, {\Delta \tau} \gtrsim 0.15$. On the other hand, at $\kappa \, {\Delta \tau} \lesssim 0.1333$ numerical results in the weak-coupling regime agree nicely with the analytical result obtained in the non-interacting model.

 We also note that in the weak-coupling limit the extrapolated conductivities $\bar{\sigma}\lr{m \rightarrow 0}$ differ quite significantly for different temperatures, in contrast to the result (\ref{free_smeared_conductivity}) which was obtained in Appendix \ref{appsec:free_spectral_func} in the Dirac approximation. This indicates that the deviations from the linear dispersion law $E = v_F \, k$ at $E \gtrsim \kappa$ and the finite width of the valence band of graphene (with $E_{max} = \sqrt{9 \kappa^2 + m^2}$) are still important at our values of the temperature.

\section{Discussion and conclusions}
\label{sec:conclusions}

 In this paper we have presented the results of numerical studies of the tight-binding model of graphene with Coulomb interaction. We have assumed that the strength of Coulomb interaction is controlled by substrate dielectric permittivity $\epsilon$, so that the QED coupling constant $\alpha_0 \approx 1/137$ is multiplied by the factor $2/\lr{\epsilon + 1}$.

 Our results indicate that for sufficiently strong Coulomb interaction, that is, at substrate dielectric permittivities $\epsilon \lesssim 4$, and at sufficiently small temperatures ($T \lesssim 0.28 \, \kappa$) the symmetry between the two simple sublattices of the hexagonal lattice of graphene is spontaneously broken by a nonzero expectation value $\vev{ \Delta_N }$ of the difference of the numbers of particles localized on the sites of each sublattice. At the critical value $\epsilon \approx 4$, the susceptibility $\chi_N$ of $\vev{ \Delta_N }$ as well as the connected part of the dispersion of $\Delta_N$ $\cev{\Delta_N^2}_{conn.}$ have distinct peaks indicative of a second-order phase transition. This result agrees with the results of simulations of graphene effective field theory with staggered Dirac fermions \cite{Lahde:09:1, Lahde:09:2, Lahde:09:3, Lahde:10:1, Lahde:11:1, Hands:08:1, Hands:10:1, Hands:11:1, Giedt:11:1, Ulybyshev:12:1}. Our estimate of the critical value $\epsilon_c = 4 \pm 1$ corresponds to the effective QED coupling constant $\alpha = \frac{e^2}{v_F} \, \frac{2}{\epsilon + 1} = 0.9 \pm 0.2$, which agrees with the value $\alpha_c = 1.11 \pm 0.06$ obtained in \cite{Lahde:09:1, Lahde:09:2, Lahde:09:3, Lahde:11:1, Ulybyshev:12:1}. This fact suggests that lattice artifacts of staggered fermions have no significant effect on the position of the semimetal-insulator phase transition.

 At higher temperature ($T = 0.42 \, \kappa$) sublattice symmetry is still broken, but the transition to the broken phase becomes softer and looks more like a crossover. In particular, in this case only $\cev{\Delta_N^2}_{conn.}$ has a characteristic peak, and $\chi_N$ is a monotonic function of $\epsilon$. At even higher temperature, $T = 0.56 \, \kappa$, there are no signatures of spontaneous symmetry breaking. It seems therefore that at $T = 0.42 \, \kappa$ we are in the vicinity of the finite-temperature phase transition at which sublattice symmetry is restored.

 In order to quantify the electronic transport properties of graphene, we have considered the low-frequency conductivity $\bar{\sigma}$ smeared over frequencies $w \lesssim T$ according to (\ref{conductivity_midpoint}). We have found that for all temperatures which we have considered the conductivity $\bar{\sigma}$ quickly decreases with $\epsilon$ at $\epsilon \lesssim 4$ down to a finite value which is around $20-30 \%$ of its weak-coupling limit for the strongest coupling ($\epsilon = 1$, which corresponds to suspended graphene). It turns out that for the tight-binding model the low-frequency conductivity decreases somewhat slower than for the graphene effective field theory, which suggests that the insulator-semiconductor phase transition is somewhat weaker in our case. At $\epsilon \gtrsim \epsilon_c$, $\bar{\sigma}$ practically does not depend on $\epsilon$ and gradually decreases with temperature. Such behavior of the conductivity indicates that the phase transition associated with spontaneous symmetry breaking might persist at higher temperatures, but become weaker. For example, there could be a second-order phase transition at small temperatures ($T \lesssim 0.28 \, \kappa$) and a crossover at higher temperatures. A more accurate analysis of finite-temperature and finite-volume effects is required in order to classify the order of the observed phase transition and to obtain the corresponding critical exponents.

 Finally, let us comment on the obvious discrepancy between the results of lattice simulations (reported in the works \cite{Lahde:09:1, Lahde:09:2, Lahde:09:3, Lahde:10:1, Lahde:11:1, Giedt:11:1, Ulybyshev:12:1} and in this paper), which suggest that suspended graphene should be an insulator, and most recent experimental results \cite{Elias:11:1, Elias:12:1}, which find no signature of an insulating state in suspended graphene. As discussed in \cite{Wehling:11:1, Katsnelson:11:1}, Coulomb interactions in graphene can be additionally screened both by valence electrons and by electrons on other orbitals of carbon atoms. This screening becomes stronger at small momenta and can effectively decrease the electromagnetic coupling constant by a factor of two or larger. Due to the screening, the critical value of the substrate dielectric permittivity might eventually become less than $\epsilon_c = 1$, which would make the insulator-semimetal phase transition inobservable in the real world. Since our simulations automatically take into account the screening of Coulomb interactions by valence electrons, this effect might be explained by the influence of electrons on other orbitals. As well, the size of the lattices which we use for our simulations might be too small, so that the momentum scale at which the screening becomes sufficiently strong is not yet reached \footnote{We thank Prof. Dr. M. I. Katsnelson for pointing this to us.}. Another possible cause of this discrepancy is the wrong value of the on-site interaction potential $u_0$, which is also a free parameter of the tight-binding model. If we couple the tight-binding model to the lattice gauge field, as in (\ref{tb_hamiltonian_peierls}), the value of $u_0$ is fixed by the choice of the lattice action and might be quite different from its physical value. All these conjectures require separate investigations.

\begin{acknowledgments}
 The authors are much obliged to Dr.~M.~I.~Katsnelson, Dr.~T.~Lahde, Dr.~O.~V.~Pavlovsky, Dr.~B.~Rosenstein, Dr.~M.~V.~Ulybyshev and Dr.~M.~A.~Zubkov for interesting and useful discussions. The work was supported by the Russian Ministry of Science and Education under contract No. 07.514.12.4028 and by the Grant RFBR-11-02-01227-a of the Russian Foundation for Basic Research. Numerical calculations were performed at the ITEP computer systems "Graphyn" and "Stakan" (authors are much obliged to A.~V.~Barylov, A.~A.~Golubev, V.~A.~Kolosov, I.~E.~Korolko, M.~M.~Sokolov for the help), the MVS 100K at Moscow Joint Supercomputer Center and at Supercomputing Center of the Moscow State University. The work of P.V.B. was also supported by the S. Kowalewskaja award from the Alexander von Humboldt foundation.
\end{acknowledgments}

\appendix

\section{Eigenspectrum of the tight-binding Hamiltonian on the finite lattice}
\label{appsec:spectrum}

 In this Appendix we discuss the spectra of the free single-particle Hamiltonian (\ref{one_part_ham_action}) and the fermion hopping matrices $M_{\sigma}$ introduced in (\ref{fermion_lattice_action}) on lattices of finite size.

 Invariance of the one-particle Hamiltonian (\ref{one_part_ham_action}) under translations implies that its eigenfunctions take the form
\begin{eqnarray}
\label{eigenfunctions}
  \psi_{\zeta}\lr{\alpha, \xi; q} = \mathcal{N}_{\alpha, \zeta}\lr{q} \, \expa{i \, q \xi},
 \nonumber \\
  \psi_{\zeta}\lr{\beta, \xi; q}  = \mathcal{N}_{\beta, \zeta}\lr{q}  \, \expa{i \, q \xi}  ,
\end{eqnarray}
where we have introduced an additional label $\zeta$ to distinguish between states with equal momenta but different energies. The components of the wave vector $q$ in Cartesian coordinates are
\begin{eqnarray}
\label{wave_vector_cartesian}
  k_x = \frac{q_1}{\sqrt{3} \, a},
 \quad
  k_y = \frac{2 \, q_2}{3 \, a} - \frac{q_1}{3 \, a} .
\end{eqnarray}

 It is easy to check that the functions (\ref{eigenfunctions}) are the eigenfunctions of the one-particle Hamiltonian (\ref{one_part_ham_action}) with the eigenvalues
\begin{eqnarray}
\label{eigenvalue}
 E_{\zeta}\lr{q} \equiv \zeta \, E\lr{q} = \zeta \sqrt{m^2  +  \kappa^2 \, |\Phi\lr{q}|^2}  ,
\end{eqnarray}
where $\zeta$ takes values $\zeta = \pm 1$ and
\begin{eqnarray}
\label{Phi_def}
 \Phi\lr{q} = \sum\limits_{b} e^{i q \rho_b}
 =
 1 + e^{-i q_1 + i q_2} + e^{-i q_1}  .
\end{eqnarray}
The ratio of the normalization coefficients $\mathcal{N}_{\alpha, \zeta}\lr{q}$ and $\mathcal{N}_{\beta, \zeta}\lr{q}$ is:
\begin{eqnarray}
\label{norm_coeff_ratio}
 \mathcal{N}_{\beta, \zeta}\lr{q}/\mathcal{N}_{\alpha, \zeta}\lr{q} = \frac{m - E_{\zeta}\lr{q}}{\kappa \, \Phi\lr{q}}
\end{eqnarray}
This equation and the normalization condition $|| \psi\lr{q} ||^2 = L_x \, L_y \, \lr{ |\mathcal{N}_{\alpha, \zeta}\lr{q}|^2 + |\mathcal{N}_{\beta, \zeta}\lr{q}|^2 }$ fix the values of the normalization coefficients:
\begin{eqnarray}
\label{norm_coeff}
 \mathcal{N}_{\alpha, \zeta}\lr{q} =
 \sqrt{\frac{E\lr{q} + \zeta \, m}{2 \, E\lr{q} \, L_x \, L_y }}  ,
\nonumber \\
 \mathcal{N}_{\beta, \zeta}\lr{q} = -\zeta \, e^{-i \vartheta\lr{q}} \,
 \sqrt{\frac{E\lr{q} - \zeta \, m}{2 \, E\lr{q} \, L_x \, L_y }}  ,
\end{eqnarray}
where $\vartheta\lr{q} = \arg \, \Phi\lr{q}$.

 Dirac points correspond to lattice momenta $q$ with $\Phi\lr{q} = 0$. This condition is equivalent to the two equations
\begin{eqnarray}
\label{dirac_point}
 \cos{q_1} + \cos{q_2} + 1 = 0, \quad \sin{q_1} = - \sin{q_2} .
\end{eqnarray}
Solving these equations, we find two Dirac points $q^{\lr{\pm}}$ with $q_1^{\lr{\pm}} = \pm \frac{2 \pi}{3}$, $q_2^{\lr{\pm}} = \mp \frac{2 \pi}{3}$. Linear expansion of the dispersion relation (\ref{eigenvalue}) with $m = 0$ near these points leads to the well-known result $E\lr{k} = v_F |k|$ with the Fermi velocity $v_F = 3/2 \, \kappa \, a$. The largest eigenvalue of the single-particle Hamiltonian (\ref{one_part_ham_action}) is
\begin{eqnarray}
\label{max_energy}
 E_{max} \equiv E_{+}\lr{0} = \sqrt{m^2 + 9 \kappa^2}  .
\end{eqnarray}

 We should also take into account the boundary conditions (\ref{rhombic_coords_bc}), which constrain possible values of $q$:
\begin{eqnarray}
\label{spectrum_constraints1}
 q_1 L_x = 2 \pi \, m_1, \quad m_1 \in \mathbb{Z}
 \nonumber \\
 q_2 L_y - q_1 L_y/2 = 2 \pi \, m_2, \quad m_2 \in \mathbb{Z}  .
\end{eqnarray}
Expressing $q_1$, $q_2$ in terms of $m_1$ and $m_2$ we find
\begin{eqnarray}
\label{spectrum_constraints}
 q_1 = \frac{2 \pi \, m_1}{L_x},
 \quad 
 q_2 = \frac{2 \pi \, m_2}{L_y} + \frac{2 \pi \, m_1}{2 L_x}  ,
\end{eqnarray}
or, in Cartesian coordinates,
\begin{eqnarray}
\label{spectrum_constraints_cartesian}
  k_1 = \frac{2 \pi \, m_1}{\sqrt{3} \, a \, L_x} ,
 \quad
  k_2 = \frac{2 \pi \, m_2}{3 \, a \, L_y/2} .
\end{eqnarray}
The filling of the graphene Brillouin zone with discrete lattice momenta is illustrated on Fig. \ref{fig:spectrum} in Subsection \ref{subsec:action_defs}. To obtain the complete system of eigenfunctions, it is sufficient to take $m_1 = 0 \ldots L_x - 1$, $m_2 = 0 \ldots L_y - 1$. It is clear from (\ref{spectrum_constraints}) that the Dirac points are only matched by discrete lattice momenta only if the lattice size $L_x$ is a multiple of $3$ and $L_y$ is a multiple of $2$.

 The eigenvalues and the eigenfunctions of the fermionic hopping matrices $M_{\sigma}$ in the absence of interactions are
\begin{eqnarray}
 \label{fermion_hopping_eigenfunc1}
 \psi_M\lr{s, \xi, \tau; \zeta, q, w} = e^{i w \tau} \, \psi_{\zeta}\lr{s, \xi; q}  ,
 \nonumber \\
 \label{fermion_hopping_eigenval}
 \lambda_M\lr{\zeta, q, w} = 1 - e^{i w {\Delta \tau}} \, \lr{1 - E_{\zeta}\lr{q} \, {\Delta \tau}},
\end{eqnarray}
where $\psi_{\zeta}\lr{s, \xi; q}$ are the eigenfunctions of the one-particle Hamiltonian (\ref{eigenfunctions}). We note that for zero-energy states with $E_{\zeta}\lr{q} = 0$ there is only one zero eigenvalue $\lambda_M\lr{\zeta, q, w}$ of $M_{\sigma}$ within the Brillouin zone $w \in \lrs{0, \frac{2 \pi}{\Delta \tau}}$, namely, at $w = 0$. Correspondingly, fermion propagator in momentum space has only one pole. Thus the fermionic action (\ref{fermion_lattice_action}) indeed describes a single fermion for each spin component $\sigma = \uparrow, \downarrow$, and there is no fermion doubling problem.

\section{Hopping expansion for the path integral representation of the partition function of the tight-binding model}
\label{appsec:hopping}

 In this Appendix we consider the representation of the partition function (\ref{partition_function1}) of the tight-binding model in terms of fermion worldlines in Euclidean space and discuss in more details the meaning of different approximations made in the derivation of the fermionic lattice action (\ref{fermion_lattice_action}). With the help of the world-line representation, we will prove that fluctuations of the electrostatic potential $\phi\lr{s, \xi, \tau, z}$ cannot close the gap in the energy spectrum of the tight-binding model (\ref{tb_hamiltonian_peierls}). In addition, we will show that the lattice action (\ref{fermion_lattice_action}) satisfies reflection positivity \cite{MontvayMuenster}, which is important for the self-consistency of the lattice regularization.

 Our starting point is the fermionic path integral with the weight which is a product of the factors (\ref{ferm_matrix_el1}) and the weights $\expa{ -\sum\limits_{s, \xi, \sigma} \bar{\eta}_{\sigma}\lr{s, \xi} \, \eta_{\sigma}\lr{s, \xi} }$ which come from the integral over the fermionic coherent states (\ref{fermionic_coherent_states}) in the decomposition of identity (\ref{fermionic_identity_decomposition}). The fermionic action can be then written as
\begin{eqnarray}
\label{ferm_action_appendix}
  S_{tb} = \sum\limits_{\sigma=\uparrow, \downarrow} \,
  \sum\limits_{\tau, \tau'}
  \bar{\eta}_{\sigma}\lr{\tau} \,
  M_{\sigma}\lr{\tau, \tau'} \,
  \bar{\eta}_{\sigma}\lr{\tau'}  ,
\end{eqnarray}
where
\begin{eqnarray}
\label{ferm_matrix_appendix}
 M_{\sigma}\lr{\tau, \tau'} = \delta_{\tau, \tau'}
 - \delta_{\tau, \tau'-1} \, \expa{-h \, {\Delta \tau} \pm i \phi_{\tau}}
\end{eqnarray}
and we have omitted the spatial coordinates $s, \xi$, treating the blocks of the fermion hopping matrix $M_{\sigma}\lr{\tau, \tau'}$ at fixed $\tau, \tau'$ as operators which act on the space of single-particle wave functions $\psi\lr{s, \xi}$. In order to account for the anti-periodic boundary conditions, the sign of the term proportional to $\delta_{\tau, \tau'-1}$ should be changed at, say, $\tau = L_{\tau} - 1$.

\begin{figure}[h!tb]
  \includegraphics[width=7cm]{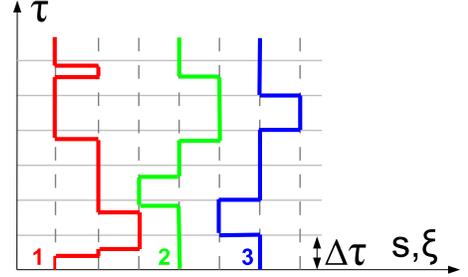}\\
  \caption{Fermionic worldlines which contribute to the partition function (\ref{partition_function1}) with different approximations for the fermionic path integral. World-line 1 corresponds to the full single-particle transfer matrix $\expa{-h \, {\Delta \tau} \pm i \phi\lr{\tau}}$. Expansion up to the first order in ${\Delta \tau}$ (\ref{ferm_matrix_exp1}) allows only world-lines which hop once per Euclidean time interval ${\Delta \tau}$, such as world-line 2. If the integration over $u$ in (\ref{ferm_matrix_exp1}) is omitted, the worldlines can only hop at $\tau = n {\Delta \tau}$, $n \in \mathbb{Z}$, as the world-line 3.}
  \label{fig:hopping_expansion}
\end{figure}

 After integrating over the fermions, the partition function (\ref{partition_function1}) can be represented in the form (\ref{path_int_final}). Consider now the logarithm of the determinant of $M_{\sigma}$ in (\ref{path_int_final}), assuming that it has the form (\ref{ferm_matrix_appendix}). Using the identity $\log \det{M_{\sigma}} = \tr\log\lr{M_{\sigma}}$ and taking into account the special form of the fermion hopping matrix $M_{\sigma}\lr{\tau, \tau'}$ with respect to $\tau$ and $\tau'$, we obtain
\begin{eqnarray}
\label{hopping_expansion1}
 \log \det{M_{\sigma}}
 = \nonumber \\ =
 \log \det{1 + \prod\limits_{\tau/{\Delta \tau} = 0}^{L_{\tau}-1}
  \expa{-h \, {\Delta \tau} \pm i \phi\lr{\tau}} }
 = \nonumber \\ =
\sum\limits_{n=0}^{+\infty} \frac{\lr{-1}^n}{n}
 \tr\lr{
  \prod\limits_{\tau/{\Delta \tau} = 0}^{L_{\tau}-1}
  \expa{-h {\Delta \tau} \pm i \phi\lr{\tau}}
 }^{n}  .
\end{eqnarray}
The last expression can be interpreted as a sum over all possible configurations of a single fermionic world-line which wraps $n$ times on the Euclidean torus with period $\lr{k T}^{-1}$ \cite{Wiese:96:1, Gutmann:92:1}, as illustrated on Fig.~\ref{fig:hopping_expansion}. For the world-line which originates from the site $s, \xi$ at time $\tau$ and goes to the site $s', \xi'$ at time $\tau + {\Delta \tau}$ the weight is multiplied by the element $\lrs{\expa{-h \, {\Delta \tau} \pm i \phi\lr{\tau}}}\lr{s, \xi; s', \xi'}$ of the single-particle transfer matrix. The factor $\lr{-1}^n/n$ accounts for the Fermi statistics and compensates the over-counting of the world-line configurations due to $\sim n$ different ways to choose the starting time of the path. Obviously, the full determinant $\det{ M_{\sigma} } = \expa{ \log\det{ M_{\sigma} } }$ can be represented as a sum over any number $N$ of fermionic world-lines, each coming with the weight (\ref{hopping_expansion1}). As usual, an additional factor $1/N!$ coming from the expansion of the exponent then compensates for $N!$ permutations of identical world-lines.

 If we expand the exponential $\expa{-h \, {\Delta \tau} \pm i \phi}$ to the first order in ${\Delta \tau}$, as in (\ref{ferm_matrix_exp1}), the worldlines are allowed to hop only once in the interval $\lrs{\tau, \tau + {\Delta \tau}}$. Each such hop changes the weight of the world-line by $ - h\lr{s, \xi; s', \xi'} \, {\Delta \tau}$. For the tight-binding Hamiltonian (\ref{tb_hamiltonian}), this means that only hops to nearest-neighbor sites on the hexagonal lattice are allowed with the weight $\kappa \, {\Delta \tau}$. Time-like segments of worldlines contribute with the weight $1 \pm m \, {\Delta \tau}$ per time ${\Delta \tau}$ depending on sublattice index $s$. In addition, each world-line $s\lr{\tau}, \xi\lr{\tau}$ acquires the complex phase $\expa{\frac{\pm i}{\Delta \tau} \, \int d\tau \phi\lr{s\lr{\tau}, \xi\lr{\tau}, \tau}}$ due to the presence of the electrostatic potential.

 Integration over $u$ in (\ref{ferm_matrix_exp1}) means that hops can happen at any time $\tau + u \, {\Delta \tau}$, $u \in \lrs{0, 1}$ (see Fig.~\ref{fig:hopping_expansion}, world-line 2). Correspondingly, before the moment $\tau + u \, {\Delta \tau}$ the fermion interacts with the electrostatic potential $\phi\lr{s, \xi, \tau, z=0}$, and after that - with the potential $\phi\lr{s', \xi', \tau, z=0}$. A simple estimate shows that if we neglect integration over $u$ for two fermionic world-lines at distance $r$ and replace the factors in (\ref{ferm_matrix_exp1}) either by $e^{i \phi\lr{s, \xi, \tau, z=0}}$ or $e^{i \phi\lr{s', \xi', \tau, z=0}}$, this results in the correction of order $O\lr{\kappa {\Delta \tau} \, e^2 a {\Delta \tau}/r^2}$ to the path integral weight. Since we keep only the leading-order terms in $\kappa {\Delta \tau}$ and $e^2 a/r$, we can discard this term. However, if one uses some sort of improved action for the electromagnetic field which reproduces the Coulomb potential with higher accuracy, it might be necessary to keep the integration over $u$ in (\ref{ferm_matrix_exp1}).

 Let us now prove that the fluctuations of the electrostatic potential $\phi\lr{s, \xi, z=0, \tau}$ cannot close the gap in the spectrum of the tight-binding Hamiltonian (\ref{tb_hamiltonian_peierls}) at $m \neq 0$. According to the discussion above, the logarithm of the determinant of the fermionic hopping matrix $M_{\sigma}$ can be represented as the following sum over worldlines $C$:
\begin{eqnarray}
\label{hopping_inequality1}
 \log \det{M_{\sigma}} = \sum\limits_{C} W\lrs{C} \, e^{i \Phi\lrs{C}}  ,
\end{eqnarray}
where $W\lrs{C}$ is the positive and real weight and $e^{i \Phi\lrs{C}}$ is the phase which includes both the integral of the electrostatic potential along the world-line and the factor $\lr{-1}^n$ in (\ref{hopping_expansion1}). Due to the inequality $|\sum\limits_{i} \, a_i| < \sum\limits_{i} \, |a_i|$ one has
\begin{eqnarray}
\label{hopping_inequality2}
 |\log \det{M_{\sigma}}| \le \sum\limits_{C} W\lrs{C} .
\end{eqnarray}
By inverting the derivation above, it is easy to show that the latter sum corresponds to the bosonic partition function
\begin{eqnarray}
\label{hopping_inequality3}
\sum\limits_{C} W\lrs{C} = \log\det{1 - \expa{-h/T}}  ,
\end{eqnarray}
which is finite if the single-particle Hamiltonian $h$ has a gap. In Appendix \ref{appsec:spectrum} we have demonstrated that for $m \neq 0$ the spectrum of $h$ indeed has a gap. From the inequality (\ref{hopping_inequality2}) we see that $\log \det{M_{\sigma}}$ remains finite for any configuration of the electrostatic potential $\phi\lr{s, \xi, z, \tau}$, and thus the effective single-particle hamiltonian $h$ always has a gap.  The finiteness of $\log \det{M_{\sigma}}$ at $m \neq 0$ implies that $\det{M_{\sigma}} \neq 0$ and hence $M_{\sigma}$ is invertible, which is crucial for the Hybrid Monte-Carlo algorithm.

 Finally, let us consider the reflection positivity of our lattice action. In our case, it is equivalent to the positive definiteness of the single-particle transfer matrix $\expa{-h {\Delta \tau}}$ \cite{MontvayMuenster}. Its expansion $1 - h {\Delta \tau}$ up to the first order in ${\Delta \tau}$ is still positive-definite if the largest eigenvalue of $h$ does not exceed $\lr{\Delta \tau}^{-1}$. From (\ref{max_energy}) we see that this condition is equivalent to
\begin{eqnarray}
\label{kappa_inequality}
 \sqrt{m^2 + 9 \kappa^2} \, {\Delta \tau} < 1  .
\end{eqnarray}

\section{Current-current correlators in the free tight-binding model}
\label{appsec:free_spectral_func}

\begin{figure}[h!tb]
  \includegraphics[width=8.0cm]{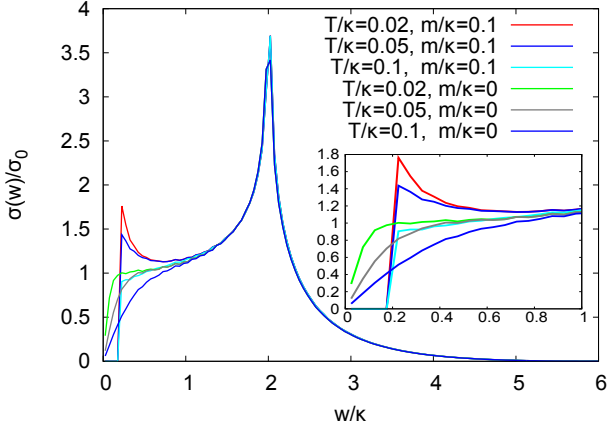}\\
  \caption{AC conductivity $\sigma^{\lr{0}}\lr{w}$ (in units of $\sigma_0 = 1/4 \, e^2/\hbar = \pi/2 \, e^{2}/h$) for the non-interacting tight-binding model of graphene (\ref{tb_hamiltonian_peierls}) with $L_x = L_y \rightarrow \infty$ at different temperatures $T$ and Dirac masses $m$. The inset shows the low-frequency region in a larger scale.}
  \label{fig:free_ac_conductivity}
\end{figure}

 In this Appendix we consider the current-current correlators (\ref{corr_euclide}) and the corresponding spectral functions for the tight-binding model (\ref{tb_hamiltonian_peierls}) without interactions with electromagnetic field. Our derivation is similar to that of \cite{Stauber:08:1}, but extends to the case of non-zero staggered potential $m$ in (\ref{tb_hamiltonian_peierls}). We start from the following general expression for the correlator of fermionic bilinear operators $\hat{J}_{A} = \sum\limits_{X, Y} j_{A; X, Y} \, \hat{\psi}^{\dag}_X \, \hat{\psi}_Y$ in a theory with bilinear Hamiltonian of the form $\hat{H} = \sum\limits_{X, Y} h_{X, Y} \, \hat{\psi}^{\dag}_X \, \hat{\psi}_Y$:
\begin{eqnarray}
\label{free_bilinear_corr_general}
 \mathcal{Z}^{-1} \tr\lr{ \hat{J}_A \, e^{-\tau \hat{H}} \, \hat{J}_B \, e^{-\lr{\beta - \tau} \hat{H} }   }
 = \nonumber \\ =
 \lr{ \tr\lr{j_A \, \frac{e^{-\beta h}}{1 + e^{-\beta h}} }} \,  \lr{ \tr\lr{j_B \, \frac{e^{-\beta h}}{1 + e^{-\beta h}} }}
 + \nonumber \\ + \,
 \tr\lr{j_{A} \, \frac{e^{-\lr{\beta - \tau} \, h}}{1 + e^{-\beta h}} \, j_{B} \, \frac{e^{-\tau \, h}}{1 + e^{-\beta h}}}
\end{eqnarray}
where $\mathcal{Z} = \tr\lr{ e^{-\beta \hat{H}} }$ and $\beta \equiv T^{-1}$. The trace on the left-hand side is taken over the full Hilbert space of the theory, and on the right-hand side - over the one-particle Hilbert space (as in (\ref{one_part_ham_action})). Applying the expression (\ref{free_bilinear_corr_general}) to the current-current correlator (\ref{corr_euclide}), we see that the contribution of the disconnected fermion diagrams (first summand on the right-hand side) is zero in this case. Evaluating the second term in the eigenbasis of the one-particle Hamiltonian (\ref{eigenfunctions}), we obtain
\begin{widetext}
\begin{eqnarray}
\label{free_bilinear_corr_eigenfunc}
 G^{\lr{0}}\lr{\tau}
 =
 \frac{1}{3 \sqrt{3} \, L_x L_y} \, \sum\limits_{\xi, \xi'} T_{bc} \,
 \tr\lr{ \hat{J}_b\lr{0} \, e^{-\tau \hat{H}} \, \hat{J}_c\lr{\xi} \, e^{-\lr{\beta - \tau} \hat{H}} }
 = \nonumber \\ =
 \sum\limits_{q, q', \zeta, \zeta'}
 \frac{2 T_{bc} \, j_{b, \zeta' \, \zeta}\lr{q', q} \, j_{c, \zeta \, \zeta'}\lr{q, q'} }{3 \sqrt{3} L_x \, L_y} \,
 \frac{e^{-\lr{\beta - \tau} \, E_{\zeta}\lr{q} } }{1 + e^{-\beta E_{\zeta}\lr{q}}}
 \frac{e^{-\tau \, E_{\zeta'}\lr{q'}}}{1 + e^{-\beta E_{\zeta'}\lr{q'}}}  ,
\end{eqnarray}
where the additional factor of two came from summation over spin indices and $j_{b, \zeta \zeta'}\lr{q, q'}$ is the matrix element of the one-particle operator $j_{\uparrow, \sigma}$ defined in (\ref{one_part_current_def}) between the eigenstates (\ref{eigenfunctions}) of the one-particle Hamiltonian (\ref{one_part_ham_action}):
\begin{eqnarray}
\label{current_matrix_element1}
 j_{b, \zeta \zeta'}\lr{q, q'} = \sum\limits_{\xi}
 i \kappa \, \bar{\psi}_{\zeta}\lr{\alpha, \xi; q} \psi_{\zeta'}\lr{\beta, \xi + \rho_b; q'}
 -
 i \kappa \, \bar{\psi}_{\zeta}\lr{\beta, \xi + \rho_b; q} \psi_{\zeta'}\lr{\alpha, \xi; q'}
 = \nonumber \\ =
 \lr{
 i \kappa \, \bar{\mathcal{N}}_{\alpha, \zeta}\lr{q} \, \mathcal{N}_{\beta, \zeta'}\lr{q} e^{i q \rho_b}
 -
 i \kappa \, \bar{\mathcal{N}}_{\beta, \zeta}\lr{q} \, \mathcal{N}_{\alpha, \zeta'}\lr{q} e^{-i q \rho_b}
 } \,
 L_x L_y \, \delta\lr{q, q'} .
\end{eqnarray}

 Since our goal is to obtain the AC conductivity (\ref{ac_conductivity}), we now contract the $b$ and $c$ indices of the correlator (\ref{free_bilinear_corr_eigenfunc}) with the matrix $T_{bc}$ introduced in (\ref{ac_conductivity}):
\begin{eqnarray}
\label{current_matrix_element2}
 \sum\limits_{b, c} T_{b c} \, j_{b, \zeta' \, \zeta}\lr{q', q} \, j_{c, \zeta \, \zeta'}\lr{q, q'}
 =
 3/2 \, \sum\limits_{b} j_{b, \zeta' \, \zeta}\lr{q', q} \, j_{b, \zeta \, \zeta'}\lr{q, q'}
 - 1/2 \sum\limits_{b, c} j_{b, \zeta' \, \zeta}\lr{q', q} \, j_{c, \zeta \, \zeta'}\lr{q, q'}
\end{eqnarray}
Explicit calculation yields the following expressions for the contracted matrix elements (\ref{current_matrix_element1}) in (\ref{current_matrix_element2}):
\begin{eqnarray}
\label{current_matrix_element3}
\sum\limits_b j_{b, \zeta \, \zeta'}\lr{q, q'} =
i/2 \, \delta\lr{q, q'} \, \lr{\zeta - \zeta'} \, \sqrt{E^{2}\lr{q} - m^2}
\end{eqnarray}
\begin{eqnarray}
\label{cme2}
\sum\limits_{b} j_{b, \zeta' \, \zeta}\lr{q', q} \, j_{b, \zeta \, \zeta'}\lr{q, q'}
= 
 \frac{\kappa^2 \, \delta\lr{q, q'}}{2 E^2\lr{q}} \, \lr{
 3 E^2\lr{q} - 3 \zeta \zeta' m^2
 -
 \lr{E^2\lr{q} - m^2} \, \zeta \zeta' \, \re\lr{
  e^{2 i \vartheta\lr{q}} \sum\limits_{b} e^{-2 i q \rho_b}
  }
 }
\end{eqnarray}
Thus the squared matrix elements of the current operator are diagonal with respect to the lattice momenta and depend only on the product $\zeta \zeta'$. Denoting
\begin{eqnarray}
\label{current_matrix_el_ansatz1}
 \frac{2}{3 \, \sqrt{3}} \, \sum\limits_{b, c} T_{b c} \, j_{b, \zeta' \, \zeta}\lr{q', q} \, j_{c, \zeta \, \zeta'}\lr{q, q'}
 =
 \begin{cases}
   F_{+-}\lr{q} \delta\lr{q, q'} & \zeta \neq \zeta' \\
   F_{++}\lr{q} \delta\lr{q, q'} & \zeta = \zeta'
 \end{cases}
\end{eqnarray}
with
\begin{eqnarray}
\label{current_matrix_el_ansatz2}
 F_{++}\lr{q} = \frac{\kappa^2 \, \lr{E^{2}\lr{q} - m^2}}{2 \sqrt{3} \, E^2\lr{q}}\,
 \lr{3 - \re\lr{
   e^{2 i \vartheta\lr{q}} \sum\limits_{b} e^{-2 i q \rho_b}
  }}
 \nonumber \\
 F_{+-}\lr{q} = \frac{\kappa^2}{2 \sqrt{3} \, E^2\lr{q}} \,
 \lr{3 E^2\lr{q} + 3 m^2 + \lr{E^2\lr{q} - m^2} \,
 \re\lr{
   e^{2 i \vartheta\lr{q}} \sum\limits_{b} e^{-2 i q \rho_b}
  }
 } - \frac{E^2\lr{q} - m^2}{3 \sqrt{3}}
\end{eqnarray}
we can represent the current-current correlator (\ref{free_bilinear_corr_general}) in the following form:
\begin{eqnarray}
\label{free_current_corr}
G^{\lr{0}}\lr{\tau}
=
\frac{1}{L_x \, L_y} \, \sum\limits_{q}
\frac{F_{+-}\lr{q} \, \cosh\lr{2 E\lr{q} \, \lr{\tau - \beta/2}} + F_{++}\lr{q}}{2 \, \cosh^2\lr{\beta E\lr{q}/2}}
\end{eqnarray}
Comparing this expression with (\ref{corr_eq}) and (\ref{kernel}), we can obtain the AC conductivity $\sigma^{\lr{0}}\lr{w}$:
\begin{eqnarray}
\label{free_spectral_func}
 \sigma^{\lr{0}}\lr{w}
 =
 \frac{2 \pi}{L_x \, L_y} \, \sum\limits_{q} F_{+-}\lr{q} \, \delta\lr{w - 2E\lr{q}}
 \, \frac{1}{2 w}  \, \tanh\lr{\frac{\beta w}{4}}
 + 
 \frac{\pi \, \beta \, \delta\lr{w}}{L_x \, L_y} \, \sum\limits_{q} \frac{F_{++}\lr{q}}{4 \, \cosh^2\lr{\beta E\lr{q}/2}}
\end{eqnarray}
\end{widetext}
One can see that the conductivity has a delta-function singularity at $w = 0$, which is a common feature of all ideal crystals due to the absence of scattering in an infinite lattice without boundaries \cite{Katsnelson:06:1}. In the limit $m \rightarrow 0$ the non-singular part of the conductivity (first summand in (\ref{free_spectral_func})) reproduces the results of \cite{Stauber:08:1}. The finite part of $\sigma\lr{w}$ is shown on Fig. \ref{fig:free_ac_conductivity}. First, one can note a distinct peak at $w = 2 \, \sqrt{\kappa^2 + m^2}$, which corresponds to the singularity in the density of states associated with the saddle point at $E = \sqrt{\kappa^2 + m^2}$ (point $M$ in Fig. \ref{fig:spectrum}) in the dispersion relation. At $m \neq 0$ and at sufficiently low temperatures, a second peak appears at $w = 2 m$. In the intermediate frequency range, $m, T \ll w \ll \kappa$, the conductivity approaches the value $\sigma = \sigma_0 = 1/4$ ($1/4 \, e^2/\hbar = \pi/2 \, e^{2}/h$ in physical units).

 Let us now explicitly evaluate the expression (\ref{free_spectral_func}) in the case when $w$ is close to the threshold value $w = 2 m$, so that only the momenta which are close to the Dirac points $q^{\lr{\pm}}$ contribute to the summation over $q$ in (\ref{free_current_corr}) and (\ref{free_spectral_func}). In this case we can approximate $E\lr{q}$ as $\sqrt{m^2 + v_F^2 \, k^2}$, where $k$ is the momentum in Cartesian coordinates (\ref{wave_vector_cartesian}). Assuming that the lattice size is sufficiently large, we can also replace summation over $q$ in (\ref{free_current_corr}), (\ref{free_spectral_func}) (that is, summation over $m_1$, $m_2$ with $q$ given by (\ref{spectrum_constraints})) by integration over $k$: $\frac{2 \pi}{L_x L_y} \, \sum\limits_{q} \approx 2 \, \frac{3 \sqrt{3} a^2}{4 \pi} \int d^2 k$, where the factor of two accounts for two Dirac points. A simple calculation shows that to the leading order in ${\delta q} = q - q^{\lr{\pm}}$ the contributions of the terms proportional to $\re\lr{e^{2 i \vartheta\lr{q}} \sum\limits_{b} e^{-2 i q \rho_b}}$ have opposite signs for $q$ close to $q^{\lr{+}}$ and $q^{\lr{-}}$ and thus cancel in the sum over $q$ in (\ref{free_current_corr}) and (\ref{free_spectral_func}) (see also \cite{Stauber:08:1}). Moreover, it is easy to show that the contribution of the last summand in the expression (\ref{current_matrix_el_ansatz2}) for $F_{+-}\lr{q}$ is suppressed by an additional power of $\kappa \, a$ and thus can be also omitted in the Dirac approximation. With such simplifications, we can change the integration variable from $k$ to $E$. Integrating out the delta-function in (\ref{free_spectral_func}), we then arrive at the following expression for the finite part of the AC conductivity (\ref{free_spectral_func}) for $0 < \lr{w - 2 m} \ll \kappa$:
\begin{eqnarray}
\label{free_spectral_func_threshold}
 \sigma^{\lr{0}}\lr{w}
 \approx
 \frac{1}{4} \, \lr{1 + \frac{4 m^2}{w^2}} \,
 \tanh\lr{\frac{\beta w}{4}}
\end{eqnarray}
If we take the limit $\beta \rightarrow \infty$ and $m \rightarrow 0$ at finite $w$ and then take $w = 0$, we obtain the limiting value of the DC conductivity of graphene $\sigma_0 = 1/4$, which is $\sigma_0 = e^2/\lr{4 \hbar} = \pi e^{2}/\lr{2 h}$ in SI units (see \cite{Katsnelson:06:1} for a discussion of the physical meaning of this result). From Fig. \ref{fig:free_ac_conductivity} one can see that when $m = 0$, this limiting value is reached at small frequencies $w \ll \kappa$ for temperatures $T = 0.05 \, \kappa$ and lower. On the other hand, the value of the conductivity at the threshold $w = 2 m$ is always two times larger than $\sigma_0$.

 Within the Dirac approximation one can also explicitly calculate the smeared low-frequency conductivity $\bar{\sigma}^{\lr{0}}$ (introduced in (\ref{conductivity_midpoint})) in the absence of interactions:
\begin{eqnarray}
\label{free_smeared_conductivity}
 \bar{\sigma}^{\lr{0}} = \frac{\beta^2}{\pi} \, G^{\lr{0}}\lr{\beta/2}
 = \nonumber \\ =
 \frac{\beta^2}{\pi \, L_x \, L_y} \,
 \sum\limits_{q} \frac{F_{+-}\lr{q} + F_{++}\lr{q}}{2 \cosh^2\lr{\beta E\lr{q}/2}}
 \approx \nonumber \\ \approx
 \frac{9 \, a^2 \, \kappa^2 \, \beta^2}{4 \pi^2} \int d^2k \, \frac{1}{\cosh^2\lr{\beta E\lr{k}/2}}
 = \nonumber \\ =
 \frac{\beta^2}{\pi^2} \, \int\limits_{m}^{+\infty} \frac{dE \, E}{\cosh^2\lr{\beta E/2}}
 = \nonumber \\ =
 \frac{4}{\pi^2} \, \log\lr{1 + e^{-\beta m}} + \frac{4 \, \beta \, m}{\pi^2 \, \lr{1 + e^{\beta m}}}
\approx \nonumber \\ \approx
\frac{4 \log 2}{\pi^2} - \frac{\lr{\beta m}^2}{2 \pi^2} + O\lr{\lr{\beta m}^4}
\end{eqnarray}
It is interesting to note that the value of $\bar{\sigma}^{\lr{0}}$ in the limit $m \rightarrow 0$ $\bar{\sigma}^{\lr{0}} = \frac{4 \log 2}{\pi^2} = 0.281$ is quite close to the commonly quoted value $\sigma_0 = 1/4$ and is also temperature-independent.


\end{document}